\documentclass[12pt]{article}
\usepackage[utf8]{inputenc}
\usepackage{amsmath,amssymb,amsthm,enumitem,xcolor,caption,subcaption}
\usepackage{jheppub}

\newcommand{\be}{\begin{equation}}
\newcommand{\ee}{\end{equation}}
\newcommand{\bi}{\begin{itemize}}
\newcommand{\ei}{\end{itemize}}

\newcommand{\R}{\mathbf{R}}

\newcommand{\GN}{G_{\rm N}}

\newcommand{\vr}{\vec{r}}

\title{Evaporating universes}
\author{Divij Gupta}

\affiliation{Department of Physics, University of Illinois, Urbana IL 61801, USA}
\affiliation{Martin Fisher School of Physics, Brandeis University, Waltham MA 02453, USA}

\emailAdd{divijg3@illinois.edu}

\abstract{Recent work by Headrick, Sasieta and myself provides an extension of the HRT formula for asymptotically flat spacetimes. I use this formula to construct a holographic model of black hole evaporation in four-dimensional asymptotically flat spacetimes using Brill-Lindquist (BL) wormholes. The wormhole is interpreted via ER=EPR to represent the entanglement geometry between an evaporating black hole and baths into which the Hawking radiation is collected. Applying HRT, I compute the entanglement entropy by numerically computing the areas of the minimal surfaces, which is shown to obey the Page curve, consistent with information conservation. Numerical analysis is done for both three and four-boundary BL wormholes ($n=3,4$). Index-1 surfaces in the wormhole interior are interpreted as the candidate bulges involved in the python's lunch conjecture (PLC), and their areas are used to compute the restricted complexity $\mathcal{C}$ of decoding the Hawking radiation. The results are compared to the time-dependent predictions of the PLC.
\newline
\newline
A video abstract is available at https://youtu.be/6-B9zFgx40U}
\begin{document}

\maketitle
\section{Introduction}
\label{sec:introduction}

The development of the AdS/CFT dictionary \cite{Maldacena:1997re} has yielded many fruitful insights into subjects ranging from entanglement wedge reconstruction to the emergence of spacetime \cite{Czech:2012bh,Almheiri:2014lwa,Jafferis_2016,Dong_2016,VanRaamsdonk:2010pw}. Tying together many of these ideas is the Ryu-Takayanagi (RT) formula \cite{Ryu:2006bv}, which formalizes the connection between the bulk AdS geometry and quantum entanglement in the CFT. The formula reads
\begin{equation}
    \label{eq:RT}
    S(A) = \frac{|\gamma_{\rm RT} (A)|}{4 G_{\rm N}},
\end{equation}
where $S(A) = - \rm {Tr}(\rho_A \log{\rho_A})$ is the entanglement entropy of the CFT restricted to the spatial subregion $A$. $\gamma_{\rm RT}$ is the RT surface, defined to be the minimal-area surface homologous to the boundary subregion $A$, with $|\cdot |$ denoting its area.

The RT formula, along with its covariant extension the HRT formula \cite{Hubeny:2007xt}, can also be thought of as a particular manifestation of ER=EPR \cite{erEpr}. ER=EPR however is expected to hold beyond asymptotically AdS spacetimes, and thus so should the HRT formula. This motivated a recent proposal extending the HRT formula to asymptotically flat spacetimes \cite{gupta2025entangleduniverses}, which can be used to study flat spacetime versions of systems that were previously analyzed via AdS/CFT. One such application is to model black hole evaporation. Here the entangled black hole-radiation system is dual to a connected wormhole, referred to as a ``quantum octopus". More concretely, one may take the entangled system at an instant of time (a snapshot) and collapse the radiation quanta into black holes. These `radiation black holes' are entangled with the evaporating black hole and via ER=EPR form a connected wormhole geometry.

\begin{figure}[ht]
    \centering
    \includegraphics[scale = 0.45]{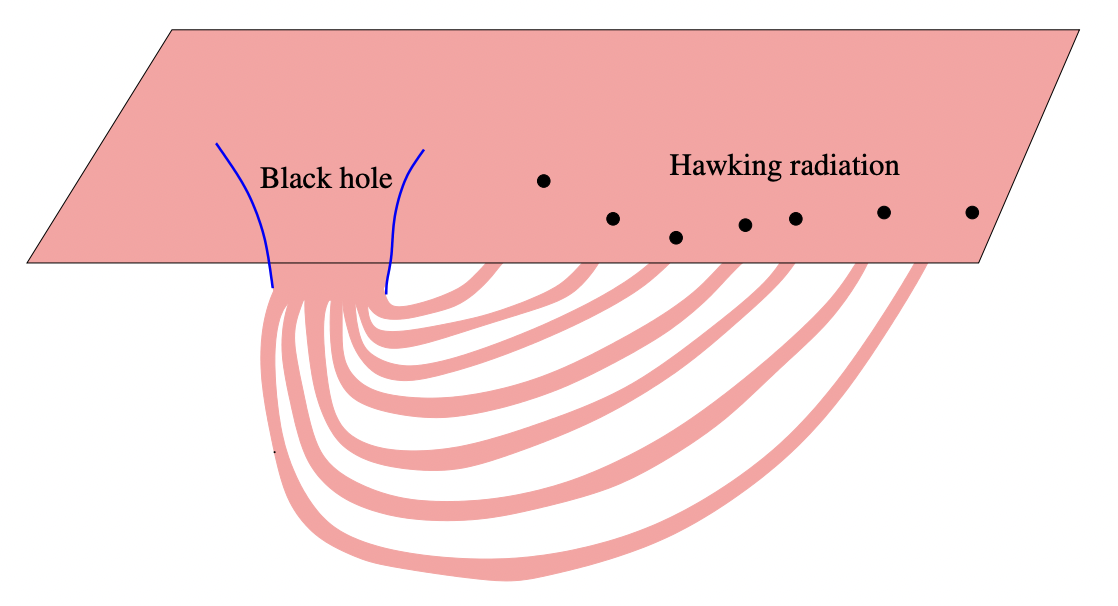}
    \caption{Quantum octopus for an evaporating black hole. Though the radiation and black hole exist in the same connected region, we take them to be sufficiently far apart such that they are approximated to be living on disconnected sheets (as in a wormhole). \textit{Reproduced from \cite{erEpr}}.}
    \label{fig:er=epr}
\end{figure}

The utility of this geometrical description of the black hole-radiation system lies in the fact that the entanglement entropy between the black hole and radiation can now be computed via the RT formula by finding the minimal surfaces in the geometry. The entanglement entropy is important as its behavior is at the heart of the black hole information paradox. Hawking's original computation using semiclassical methods \cite{hawkingradiation} opened the question of whether information was conserved or destroyed in the evaporation process. Hawking himself later argued that the evaporation process violates unitarity, which would result in the entanglement entropy increasing monotonically with time \cite{Hawking:1976ra}. Page instead argued the process should be consistent with unitarity, and thus the entanglement entropy must have a turning point, known as the Page time, when the entropy begins to decrease \cite{page1,page2}. Recent work suggests that applying the quantum extremal surface (QES) formula \cite{Engelhardt_2015} to models of black hole evaporation in AdS/CFT yields the Page curve for the entropy \cite{Almheiri:2019psf,Penington:2019npb,Almheiri_2020}, which is taken to be the litmus test for unitary evaporation.

\begin{figure}[ht]
    \centering
    \includegraphics[width=0.98\textwidth]{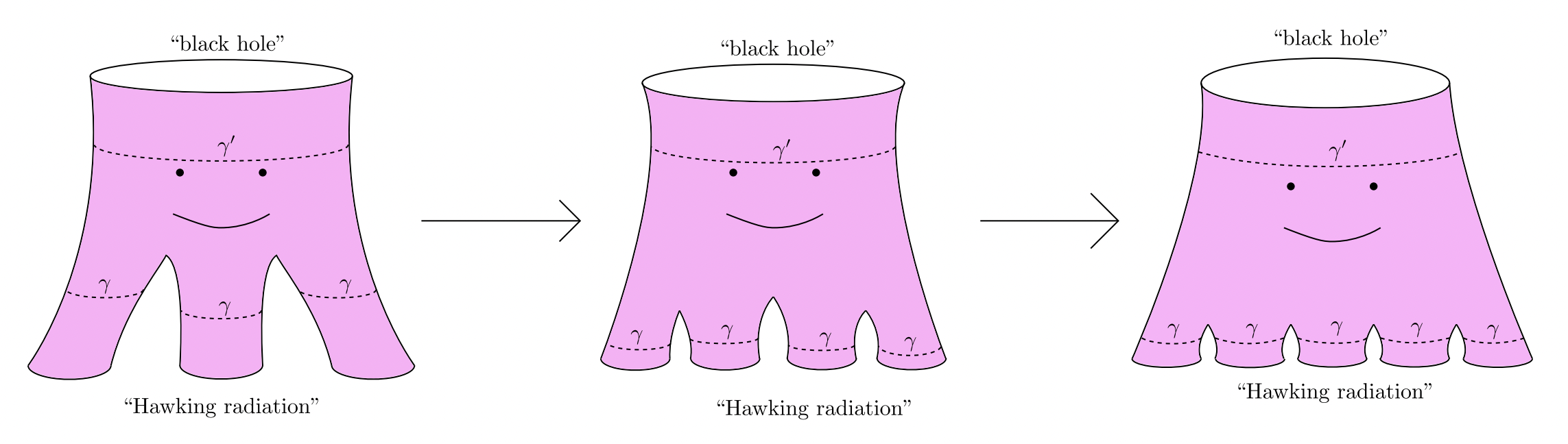}
    \caption{Entanglement wormhole structure in \cite{Akers:2019nfi}. With time, more black holes are emitted and entangled to the existing `octopus' geometry, which grows $n$ legs. \textit{Reproduced from \cite{Akers:2019nfi}.}}
    \label{fig:akers model}
\end{figure}

The models \cite{Almheiri:2019psf,Penington:2019npb,Almheiri_2020} however all involve a semi-classical AdS bulk theory describing the evaporating black hole which is coupled to an auxilliary system. A setup closer to the quantum octopus of ER=EPR is the \textit{classical octopus} of \cite{Akers:2019nfi} in AdS$_3$. The octopus, shown in Fig. \ref{fig:akers model}, represents the entanglement between the evaporating black hole (the `head' of the octopus) and the emitted radiation (the `legs' of the octopus). As the black hole evaporates, the head becomes smaller and the octopus `grows' legs. The candidate RT surfaces are the evaporating black hole horizon $\gamma'$ and the union of the radiation horizons $n\gamma$. \cite{Akers:2019nfi} showed that this model recovered the Page curve, thus representing a unitary model of black hole evaporation.

I describe this setup to introduce my own model of black hole evaporation inspired by \cite{Akers:2019nfi}. In this paper, I propose using the HRT conjecture of \cite{gupta2025entangleduniverses} to construct a toy model of black hole evaporation in Mink$_4$. The key objects used for the construction are four-dimensional Brill-Lindquist (BL) wormholes, which describe initial data with an arbitrary number of asymptotically flat regions. In section \ref{sec:preliminaries}, I discuss important preliminaries, including a brief overview of HRT in asymptotically flat spacetimes and minimal surfaces in BL wormholes, along with basic aspects of the (generalized) python's lunch conjecture. Section \ref{sec:model} introduces the model and describes the dynamics for BL wormholes with both three and four asymptotic regions ($n = 3, 4$), including an interpretation of index-1 surfaces as candidate bulge surfaces for the python's lunch conjecture. Finally in section \ref{sec:discussion}, I conclude by discussing additional points and future directions, including tensor network descriptions and extensions of the model.

\section{Preliminaries}
\label{sec:preliminaries}
\subsection{HRT in asymptotically flat spacetimes}
\label{sec:HRT flat}
I will begin by reviewing an extension of the HRT formula proposed for asymptotically flat multi-boundary wormholes by Headrick, Sasieta and myself (see conjecture 3 in \cite{gupta2025entangleduniverses}). In this extension, one assumes the existence of a quantum theory of gravity that admits a $D$-dimensional Minkowski (Mink$_D$) vacuum $V$ and associated Hilbert space $\mathcal{H}^V$. Consider the theory on $M$, a classical connected solution with $n$ asymptotically Mink$_D$ regions $a_1, \cdots, a_n$. \cite{gupta2025entangleduniverses} conjectures that $M$ represents a set $\mathcal{H}_M \subset \mathcal{H}^{a_1} \otimes \cdots \otimes \mathcal{H}^{a_n}$ of entangled states in the tensor product of the Hilbert spaces $\mathcal{H}^{a_i}$, each of which is given by quantizing the theory on region $a_i$ with Mink boundary conditions. Furthermore for any subset $A \subseteq \{a_1,\cdots,a_n\}$, the entanglement entropy between $A$ and $A^c$ is given by
\begin{equation}
    \label{eq:flat RT}
    S(A) = S(A^c) = \frac{|\gamma_{\rm RT}(A)|}{4 G_{\rm N}},
\end{equation}
where the entropy is computed for the entangled state $\rho_A \in \mathcal{H}_M$ and the RT surface $\gamma_{\rm RT}$ is defined the same way as in the original RT formula \eqref{eq:RT}.

\cite{gupta2025entangleduniverses} justifies \eqref{eq:flat RT} in part by considering an asymptotically flat spacetime $M$, then gluing AdS boundary regions to each Mink boundary region to yield a new spacetime $M'$. $M'$ is asymptotically AdS, thus allowing the application of the HRT formula. Strictly speaking, the HRT formula would now compute the entanglement entropy in the dual CFT Hilbert space $\mathcal{H}_{M'} \cong \mathcal{H}_{\rm CFT}$, but if the gluing procedure does not introduce any new minimal surfaces, the HRT surface would still lie in the unchanged region of the original spacetime $M$. The entanglement entropy computed via the HRT formula in AdS thus yields the same value as \eqref{eq:flat RT}. In other words, even though the ultraviolet structure of $\mathcal{H}_M, \mathcal{H}_{M'}$ may be different, the infrared structure remains unchanged with the gluing and we can conclude that the entanglement entropies of the two Hilbert spaces are equal, which motivates the use of \eqref{eq:flat RT} to compute the entanglement entropy in $\mathcal{H}_M$.

It should be noted that this conjecture does not rely on the existence of a dual theory $\mathcal{H}_{V'}$, whether that is a quantum field theory or another system. This can be seen by the fact that many of the non-trivial properties of the RT surface such as existence and causal wedge inclusion can be proved without reference to the dual theory, just as in the AdS case. Similarly, its area obeys non-trivial properties expected of entanglement entropies, such as the strong subadditivity and MMI inequalities, with the proofs once again being identical to the AdS case (see \cite{gupta2025entangleduniverses} for a more complete discussion of these consistency checks; see \cite{Wall:2012uf} for details of these proofs in AdS).

This is in contrast to many other proposed holographic dualities involving asymptotically flat spacetimes, such as Carrollian, Lifshitz, Galilean, celestial holography and many more \cite{Li:2010dr,Apolo:2020bld,Gentle:2015cfp,Bagchi:2014iea,Sato:2015tta,Doi:2022iyj,Narayan:2012ks,Narayan:2015vda,Narayan:2017xca,Narayan:2020nsc,Narayan:2022afv,Akal:2020wfl,Ogawa:2022fhy,Jiang:2017ecm,Wen:2018mev}. These proposals rely on the existence of a dual theory, usually endowed with a Hilbert space that admits spatial tensor factorizations. This factorization property is used to justify assigning \textit{subregions} of the boundary an entropy via a more powerful form of \eqref{eq:flat RT}, where the region $A$ may include boundary subregions. \cite{gupta2025entangleduniverses} instead aims to be a more conservative generalization of HRT by not making any reference to a dual theory, and as a result the formula \eqref{eq:flat RT} is restricted to only being applicable to (unions of) \textit{entire} boundary components.

\subsection{Brill-Lindquist wormholes}
\label{sec:BL}
Here I will review some properties of the Brill-Lindquist solution that are necessary for the model proposed in Section \ref{sec:model}. For a more in-depth review, see \cite{gupta2025entangleduniverses}. Note that in the rest of the paper I set $G_{\rm N} = 1$, and the factors can be recovered by multiplying all masses and entropies by $G_{\rm N}$.

The Brill-Lindquist (BL) solutions \cite{brill-lindquist} describes time-symmetric initial data that satisfy the constraint equations of the Einstein-Maxwell system with $\Lambda = 0$ and an arbitrary number of asymptotically flat regions. The solutions are valid for arbitrary dimension; here we restrict to the four-dimensional solutions. Time evolution of the initial data is not known analytically, but since the HRT surface in a time-symmetric spacetime is the minimal surface on the symmetric slice (RT formula), the initial data is sufficient to compute entropies.

The solution to the constraint equations for neutral black holes is the conformally flat metric
\be \label{eq:basic metric}
  ds^2_3 = \psi(\vec r)^4 d\vec{r}\,^2\,,
\ee
where $\vec r$ is a 3-dimensional vector, $d\vec{r}^2$ is the flat 3-dimensional Euclidean metric. $\psi$ is a positive harmonic function on $\R^3$ with $n-1$ points removed at $\vec{r}=\vec{r}_i$ ($i=1,\ldots,n-1$)\footnote{In the notation of \cite{brill-lindquist}, this corresponds to the case $\beta_i=\alpha_i$ and $\chi=\psi$, with $N=n-1$.}:
\begin{equation}
    \label{eq:conformal factor}
    \psi = 1 + \sum^{n-1}_{i=1} \frac{\alpha_i}{\|\vec{r}-\vec{r_i}\|}\,,
\end{equation}
where $\alpha_i$ are arbitrary positive coefficients (with units of length) and $\|\cdot\|$ denotes the 3-dimensional Euclidean norm. The change of coordinates
\be
\vec{r}'=\alpha_i^2\frac{\vr-\vr_i}{\|\vr-\vr_i\|^2}
\ee
shows that the region close to each puncture $\|\vec{r}-\vec{r}_i\|\ll \alpha_i$ is also asymptotically flat, in addition to $\|\vec{r}\| \rightarrow \infty$ i.e. the metric has $n$ asymptotically flat regions. For $n=2$, the metric reduces to the initial data for the two-sided Schwarzschild black hole with radius $\alpha_1$ (in isotropic coordinates, including both exterior regions); the bifurcation surface is the sphere $\|\vec{r}-\vec{r}_1\|=\alpha_1$.

The ADM mass \( m_i \) associated with the asymptotic region \( a_i \) (in the limit \( \vec{r} \to \vec{r}_i \)) is obtained by expanding the metric near \( \vec{r} = \vec{r}_i \) to leading order and comparing it to the Schwarzschild solution. This yields,
\begin{align}
\label{eq:M_i_ADM}
m_i = 2\alpha_i \left(1 + \sum_{ j \neq i}^{n-1} \frac{\alpha_j}{\|\vec{r}_i - \vec{r}_j\|} \right) \qquad (i = 1, \ldots, n-1)\,.
\end{align}
Similarly, the ADM mass \( m_n \) corresponding to the asymptotic region \( a_n \) (as \( \|\vec{r}\| \to \infty \)) is found by expanding the metric at spatial infinity and matching to the Schwarzschild form,
\begin{equation}
\label{eq:M_inf_ADM}
m_n = 2 \sum_{i=1}^{n-1} \alpha_i\,.
\end{equation}
With the following alternative ansatz,
\begin{equation}
ds^2 = \phi(\vec{s})^4\,d\vec{s}^{\,2}\,, \qquad
\phi(\vec{s}) = \sum_{i=1}^n \frac{\mu_i}{\|\vec{s} - \vec{s}_i\|}\,,
\end{equation}
the $n$ asymptotic regions are treated on equal footing. Since there is no ``1+'' in the expression for $\phi$, the limit $\|\vec{s}\| \to \infty$ does not correspond to an asymptotically flat region; instead, the point $\|\vec{s}\| = \infty$ lies at a finite distance and must be appended to $\R^3$, giving an $n$-punctured 3-sphere. The ADM mass associated for region $a_i$ is then given by,
\begin{equation}
m_i = 2\mu_i \sum_{j \ne i}^n \frac{\mu_j}{\|\vec{s}_i - \vec{s}_j\|} \qquad (i = 1, \ldots, n)\,.
\end{equation}
The coordinates $\vec{s}$ are referred to as ``inverted coordinates'' because they are related to the original $\vec{r}$ coordinates via an inversion transformation
\begin{equation}
\vec{s} = \frac{\mu_n^2\,\vec{r}}{r^2}\,,
\end{equation}
where $\mu_n$ is an arbitrary length scale. The parameters in the two coordinate systems are related by
\begin{equation}
\frac{\mu_i}{\mu_n} = \frac{\alpha_i}{r_i}\,, \qquad \frac{\vec{s}_i}{\mu_n^2} = \frac{\vec{r}_i}{r_i^2} \qquad (i = 1, \ldots, n-1)\,,
\end{equation}
with $\vec{s}_n = \vec{0}$.

Despite the solution being explicitly provided, the minimal surfaces are not known analytically and must be found via a numerical shooting method (see \cite{gupta2025entangleduniverses} for an overview of the procedure). Here I provide a qualitative characterization of the surfaces. \cite{brill-lindquist} found a transition in the topology of the (locally) minimal surfaces as the separation between the points $\vec{r}_i$ is varied. Specifically, they focused on the case $n=3$ with equal parameters $\alpha_1=\alpha_2=:\alpha$. For each puncture $\vec{r}_{1,2}$, there is a minimal surface $\gamma_{1,2}$ of spherical topology enclosing that puncture; this is the throat of the corresponding Einstein-Rosen bridge.

\begin{figure}[ht]
    \centering
    \includegraphics[scale = 0.35]{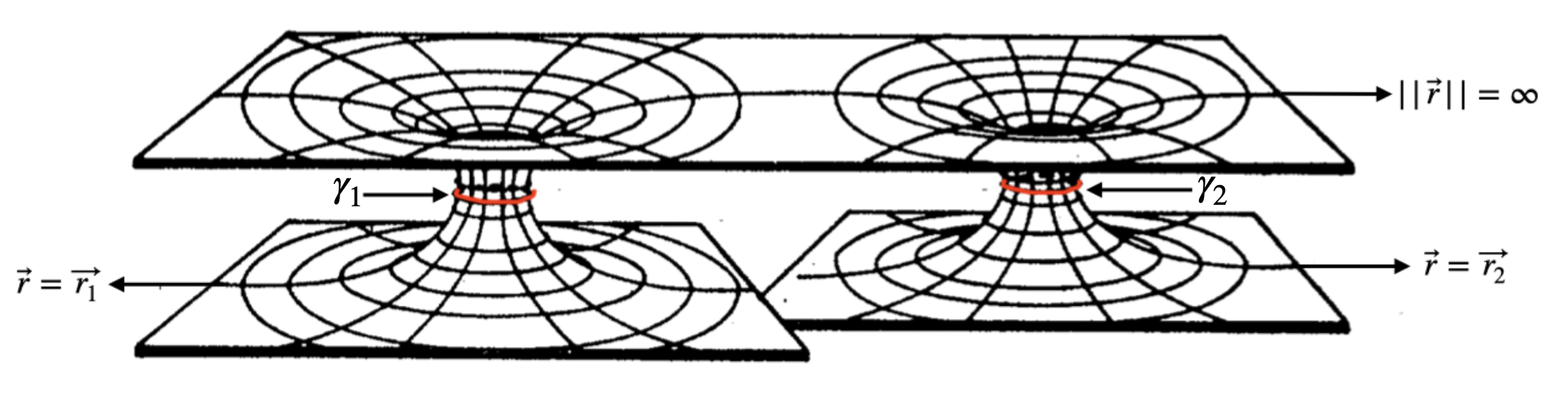}
    \caption{Embedding diagram of BL initial data geometry for $n=3$, $\alpha_1=\alpha_2=:\alpha$, and large separation, $\|\vec{r}_1-\vec{r}_2\|\gg \alpha$. There are two Einstein-Rosen bridges and two minimal surfaces $\gamma_1$, $\gamma_2$. \textit{Adapted from \cite{brill-lindquist}.}}
    \label{fig:2 ER}
\end{figure}

For large separation $r_{12}:=\|\vec{r}_2-\vec{r}_1\|\gg\alpha$, $\gamma_{1,2}$ are the only minimal (or extremal) surfaces; as $r_{12}$ is increased, they approach the spheres $\|\vec{r}-\vec{r}_i\|=\alpha$ (see Fig.\ \ref{fig:2 ER}). The minimal surface homologous to the sphere at infinity in region $a_3$ is $\gamma_1\cup \gamma_2$.

\begin{figure}
    \centering
    \includegraphics[width=0.6\linewidth]{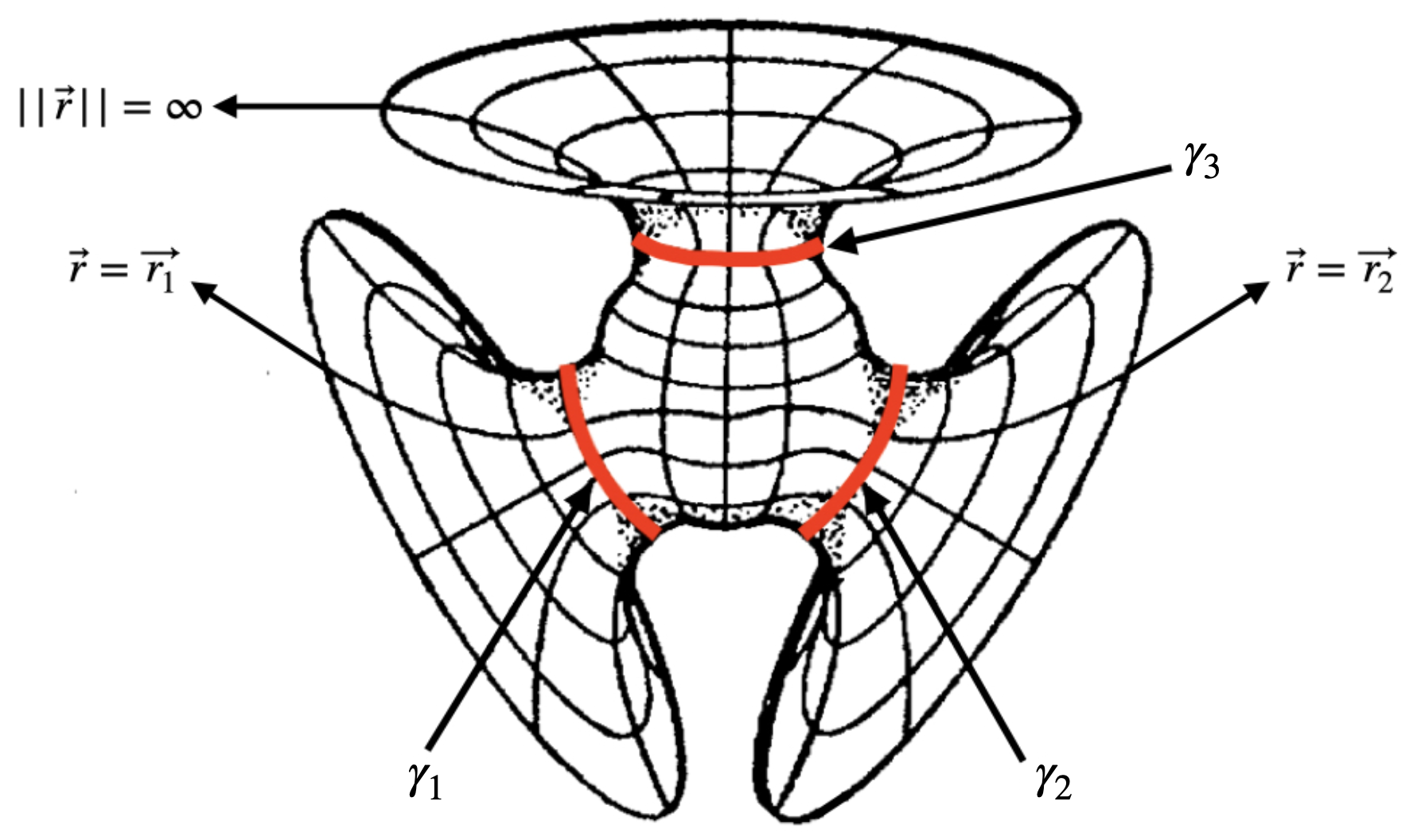}
    \caption{Embedding diagram of BL initial data geometry for small separation, $r_{12}\ll \alpha$. \textit{Adapted from \cite{brill-lindquist}.}}
    \label{fig:3 ER}
\end{figure}

As the separation \( r_{12} \) decreases, a critical value \( r_{12}^c \approx 3.1\alpha \) is reached \cite{crit} at which a new (locally) minimal surface \( \gamma_3 \) with spherical topology emerges, enclosing both punctures. The original surfaces \( \gamma_{1,2} \) become hidden behind \( \gamma_3 \), so an observer in region \( a_3 \) perceives a single black hole. The spatial geometry resembles a three-boundary AdS wormhole, with each asymptotic region connected via a neck to a central region (see Fig.~\ref{fig:3 ER}).

\begin{figure}[ht]
    \centering
    \begin{subfigure}[b]{0.49\textwidth}
        \centering
        \includegraphics[width=\textwidth]{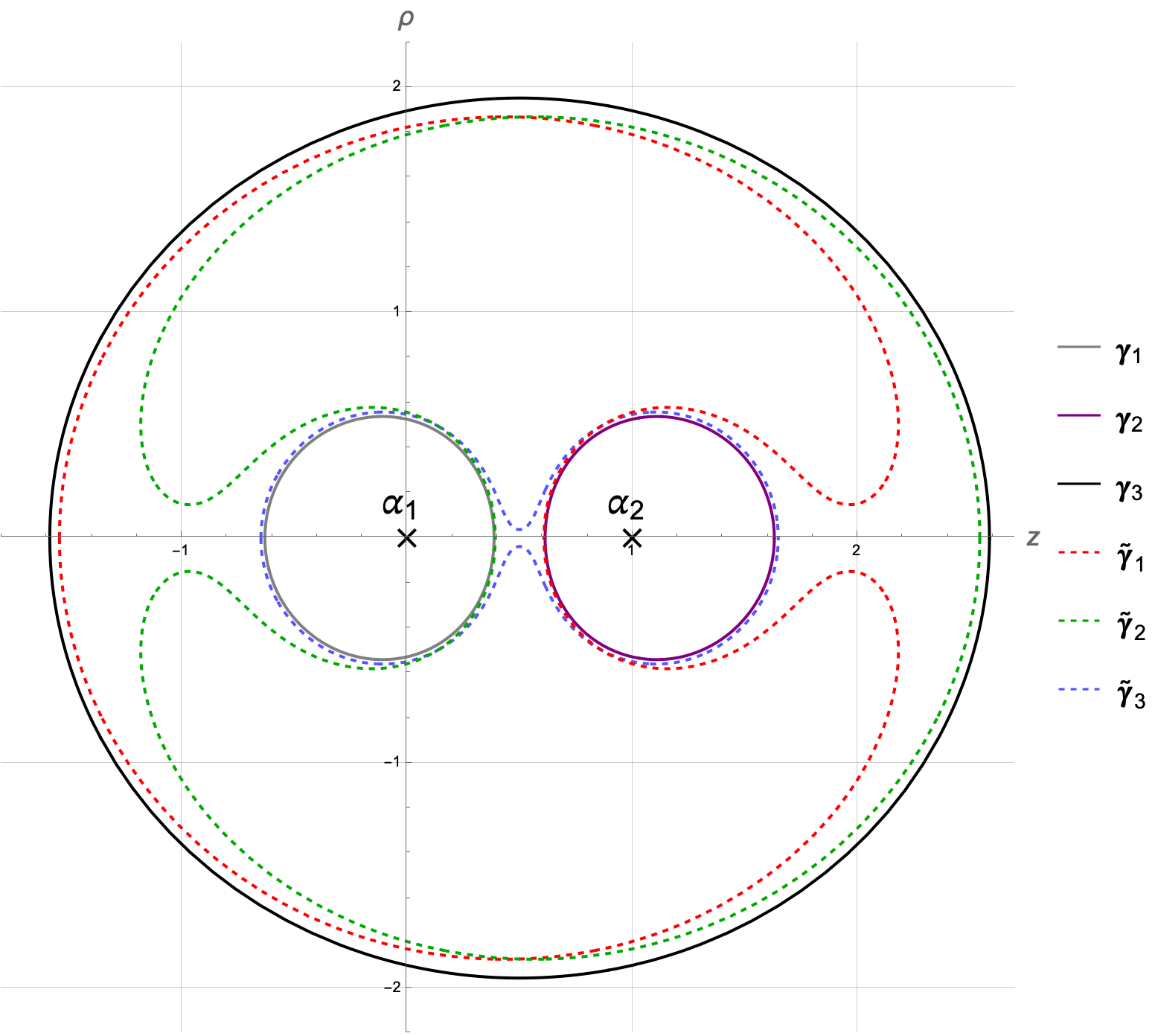}
    \end{subfigure}\hfill
    \begin{subfigure}[b]{0.49\textwidth}
        \centering
        \includegraphics[width=\textwidth]{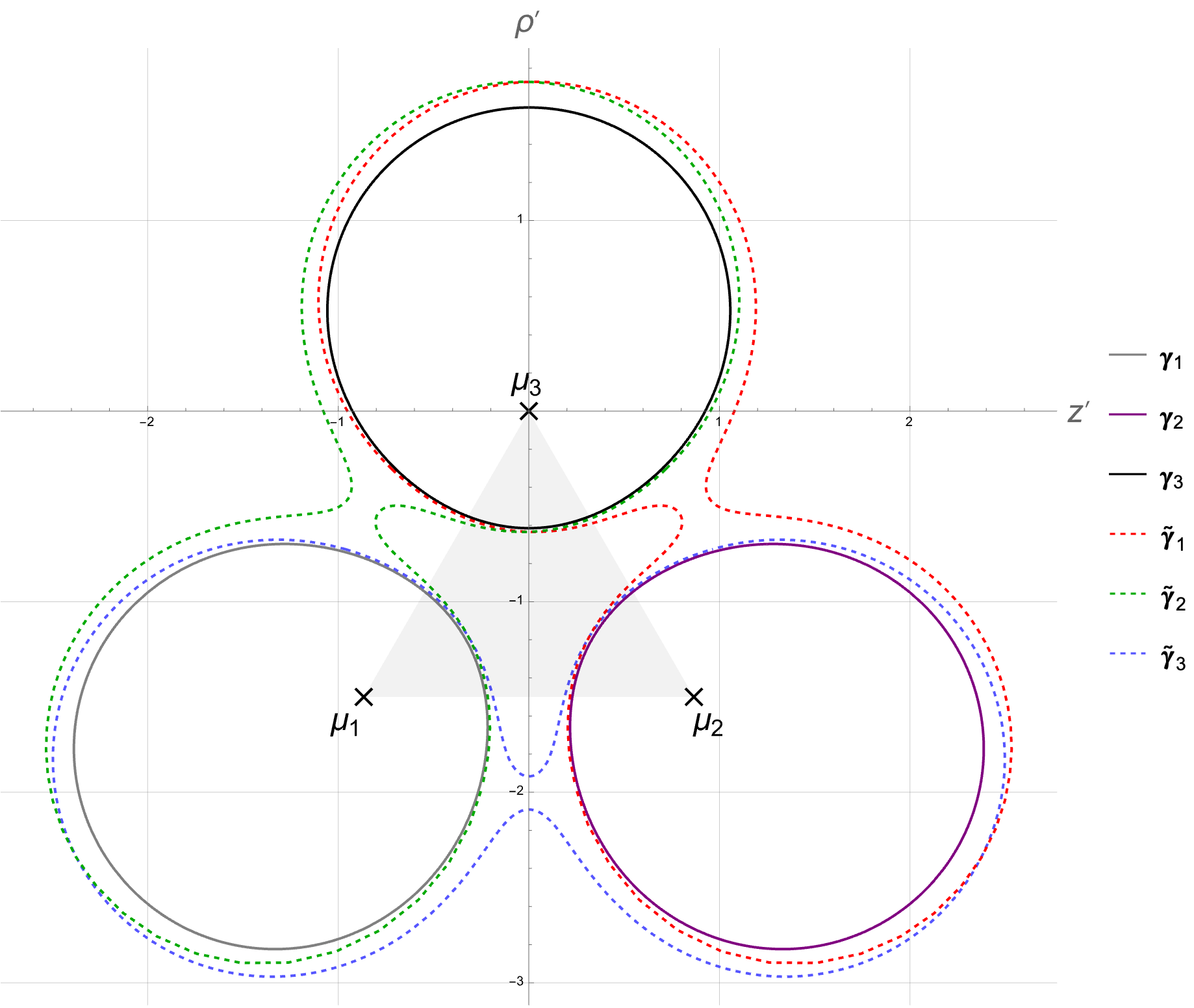}
    \end{subfigure}
\caption{\label{fig:surfaces}Planar sections of minimal surfaces $\gamma_{1,2,3}$ (solid curves) and index-1 extremal surfaces $\tilde\gamma_{1,2,3}$ (dashed curves) in (left) original $\vec{r}$ coordinates and (right) inverted $\vec{s}$ coordinates}
\end{figure}

Morse theory suggests that extremal surfaces in a given homology class generally appear or disappear in pairs under smooth metric deformations, with their Morse indices differing by one. Thus, an index-1 surface is expected to accompany the index-0 surface \( \gamma_3 \) at the critical separation. Indeed, a second surface \( \tilde{\gamma}_3 \), appearing at \( r_{12}^c \), likely has index-1.\footnote{While not directly verified, the narrow neck of \( \tilde{\gamma}_3 \) suggests a negative mode; near the neck, the surface resembles a catenoid in flat space, which has one such mode.} By symmetry, index-1 surfaces \( \tilde{\gamma}_1 \) and \( \tilde{\gamma}_2 \), homologous to \( \gamma_1 \) and \( \gamma_2 \), respectively, also appear (Fig.~\ref{fig:surfaces}).\footnote{See \cite{gupta2025entangleduniverses} for a more comprehensive discussion of the $S_3$ exchange symmetry in the $n=3$ BL metric.}

Henceforth, index-0 surfaces are referred to as minimal and play the role of candidate RT surfaces (I will discuss the surfaces for $n=4$ in Sec. \ref{sec:n4}). Index-1 surfaces (denoted with a tilde) are referred to as extremal and are involved in the python's lunch conjecture, discussed below. I will also use the convention where the subscript denotes the homology class of the surface; $\tilde{\gamma}_1$ is homologous to $\gamma_1$, $\tilde{\gamma}_2$ to $\gamma_2$ and so on. It should be noted for example that $\gamma_3$ is homologous to $-(\gamma_1\cup \gamma_2) = (-\gamma_1)\cup(-\gamma_2)$ where the minus sign denotes the orientation reversed surface; however I will not keep track of orientations and I will simply refer to $
\gamma_1 \cup \gamma_2$ as being homologous to $\gamma_3$. I will also work with BL wormholes with cylindrical symmetry, where all punctures (in $\vec{r}$ coordinates) are collinear; thus the only physically meaningful parameters are $\alpha_i$ and the separations $r_{ij} := \|\vec{r}_i-\vec{r}_j \|$.

\subsection{Python's lunch}
\label{sec:python}
An important consequence of the RT formula is that the RT surface delimits $W(A)$, the entanglement wedge of $A$, which is the bulk region whose quantum information is determined by the boundary region $A$. Entanglement wedge reconstruction then allows the reconstruction of bulk operators in $W(A)$ via boundary operators with support restricted to $A$. The bulk reconstruction map between the entanglement wedge $W(A)$ and boundary region $A$ admits a recovery map $\mathcal{R}$ (see \cite{Jafferis_2016,Dong_2016,Harlow_2017,Ohya_Petz,Cotler_2019} for a review). For a two-boundary wormhole with distinct minimal surfaces $\gamma_A, \gamma_{A}^c$ for each throat, there exists an extremal bulge surface $\gamma^b_A$ between them homologous to $A$. The python's lunch conjecture (PLC) \cite{Brown:2019rox} relates the (restricted) computational complexity associated to the recovery map $\mathcal{C}(\mathcal{R})$ to the area of the bulge surface $|\gamma^b_A|$. The bulge surface is known as the python's lunch, and the conjecture states
\begin{equation}
\label{eq:complexity conjecture}
    \mathcal{C}(\mathcal{R}) \sim \exp\left(\frac{|\gamma^b_A| - |\gamma_{A}^c|}{8G_{\rm N}}\right),
\end{equation}
where $\gamma_{A}^c$ is the larger minimal surface, often known as the constriction (assuming $\gamma_A = \gamma_{\rm RT} (A)$ is the RT surface). Note that the complexity has polynomial factors dependent on the wormhole volume which are omitted above to only show the leading order (exponential) behavior.

In the case where multiple minimal and bulge surfaces are present, the situation is more complicated. I will start by reviewing the construction of the lunch in \cite{Brown:2019rox} as it relates to Almgren-Pitts min-max theory. We start with a two-sided wormhole geometry with index-0 (minimal) surfaces $\gamma_1, \gamma_2$ homologous to the boundary region $A$, that bound a subregion $\sigma_1$ of the Cauchy slice $\sigma$ (since the homology class will remain fixed, I will henceforth drop the subscript $A$). The bulge surface is defined as the surface with area solving the minimax problem
\begin{equation}
    \min_{\phi} \max_{x} |\phi^{-1}(x)|,
\end{equation}
where $\phi^{-1}$ are the level sets for the function $\phi:\sigma \rightarrow [0,1]$ with $\phi^{-1}(0) = \gamma_1, \phi^{-1}(1) = \gamma_2$. The solution is the area of an index-1 surface between $\gamma_1$ and $\gamma_2$, which we refer to as the bulge $\gamma^b$.\footnote{Rigorously proving the existence of this surface from the minimax procedure is beyond the scope of this paper; I will assume such a surface always exists for the spacetimes considered here.}

\cite{Arora:2024edk} points out that directly following this minimax procedure is not necessarily practical as it involves minimizing over an infinite-dimensional space of functions $\phi$. Instead, we can restrict to looking for index-1 surfaces in the relevant homology class (i.e. homologous to $A$). When there are multiple index-1 surfaces $\{\gamma^b_{1,j}\}$ between $\gamma_1, \gamma_2$, also known as candidate bulges, following the above minimax procedure shows that the bulge $\gamma^b_1$ is given by the smallest area surface.

It is also possible to have multiple, non-intersecting\footnote{The surfaces may `intersect' on connected components.} index-0 surfaces $\{\gamma_1,\cdots ,\gamma_n\}$, all homologous to $A$. On a given Cauchy slice $\sigma$, assume $\gamma_1$ is the RT surface. We also assume the minimal surfaces $\gamma_j, \gamma_{j+1}$ are \textit{adjacent}, meaning there exist no index-0 surfaces between them. \cite{Engelhardt_2022} considers the scenario where between each pair of adjacent minimal surfaces $\gamma_j,\gamma_{j+1}$, there exists a unique index-1 (bulge) surface $\gamma^b_{j}$. The (generalized) PLC\footnote{The maximization is written incorrectly as being over $i<j$ in \cite{Arora:2024edk}.} then computes the complexity as
\begin{equation}
    \label{eq:complexity multiple lunches}
    \mathcal{C}(\mathcal{R}) \sim \max_{i>j}\left\{\exp\left(\frac{|\gamma_j^b| - |\gamma_i|}{8G_{\rm N}}\right)\right\}.
\end{equation}
The motivation for the maximization here is that based on considerations from tensor network toy models, the complexity will be dominated by the largest area difference.

\begin{figure}[ht]
    \centering
    \includegraphics[width =0.9\textwidth]{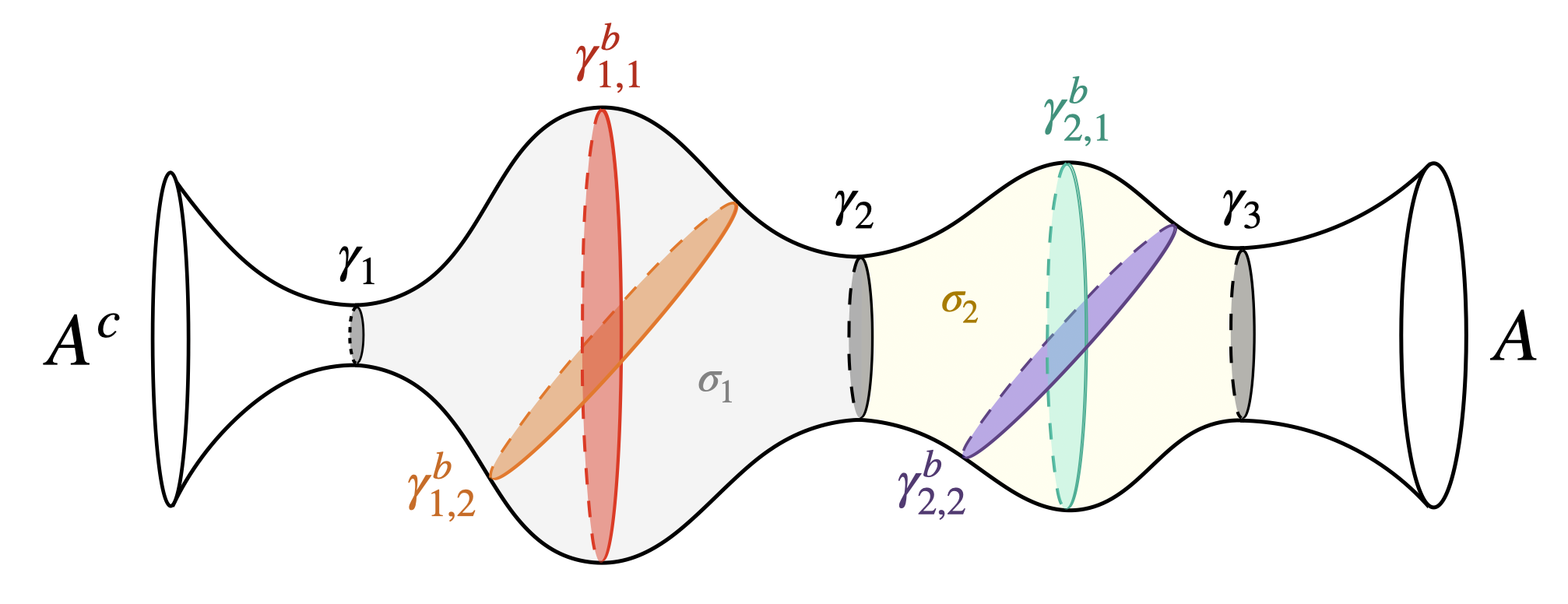}
    \caption{Setup for generalized PLC. $\gamma_1, \gamma_2, \gamma_3$ are minimal surfaces while $\gamma^b_{1,1}, \gamma_{1,2}^{b} \subset \sigma_1$, $\gamma^b_{2,1}, \gamma_{2,2}^{b}\subset \sigma_2$ are (index-1) extremal surfaces, all homologous to the boundary $A$.}
    \label{fig:multiple lunches}
\end{figure}

We now consider a slight generalization of this setup, where between each pair of adjacent minimal surfaces there is no longer a unique index-1 surface. Instead, in the Cauchy subregion $\sigma_j$ bounded by $\gamma_j, \gamma_{j+1}$ there now exist the index-1 surfaces $\{\gamma^b_{j,k}\}$, where $k$ is used to enumerate the index-1 surfaces in $\sigma_j$.\footnote{Using an index to denote the enumeration is a matter of convenience; the results discussed here continue to hold in the case of an uncountable number of index-1 surfaces within a subregion.} Combining the previous results, we then have the generalized PLC
\begin{equation}
    \label{eq:complexity multiple lunches multiple index}
    \mathcal{C}(\mathcal{R}) \sim \max_{i>j} \min_{k}\left\{\exp\left(\frac{|\gamma_{j,k}^b| - |\gamma_i|}{8G_{\rm N}}\right)\right\},
\end{equation}
where the minimization over $k$ is done for fixed $i,j$ within a subregion $\sigma_j$ (see Fig. \ref{fig:multiple lunches}). The order of the procedures can be interpreted as starting with the PLC restricted to a Cauchy subregion $\sigma_j$, where the considerations from \cite{Arora:2024edk} relating to Almgren-Pitts min-max theory tell us to find the minimal area index-1 surface within that subregion, which we call the bulge for that subregion. Once we have a unique bulge in each subregion, the decoding complexity is computed using the maximization procedure from \cite{Engelhardt_2022}. The minimization must be done first as the maximization is ill-defined without a \textit{unique} bulge in each subregion, with the answer then depending on the particular enumeration over $k$ of bulge surfaces $\{\gamma^b_{i,k}\}$ within each subregion $\sigma_i$. The case where there are multiple possible sets of non-intersecting minimal surfaces is discussed in Sec. \ref{sec:discussion}.

The PLC setup can be applied to the entanglement geometry of an evaporating black hole, where the outer minimal surfaces are the constrictions $\gamma_{\rm BH} \rightarrow \gamma_1$ and $\gamma_{\rm R} \rightarrow \gamma_n$ on the two ends of the wormhole, which represent the evaporating black hole and radiation respectively. The generalized PLC then allows us to interpret extremal (index-1) surfaces in the entanglement geometry as candidate lunches. This is important as the complexity of bulk reconstruction is now interpreted as the complexity of decoding the Hawking radiation. This is only true after the Page time, as before this the RT surface is $\gamma_{\rm R}$ and thus the entanglement wedge $W(A)$ does not include a python. Thus, we will only be interested in the computational complexity after the Page time.

An important check on the proposed model will be whether the time-dependent behavior matches the PLC predictions. Specifically, \cite{Brown:2019rox} suggests that the complexity $\mathcal{C}(t)$ undergoes 2 phase transitions: the first is at the Page time, as the entanglement wedge $W(A)$ changes at this point. The second transition is when $|\gamma_{\rm BH}| = |\gamma_0|/2$, where $|\gamma_0|$ is the initial black hole area before evaporation. The original evaporation geometry in \cite{Brown:2019rox} actually has 2 candidate lunches, which are named the forward and reverse sweep surface, and the second transition occurs when the python's lunch switches from the forward to the reverse sweep surface. Furthermore once the black hole has finished evaporating, the decoding task is no longer exponentially complex as there is no longer a bulge geometry. By \eqref{eq:complexity multiple lunches multiple index} we must thus have $|\gamma_{j,k}^b| \rightarrow |\gamma_i|$, i.e. the surfaces must approach each other in area as the black hole evaporates (after the second transition). I will check these time-dependent predictions for the model proposed below.

\section{Model}
\label{sec:model}

\begin{figure}[ht]
    \centering
    \includegraphics[width=\linewidth]{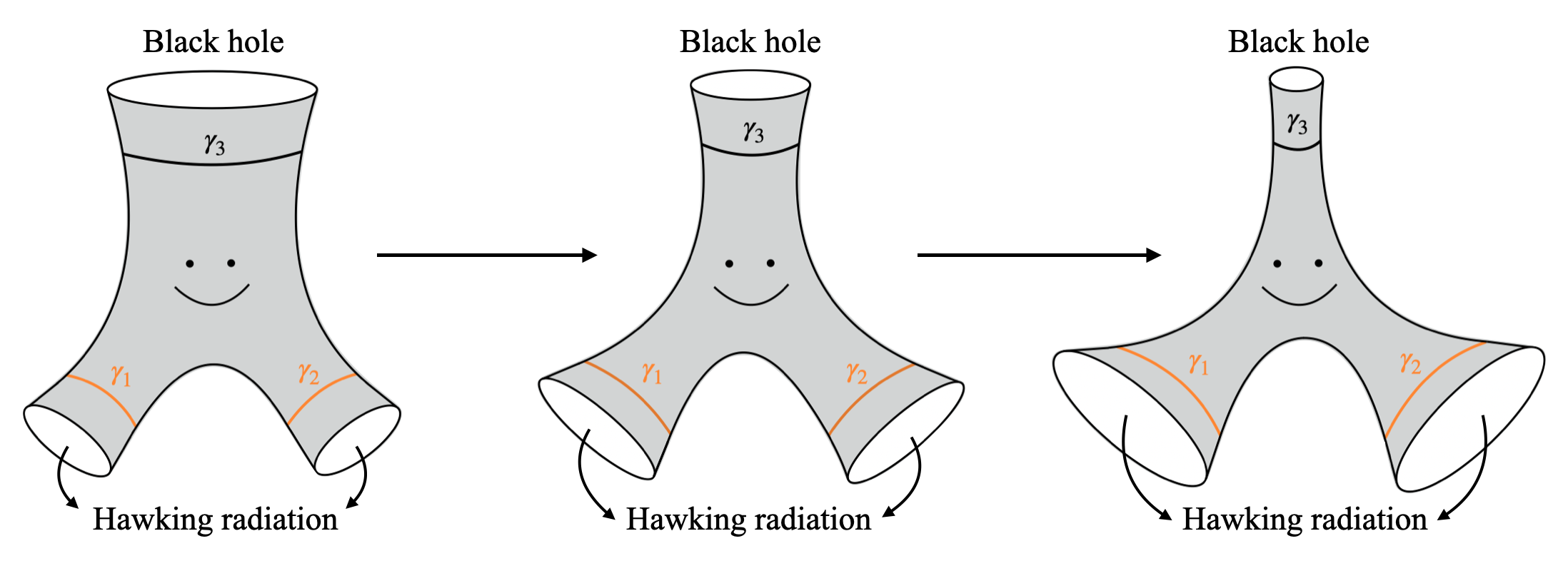}
    \caption{`Fixed' octopus geometry for the proposed model using $n=3$ BL wormholes as an example. The radiation collects into entangled baths, the legs (minimal surfaces $\gamma_1, \gamma_2$), while the black hole, the head, evaporates (minimal surface $\gamma_3$).}
    \label{fig:my octopus}
\end{figure}

I now introduce a model of black hole evaporation using the four-dimensional Brill-Lindquist wormhole with $n$ asymptotically flat boundary regions (higher dimensional analogues by considering the BL metric in higher dimensions are also possible). As shown in Fig. \ref{fig:my octopus}, I identify the region $a_n$ at asymptotic infinity ($\|\vec{r}\| \rightarrow \infty$) with the evaporating black hole (minimal surface $\gamma_n$), which is the ``head" of the octopus. The remaining $n-1$ asymptotic regions $a_i$ are associated with the Hawking radiation (minimal surfaces $\gamma_1 ,\cdots, \gamma_{n-1}$), which are the ``legs" of the octopus. Instead of the `dynamic' octopus of \cite{Akers:2019nfi}, this octopus is `fixed': it does \textit{not} grow legs as the evaporation progresses i.e the number of asymptotic boundaries remains fixed. Instead, the head shrinks (both in mass and corresponding area $|\gamma_n|$) while the legs grow, representing the black hole evaporating and radiation collecting respectively.
As the number of legs (wormhole mouths associated with the radiation) are now fixed, they can no longer be interpreted as small black holes emitted from the large evaporating black hole. Instead, it is best to think of the legs as thermal baths coupled to the evaporating black hole into which the radiation collects. The number of baths ($n-1$) is fixed, and the dynamics are represented by the black hole losing mass (the head shrinking) and the baths gaining that mass (legs growing). In the remainder of the paper I will continue to refer to the asymptotic region representing the evaporating black hole as the `head' and the asymptotic regions representing the radiation baths as `legs'.

A sensible model also requires energy conservation, which \cite{Akers:2019nfi} imposed by conserving the sum of the black hole masses. The relevant constraint here is on the ADM mass of the head, $m_{\rm BH}$, given by \eqref{eq:M_inf_ADM} and the ADM masses of the legs, $m_{\rm baths} = \sum_{i=1}^{n-1}m_i$, given by \eqref{eq:M_i_ADM}. Thus we have,
\begin{align}
    m_{\rm total} &= m_{\rm BH} + m_{\rm baths}\\
    &= m_n + \sum_{i=1}^{n-1} m_i\\
    \label{eq:conservation eqn}
    &= 2\sum_{i=1}^{n-1} \alpha_i + 2\left(\sum_{i=1}^{n-1}\alpha_i + \sum_{i=1}^{n-1} \sum_{j \neq i}\frac{\alpha_j \alpha_i}{r_{ij}}\right)
\end{align}
where we have $\|\vec{r}_i - \vec{r}_j\| = r_{ij}$. It should be noted that there are two important reasons for introducing dynamics differing from \cite{Akers:2019nfi} i.e. keeping the number of legs fixed. The first is that BL wormholes for large $n$ have many complicated extremal surfaces that are difficult to find, making models with fixed $n$ useful. The other is the use of the simplifying assumption that each bath has equal mass $m_i = m_j$. From \eqref{eq:conservation eqn} we can see then $m_{\rm baths} \geq m_{\rm BH}$, which is a problem for a dynamic octopus model. To see this, consider a model similar to \cite{Akers:2019nfi} for the BL wormholes, with the initial configuration given by the $n=3$ geometry.\footnote{The system cannot start at $n=2$ as this is the Schwarzschild initial data, where $m_{\rm BH} = m_{\rm Bath}$ (there is only one leg/bath). Growing a leg of equal mass would involve the black hole losing all of its mass.} Since we assume $m_1 = m_2$ and we have $m_1 + m_2 \geq m_{\rm BH}$, conservation of energy yields $m_1, m_2 \geq m_{\rm total}/4$. Time evolving by increasing $n$ and `emitting' a new black hole with mass equal to $m_1$ we see that we cannot evolve past $n=5$, as at that point $m_{\rm baths} = 4 m_1 \geq m_{\rm total}$. Relaxing the condition that all baths have equal mass may allow for a dynamic octopus model, but the complicated geometry of large $n$ BL wormholes makes a fixed $n$ model useful regardless.

In lieu of changing $n$, the model is instead evolved by changing the separation between the punctures. The constraints $m_i = m_j$, $m_n + \sum_{i=1}^{n-1}m_i = m_{\rm total}$ fix $n-1$ parameters, however there are $2n-3$ free parameters in the BL metric ($n-1$ parameters $\alpha_i$ and $n-2$ separations $r_{ij}$). We can additionally fix $n-3$ parameters by setting separations between adjacent punctures to be equal, i.e. $r_{ij} = \sigma |i-j|$ for fixed $\sigma$. The remaining free parameter can now be considered to be a time evolution parameter, which I define as $t := \alpha_i/\sigma$ for fixed $i$. Solving the constraint equations for a given value of $t$ fixes all free parameters and yields a definite geometry, which in this model is the state of the evaporating black hole-bath system at a given time. The choice of $t$ is a matter of convenience, as for the $n=3$ wormhole (the simplest configuration) the critical separation ($r_{12}^c \approx 3.1 \alpha$) geometry can be assigned a definite value of $t$. I will shortly explain why this geometry is relevant to this model. The particular form of the ratio is due to the system dynamics. Reducing the separation $\sigma$ between all masses causes all the $\alpha_i$ to also decrease by conservation of mass, but by a smaller amount than the decrease in $\sigma$. The black hole mass $m_{\rm BH}$ thus decreases and the mass of the baths $m_{\rm baths}$ increases while $t = \alpha_i/\sigma$ increases, making $t$ an appropriate time parameter.

The system dynamics are also sensible for early and late times. For early times $t \rightarrow 0$, the parameters $\alpha_i$ increase to a (finite) maximum while $\sigma \rightarrow \infty$. The mass of an individual bath $m_i$ can be written as: 
\begin{equation}
    \label{eq:bath mass time}
    m_i = 2\alpha_i + 2\sum_{j \neq i} \frac{\alpha_i \alpha_j}{r_{ij}} = 2\alpha_i + 2t \sum_{i \neq j} \frac{\alpha_j}{|i-j|}
\end{equation}
where we substitute $r_{ij} = \sigma |i-j|$ and $t = \alpha_i/\sigma$. As the $\alpha_i$ remain finite and non-zero, we see for early time $m_i \longrightarrow 2\alpha_i$ and thus $m_{\rm baths} \longrightarrow m_{\rm BH}$. The interpretation is that at the start of the evolution, the system comprises of a black hole coupled to baths with total mass equal to the black hole's mass. Since $m_{\rm baths} \geq m_{\rm BH}$, the system starts with this inequality saturated and time evolution increases $m_{\rm baths}$ while decreasing $m_{\rm BH}$. 

We can similarly consider the system dynamics for late time $t \rightarrow \infty$. Note that the time is unbounded as the punctures can be arbitrarily close. While both $\sigma$ and the $\alpha_i$ decrease to 0, for late times we require the relation $\sigma \sim \alpha^2$ to ensure a finite total mass. The result is the black hole mass $m_{\rm BH} \rightarrow 0$, while each radiation bath approaches a finite mass due to the finite term $\alpha_i\alpha_j/\sigma$ in \eqref{eq:bath mass time}. By conservation of mass we thus have $m_{\rm baths} = m_{\rm total}$, and at late time the black hole has entirely evaporated into the baths.

\begin{figure}[ht]
    \centering
    \includegraphics[width=0.8\linewidth]{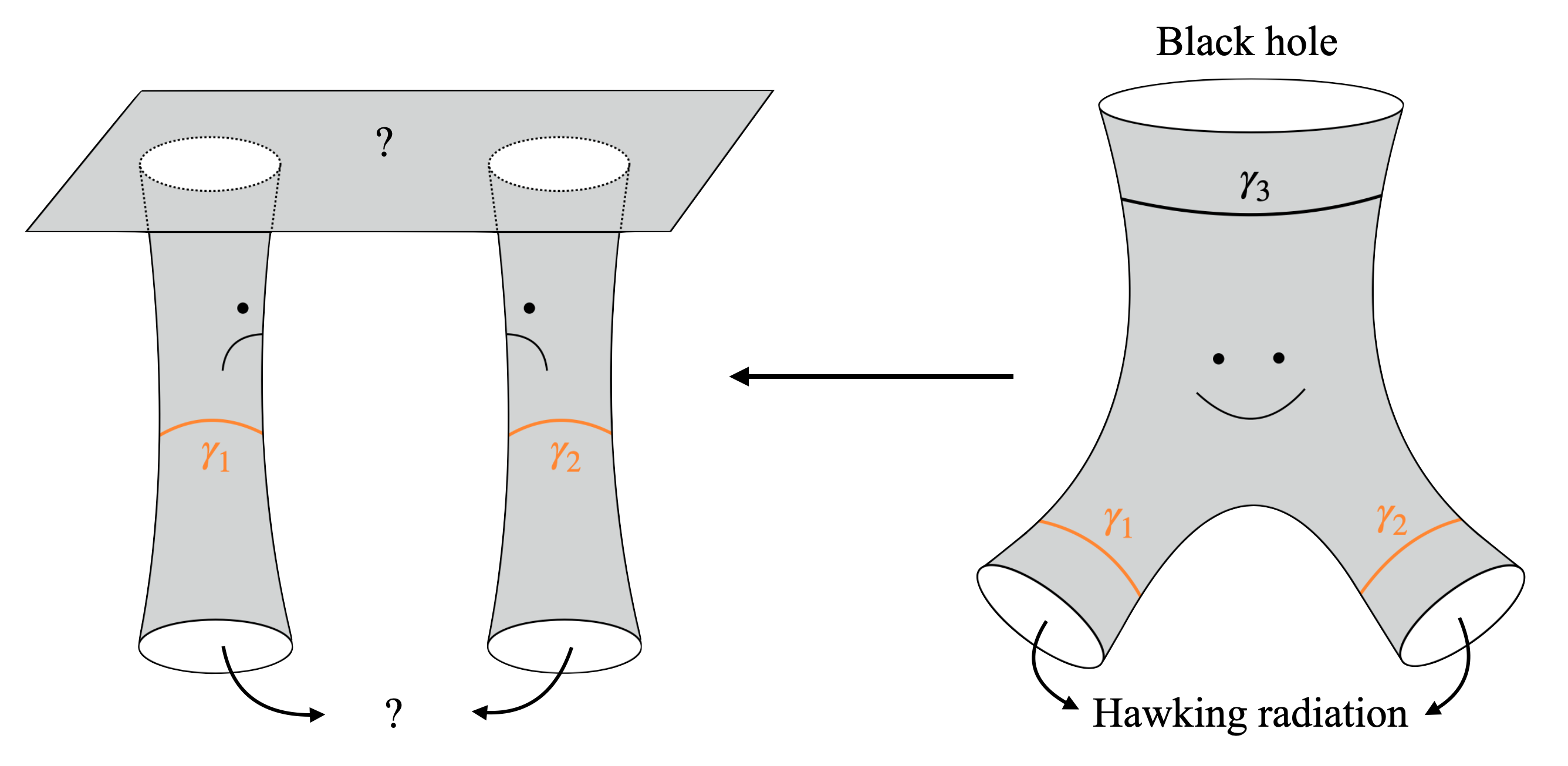}
    \caption{Starting with the geometry at the critical time $t_c$ (right), evolving back in time yields a 'broken' octopus (left), where the connected wormhole separates and the asymptotic regions no longer have the same interpretations.}
    \label{fig:broken_octopus}
\end{figure}

The early time behavior however ignores the geometric structure. Taking $n=3$ BL wormholes with $\alpha := \alpha_1 = \alpha_2$ as an example, there exists a critical time $t_c = \alpha/r^c_{12} \approx 0.323$ before which there is no connected wormhole. As shown in Fig. \ref{fig:broken_octopus}, an observer living in the region previously associated with the head of the octopus would now see two black holes instead of a single one, yielding a geometry that no longer has the interpretation of a single evaporating black hole (a `broken' octopus). Such a critical time also exists for higher $n$ BL wormholes; thus the model can only be used to represent black hole evaporation after the critical time for the given geometry. The early time asymptotics ($t \rightarrow 0$) therefore do not represent the state of the evaporating black hole under this model, but instead will require the full \textit{quantum octopus} of \cite{erEpr}. For early times $m_{\rm baths}$ will be very small, yielding a geometry where quantum effects will be important. The BL model utilizes a classical geometry, which breaks down for these times. The same argument can also be applied to late times, where the black hole will be small enough that quantum effects dominate and the classical geometry breaks down. The model described above thus has a limited regime of validity, but the asymptotic behavior of the masses for early/late times should remain the same.
\begin{equation}
    \label{eq:quantum rt}
    S_{\rm gen}(\gamma) = \frac{|\gamma|}{4\GN}+S_{\rm bulk}(\gamma)
\end{equation}
This idea of a classical connected geometry is also relevant to why the RT formula is used to compute the entropy, rather than the complete quantum corrected version of the formula \cite{Engelhardt_2015,Faulkner:2013ana}. After all, Hawking radiation is a quantum correction to classical black holes, and the entropy of the radiation  is generally accounted for by a $\mathcal{O}(G_{\rm N}^0)$ term as in \eqref{eq:quantum rt}. This term is not required for the BL models as the model should be thought of as computing a coarse-graining of the full quantum-corrected RT formula. The BL geometry represents a snapshot of the evolution, and it is formed by taking the radiation collected in each of the $n-1$ baths at a given moment and collapsing it into $n-1$ black holes. These $n-1$ black holes are entangled with the evaporating black hole and via ER=EPR can be interpreted to form a connected $n$-boundary wormhole geometry. The entanglement entropy computed for this \textit{classical} geometry should be dominated by the $\mathcal{O}(G_{\rm N}^{-1})$ term in \eqref{eq:quantum rt}, thus allowing the use of the RT formula. At early/late times the geometry may be sub-classical and quantum corrections will play a role; otherwise, a classical-connected geometry can be found and its entanglement entropy is taken to represent a coarse-graining of the full quantum-corrected entropy.

In the remainder of this section, I present exact numerics for the evaporation model for $n=3,4$ BL wormholes. The main objects of interest are the minimal surfaces, which are found numerically via the shooting method reviewed in \cite{gupta2025entangleduniverses}. While the time $t$ is a continuous parameter (unlike in \cite{Akers:2019nfi} where time evolution occurs in discrete steps), in practice the minimal surfaces are computed for a configuration with fixed $t$. Thus, a continuous curve for the minimal surface areas as functions of $t$ is difficult to obtain; instead the areas are computed for multiple discrete values of $t$ to plot the area as a function of time. I show that applying the extension of the HRT formula \eqref{eq:flat RT} to both $n=3,4$ BL geometries reproduces the Page curve for the entanglement entropy. I also compute areas of the extremal (index-1) surfaces and apply the (generalized) PLC to obtain the computational complexity, after which I compare its behavior to the predictions of \cite{Brown:2019rox}. The motivation behind applying the PLC to these models comes from the fact that the BL wormhole geometries also possess a tensor network decomposition \cite{gupta2025entangleduniverses}; this is discussed further in Sec. \ref{sec:discussion}.

\subsection[Brill-Lindquist evaporation model: n = 3]{Brill-Lindquist evaporation model: $n=3$}
\label{sec:n3}
Here I present the results for the evaporation model for $n=3$ BL wormholes. The geometry is of the kind shown in Figs. \ref{fig:surfaces}, \ref{fig:my octopus} and the minimal surfaces present are each associated to an asymptotic region. The areas of the minimal surfaces are used to compute $S(A)$, the entanglement entropy of the boundary region $A$ associated to the head of the octopus (the evaporating black hole). The surface on the head is identified with the black hole $\gamma_3 \rightarrow\gamma_{\rm BH}$ and the surfaces on the legs are identified with the radiation baths $\gamma_1 \cup \gamma_2 \rightarrow \gamma_{\rm R}$. 

I will start by checking the evolution of areas of the minimal surfaces. The constraint $m_1 = m_2$ fixes $\alpha_1 = \alpha_2 =: \alpha$ and the wormhole is thus symmetric between the exchange of $\gamma_1$ and $\gamma_2$ (and the respective punctures). I set $m_\text{total} = 10$ and start the evolution at separation $r_{12} = 3 \alpha$ (this is practically as close to $r^c_{12} \approx 3.1 \alpha$ I can still find extremal surfaces). Solving the constraint equation \eqref{eq:conservation eqn} for the start time $t_0 = 1/3$ yields the starting parameters $\alpha = 1.07145, r_{12} = 3.21429$. We consider the results for the time evolution until $t = 0.56$.

\begin{figure}[ht]
    \centering
    \includegraphics[width=0.8\linewidth]{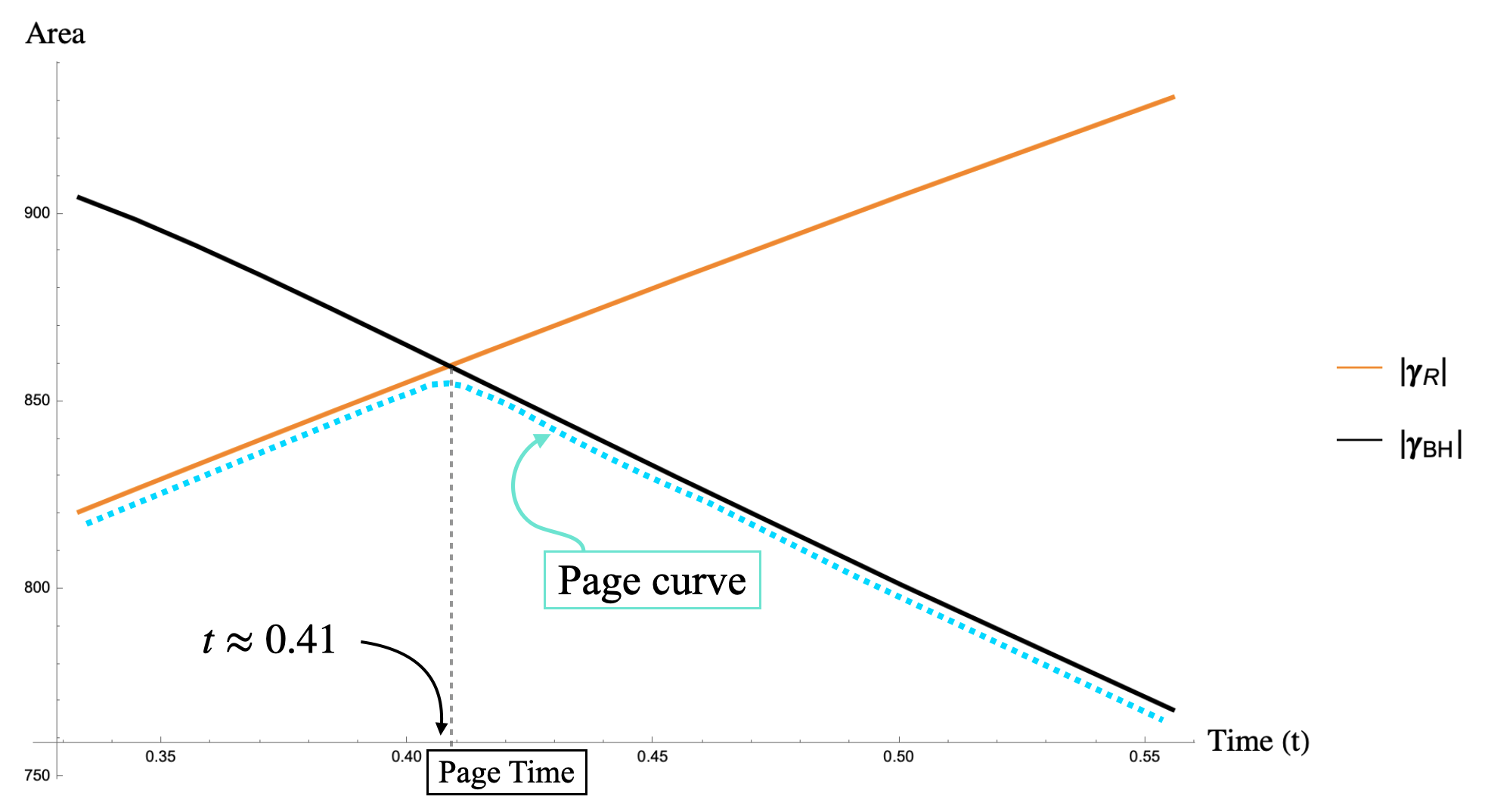}
    \caption{Areas of minimal surfaces under time evolution for $n = 3$ model. Note the identifications $\gamma_{3} \rightarrow \gamma_{\rm BH}$ and $\gamma_1 \cup \gamma_2 \rightarrow \gamma_{\rm R}$. The cyan curve is the minimal area, which follows the Page curve.}
    \label{fig:BL evaporation}
\end{figure}

As shown in Fig. \ref{fig:BL evaporation}, $|\gamma_{\rm BH}|$ decreases while $|\gamma_{\rm R}|$ increases. Applying the extended RT formula \eqref{eq:flat RT} to the wormhole, we see that before the Page time we have $S(A) = |\gamma_{\rm R}|/4$ which is increasing, and after the Page time the RT surface switches and $S(A) = |\gamma_{\rm BH}|/4$, which is decreasing. The point when the areas are equal is identified as the Page time, which for this model corresponds to $t_{\text{Page}} \approx 0.41$. The model thus reproduces the Page curve.

These areas can also be used to compute the mutual information, $I(1:2)$ between the two baths, with their respective asymptotic regions identified as $1$ and $2$. The mutual information is given by,
\begin{equation}
    4 I(1:2) = 4S(1) + 4S(2) - 4S(12).
\end{equation}
Before the Page time, $S(1) = S(2) = |\gamma_1|/4 = |\gamma_2|/4$ and $S(12) = S(A) = (|\gamma_1|+|\gamma_2|)/4$; the mutual information is thus 0. After the Page time, we have $S(12) = S(A) = |\gamma_3|/4$ and the mutual information $4I(1:2) = 2|\gamma_1| - |\gamma_3|$ is non-zero, and is in fact increasing with time. As mutual information is a measure of correlation \cite{Headrick:2019eth}, this indicates that the baths become more correlated (after the Page time) as the evaporation progresses, which is sensible as the outgoing radiation is highly entangled not just with the evaporating black hole, but among itself.

\begin{figure}[ht]
    \centering
    \begin{subfigure}[b]{0.46\textwidth}
        \centering
        \includegraphics[width =\textwidth]{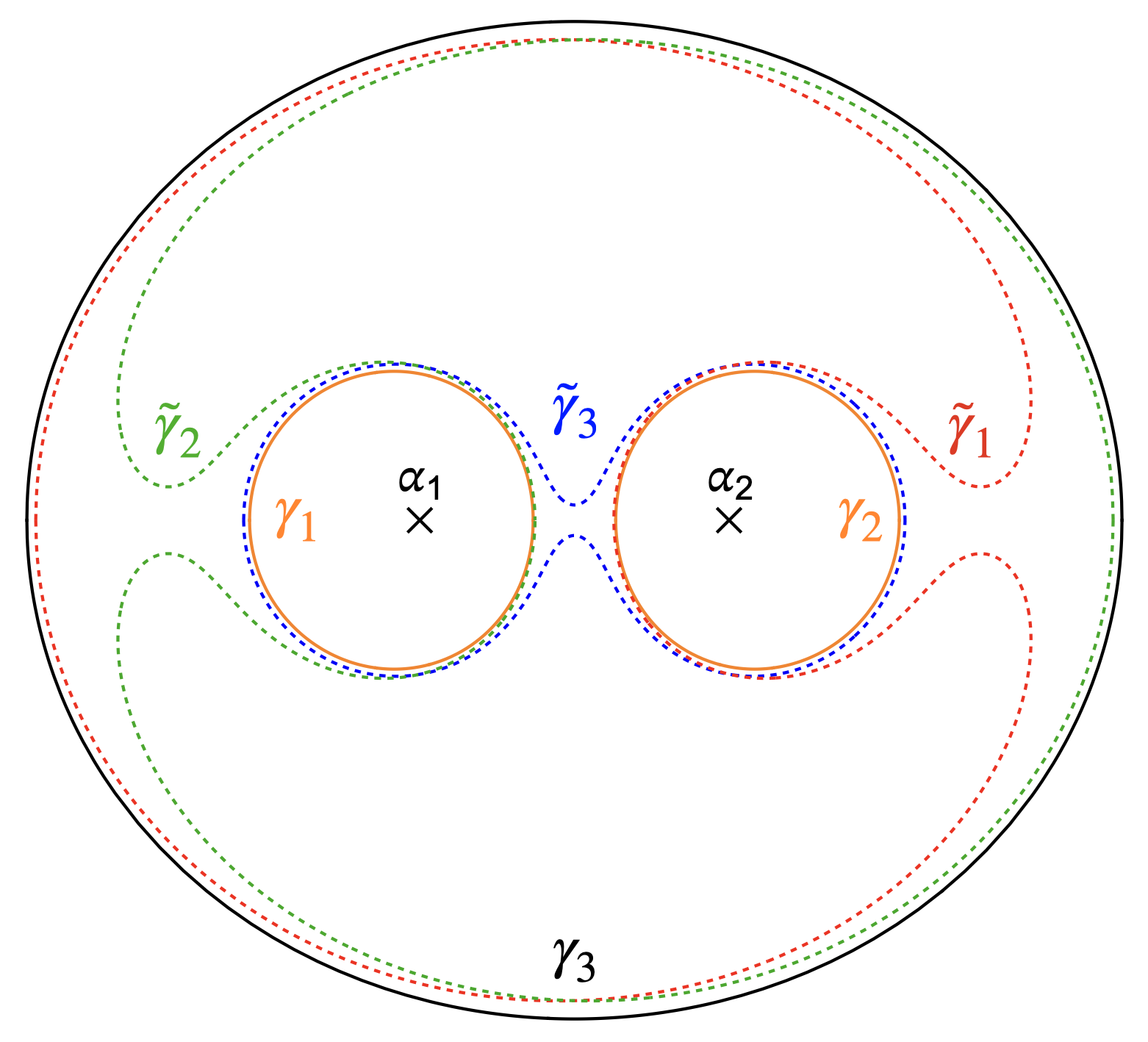}
    \end{subfigure}\hfill
    \begin{subfigure}[b]{0.46\textwidth}
        \centering
        \includegraphics[width = \textwidth]{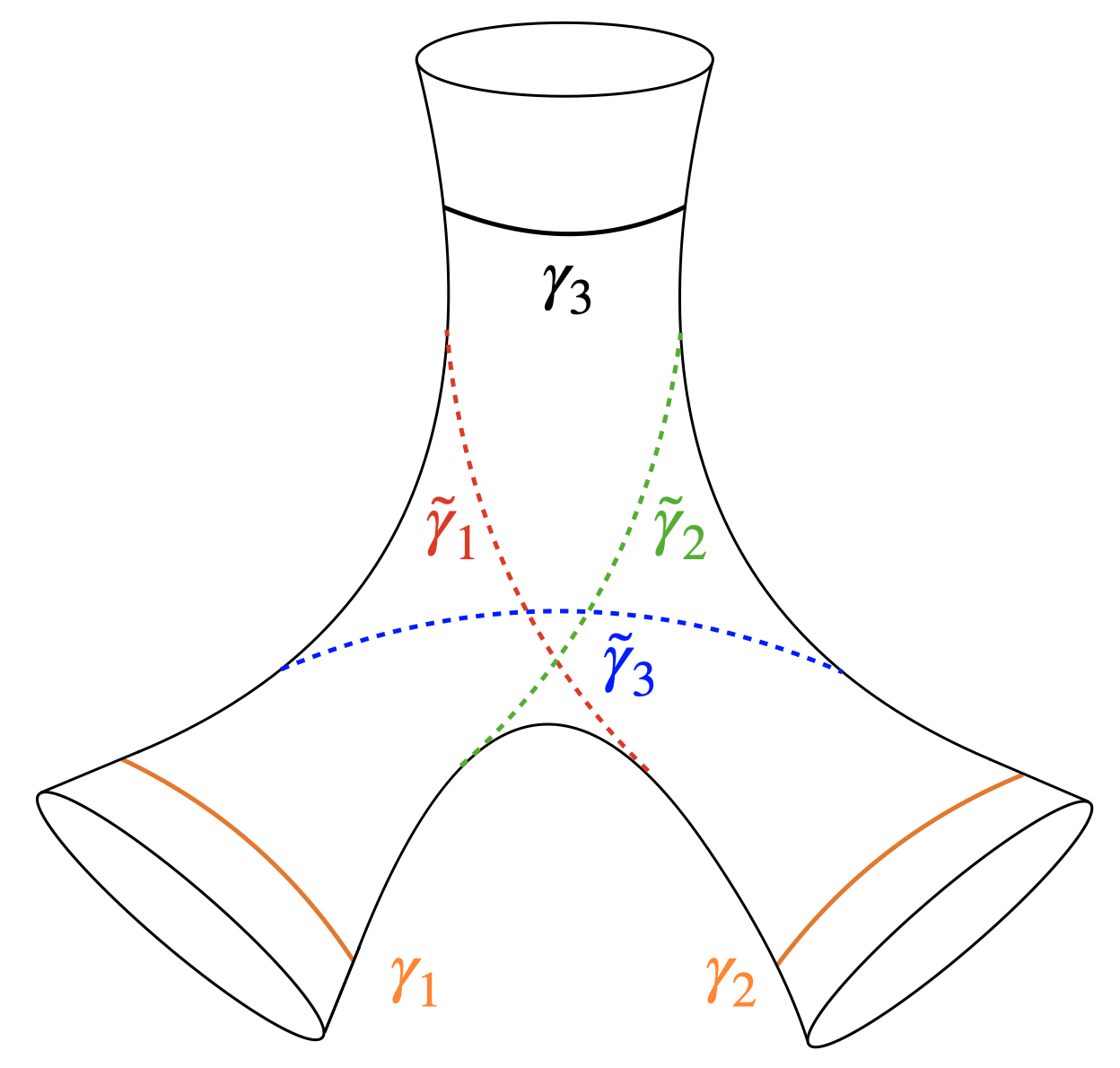}
    \end{subfigure}
    \caption{Extremal (index-1) surfaces for $n=3$ BL geometry in (left) original BL coordinates and (right) the wormhole geometry.}
    \label{fig:N=3 index 1 surfaces}
\end{figure}

Next, I will apply the PLC to compute the computational complexity $\mathcal{C}(t)$ of decoding the Hawking radiation. The region in between $\gamma_{\rm BH}, \gamma_{\rm R}$ includes multiple index-1 surfaces as shown in Fig. \ref{fig:N=3 index 1 surfaces}. Here we see the three index-1 surfaces, $\tilde\gamma_1, \tilde\gamma_2, \tilde\gamma_3$ from Fig. \ref{fig:surfaces}. Since we are interested in index-1 surfaces homologous to the boundary region $A$ (head of the octopus), the relevant index-1 surfaces are thus $\Gamma = \{\tilde
\gamma_3, \tilde\gamma_1 \cup \gamma_2, \tilde\gamma_2 \cup \gamma_1\}$. Applying \eqref{eq:complexity multiple lunches multiple index}, we see that since there are only two minimal surfaces $\gamma_{\rm BH}, \gamma_{\rm R}$, the bulge surface is the minimal index-1 surface i.e. $\min_{\tilde\gamma \in \Gamma} |\tilde\gamma| = |\gamma^b_A|$. Due to the exchange symmetry between $\gamma_1, \gamma_2$, the surfaces $\tilde\gamma_1 \cup \gamma_2, \tilde\gamma_2 \cup \gamma_1$ have the same areas and we must only consider one of them.

\begin{figure}[ht]
    \centering
    \includegraphics[width=0.8\linewidth]{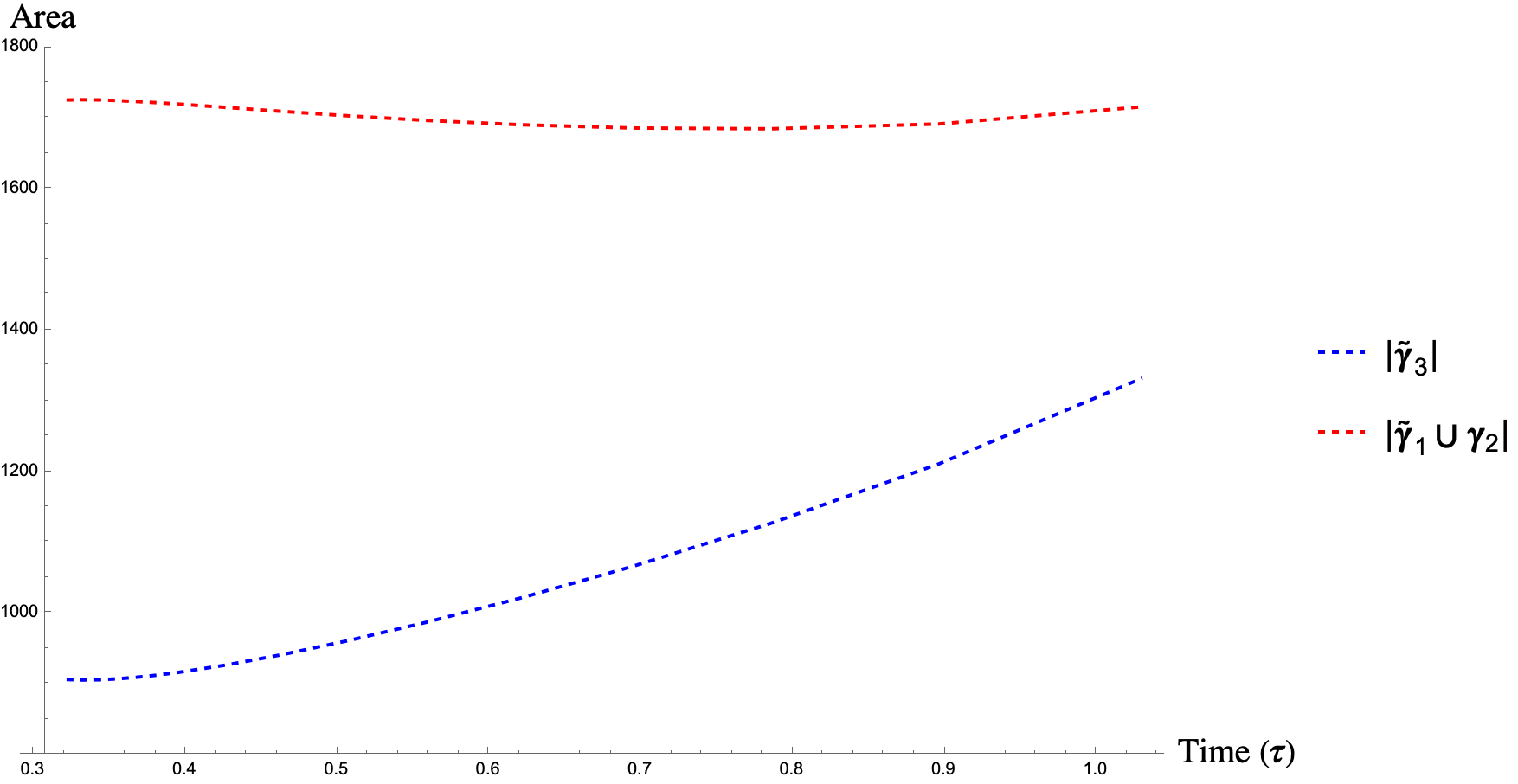}
    \caption{Areas of candidate bulge (index-1) surfaces over time ($\tau$).}
    \label{fig:n=3 index 1 areas}
\end{figure}

As shown in Fig. \ref{fig:n=3 index 1 areas}, $\tilde\gamma_3$ has smaller area throughout the evolution and is thus the bulge surface. While this relationship is only shown up to a finite time, it is expected to hold until the end of the evaporation. As $t \rightarrow \infty$, the surface $\tilde\gamma_3$ approaches the surface $\gamma_1 \cup \gamma_2$ while $\tilde\gamma_1$ approaches $\gamma_2 \cup \gamma_3$. Thus the areas for late time go as $|\tilde\gamma_3| \sim 2|\gamma_1|, |\tilde\gamma_1 \cup \gamma_2| \sim 2 |\gamma_1| + |\gamma_3|$; $\tilde\gamma_3$ is clearly smaller until $|\gamma_3| \rightarrow 0$ i.e. the black hole evaporates. Thus the PLC can be applied \eqref{eq:complexity conjecture} with the identifications $ \gamma^b_A \rightarrow \tilde\gamma_3, \gamma_A^c \rightarrow \gamma_{\rm R}$. The term controlling the exponential behavior of the complexity (after the Page time) is then the difference in areas $|\tilde\gamma_3| - |\gamma_{\rm R}|$. The time evolution of this quantity is more pronounced over a longer time scale, so we consider the evolution until $t = 1.1$. It is also important to note that $t \in [0,\infty)$ is unbounded. A finite time parameter $\tau = \arctan{(t)}$ is thus chosen in subsequent figures, where we now have $\tau \in [0,\pi/2)$.

\begin{figure}[ht]
    \centering
    \includegraphics[width=0.8\textwidth]{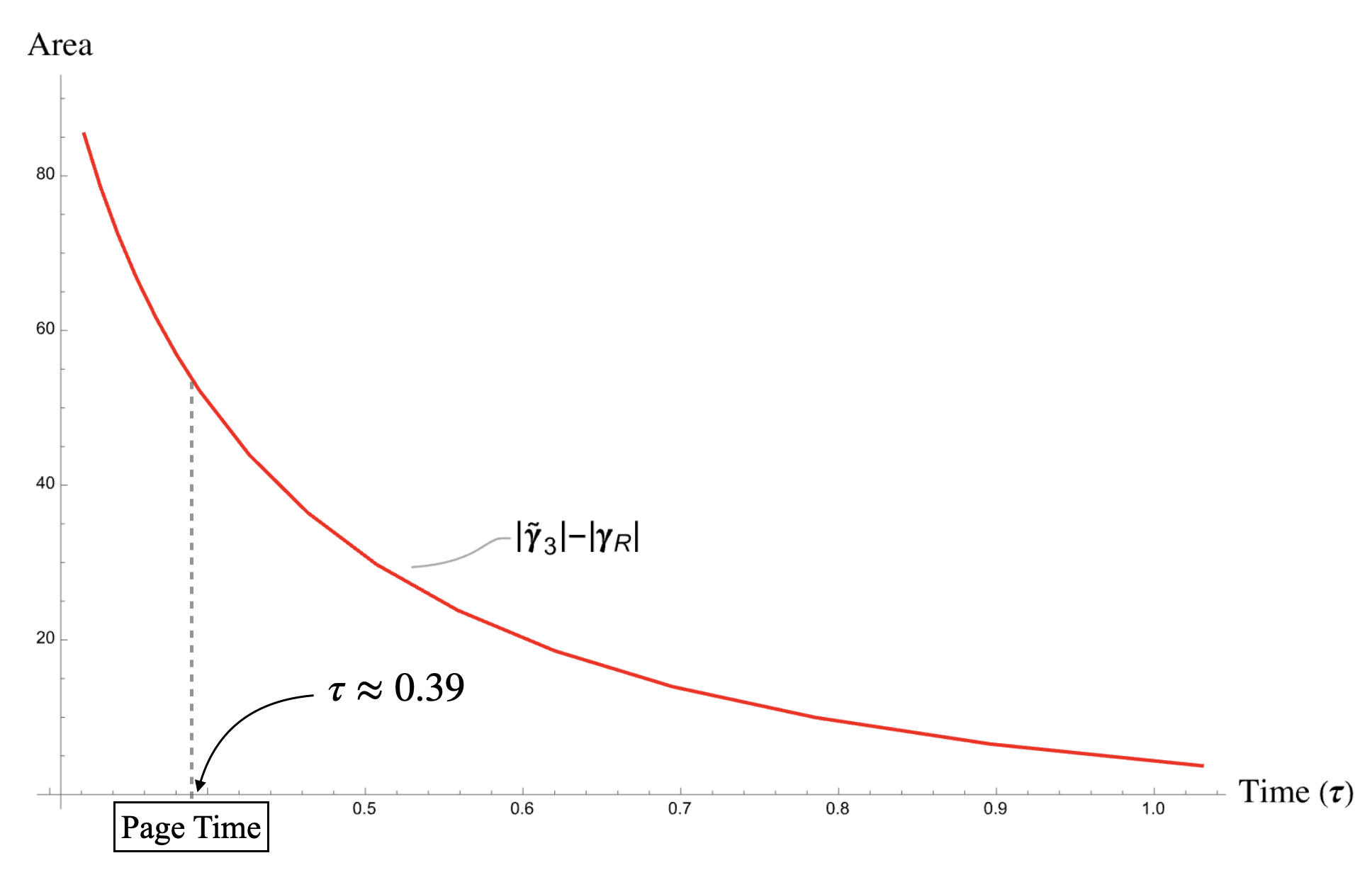}
    \caption{$|\tilde\gamma_{3}|- |\gamma_{\rm R}|$ as a function of time.}
    \label{fig:lunch evolution}
\end{figure}

Since the bulge surface remains $\tilde\gamma_3$ throughout the evolution, there is no phase transition between a forward and reverse sweep surface. More importantly, Fig. \ref{fig:lunch evolution} shows that with the new coordinate $\tau$, the difference in areas $|\tilde\gamma_{3}| - |\gamma_{\rm R}| \rightarrow 0$ as $\tau \rightarrow \pi/2$. This reproduces a basic prediction of the PLC, which is that the exponential behavior of the decoding complexity decreases as the black hole evaporates, with the complexity becoming polynomial once the black hole has evaporated ($\tau \rightarrow \pi/2$).
 
\begin{figure}[ht]
    \centering
    \begin{subfigure}[b]{0.48\textwidth}
        \centering
        \includegraphics[width = \textwidth]{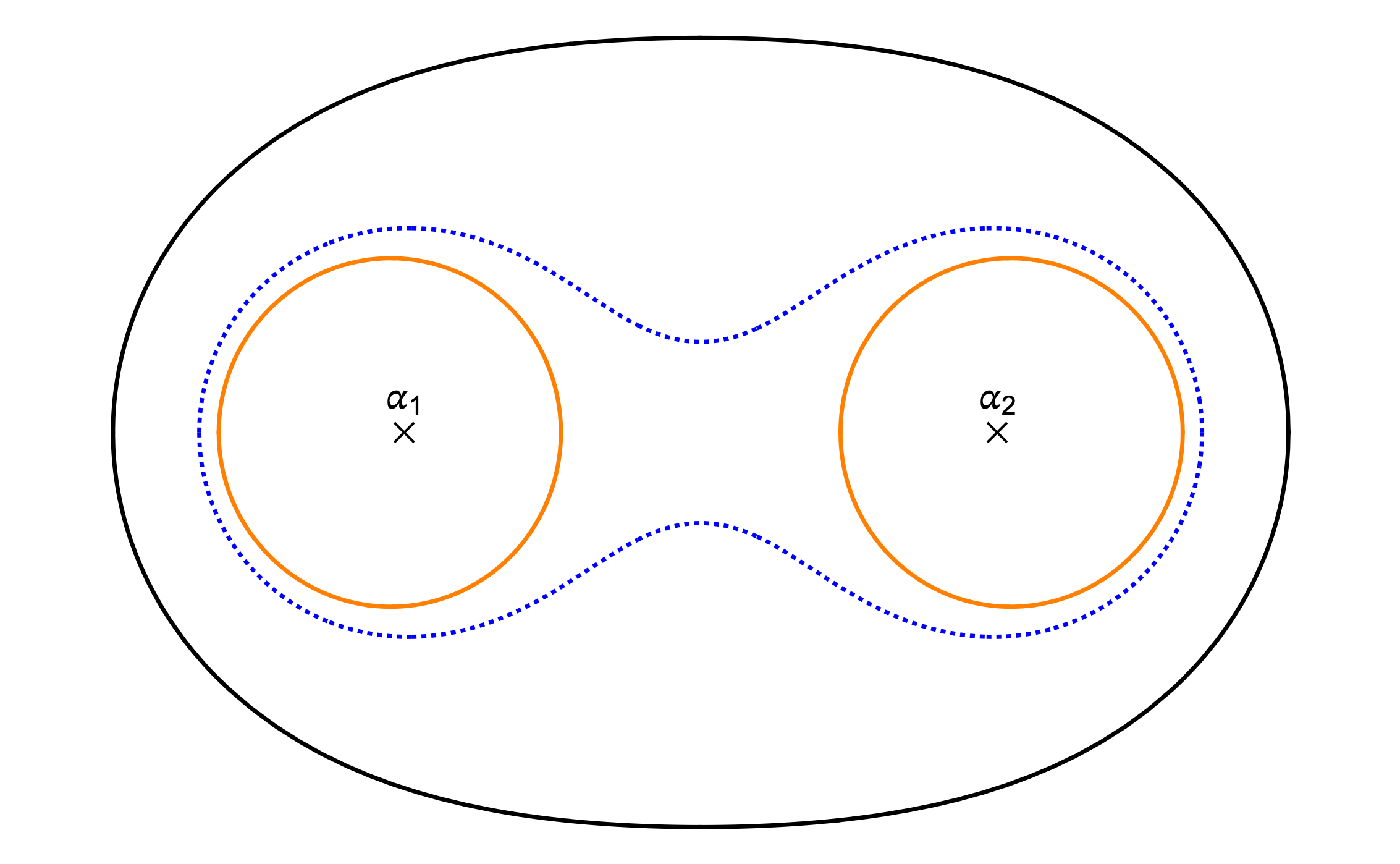}
        \caption*{$\tau = \tau_{\text{Page}} = 0.39$}
        \label{fig:a_2/m = 1.224}
    \end{subfigure}
    \begin{subfigure}[b]{0.48\textwidth}
        \centering
        \includegraphics[width =\textwidth]{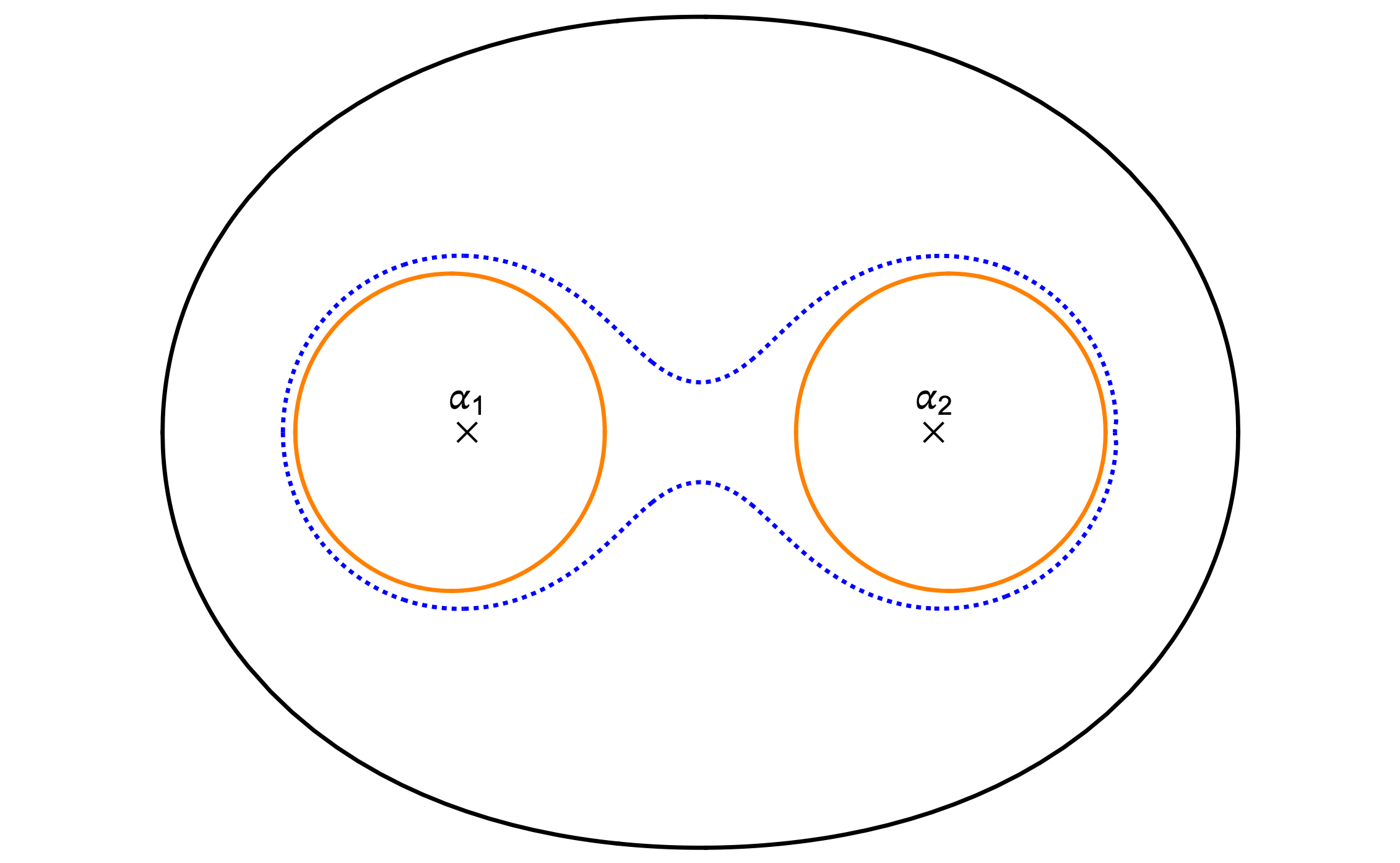}
        \caption*{$\tau = 0.46$}
        \label{fig:a_2/m = 0.79}
    \end{subfigure}\\[\smallskipamount]
    \vspace{0.3 cm}
    \begin{subfigure}[b]{0.48\textwidth}
        \centering
        \includegraphics[width = \textwidth]{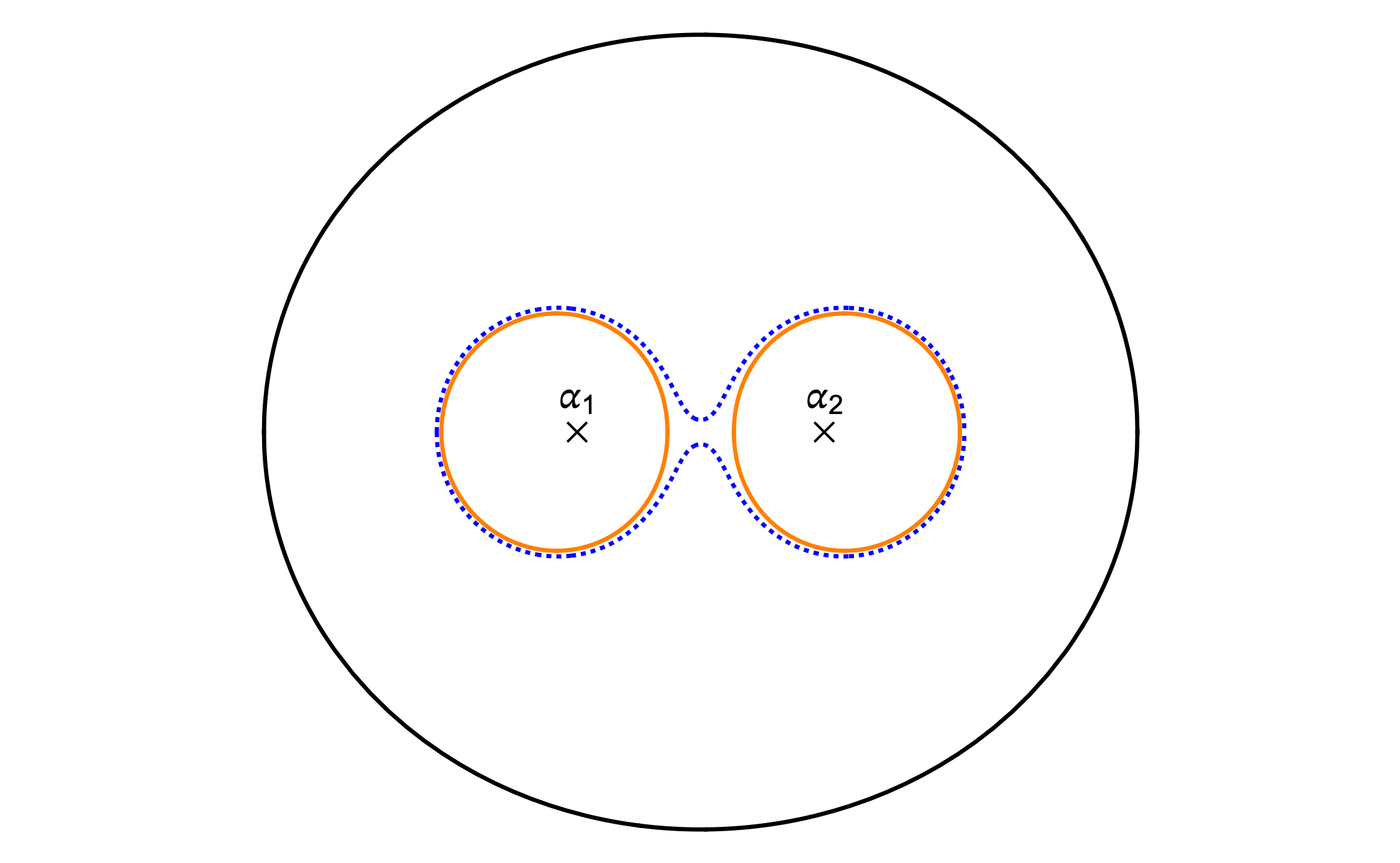}
        \caption*{$\tau = 0.7$}
        \label{fig:a_2/m = 1.03}
    \end{subfigure}
    \begin{subfigure}[b]{0.48\textwidth}
        \centering
        \includegraphics[width = \textwidth]{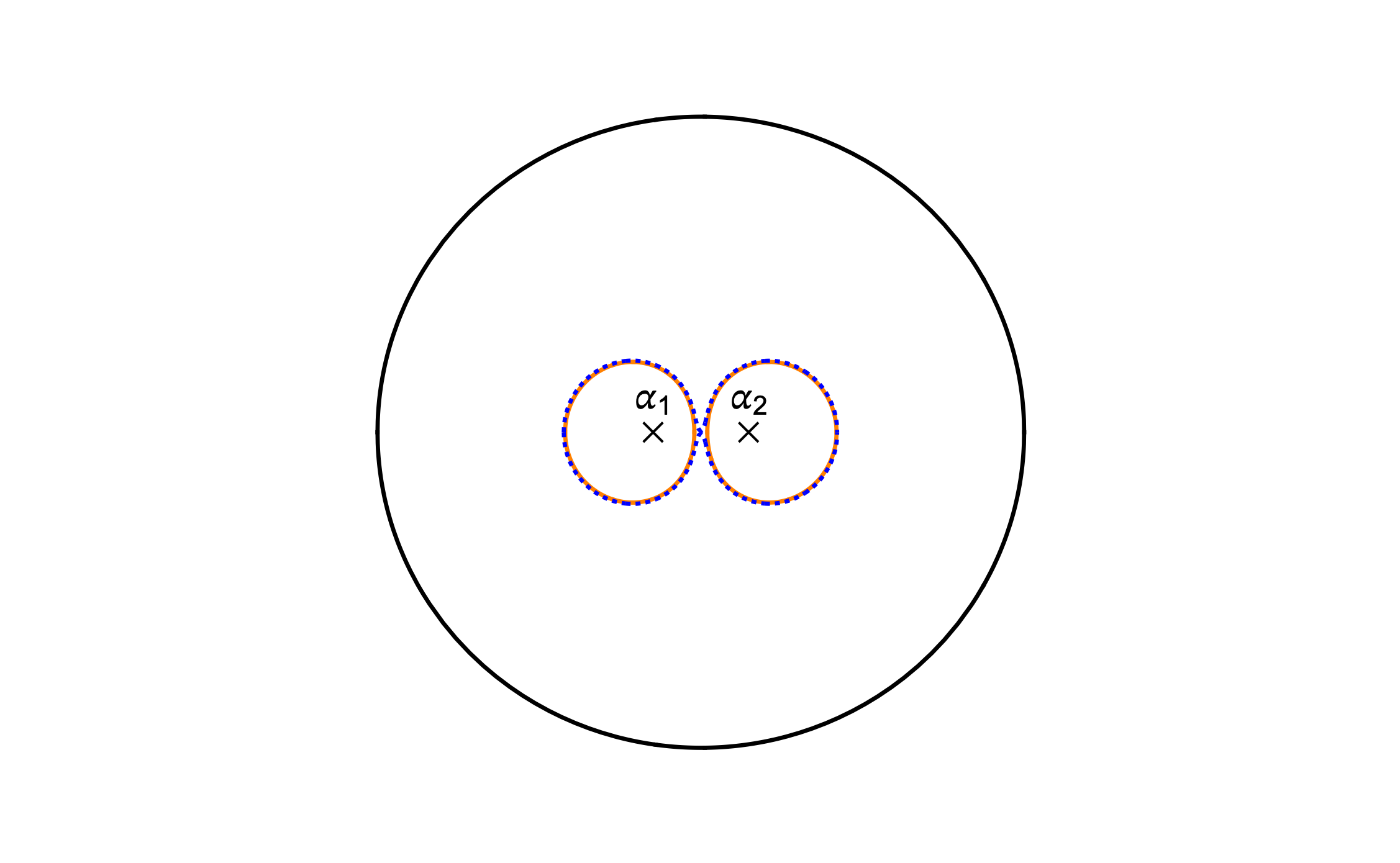}
        \caption*{$\tau =  1.03$}
        \label{fig:a_2/m = 1.28}
    \end{subfigure}
    \vspace{0.2 cm}
    \caption{Snapshots of the $n=3$ wormhole geometry after the Page time. The black surface is $\gamma_{\rm BH}$, dotted blue is $\tilde\gamma_{3}$ (bulge surface) and the left and right orange surfaces are $\gamma_1$ and $\gamma_2$ ($\gamma_{\rm R} = \gamma_1 \cup \gamma_2$). Observe how $\tilde\gamma_3$ `hugs' $\gamma_{\rm R}$ closer as the evaporation progresses.}
    \label{fig:N=3 snapshots}
\end{figure}

This late-time behavior of the areas can also be seen by looking at the form of the minimal surfaces. For late times, the surface $\tilde\gamma_3$ approaches $\gamma_{\rm R}$ as shown in Fig. \ref{fig:N=3 snapshots}; thus the areas must approach each other as well as.

As a final check, I also consider the evolution of the ADM masses.
\begin{figure}[ht]
    \centering
    \includegraphics[width = 0.8\textwidth]{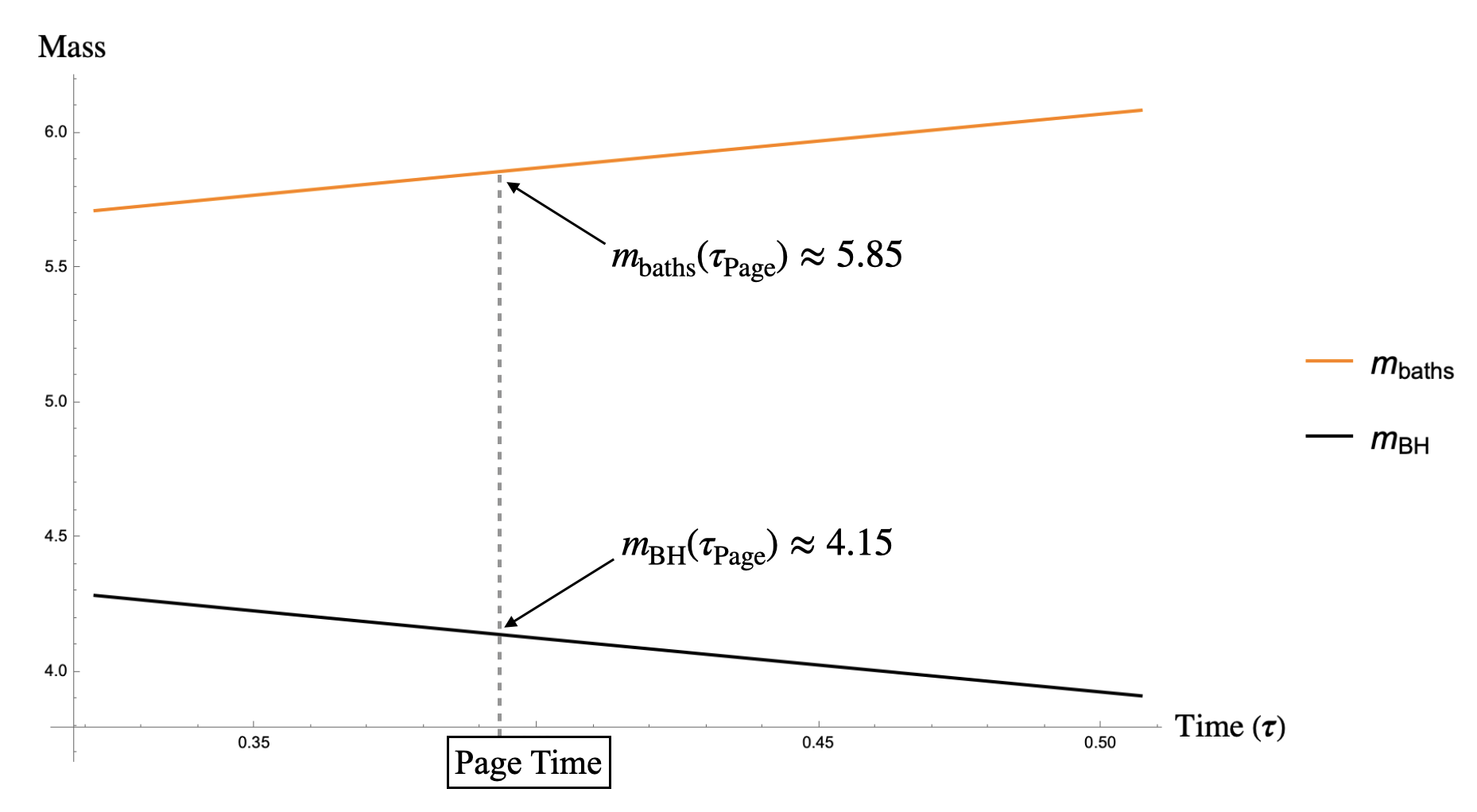}
    \caption{Evolution of black hole mass $m_{\rm BH} = m_3$ and radiation bath mass $m_{\rm baths} = m_1 + m_2$ during evaporation}
    \label{fig:Evaporation mass N=3}
\end{figure}
As shown in Fig. \ref{fig:Evaporation mass N=3}, the masses behave as we expect, with $m_{\rm BH}$ decreasing and $m_{\rm baths}$ increasing as the black hole radiates into the baths. The masses are also useful in computing the ratio of the evaporated black hole area to its initial area, $|\gamma_{\rm BH} (\tau)|/|\gamma_0|$, where $|\gamma_0|$ is the initial area. This quantity is relevant as in \cite{page2}, Page computed its value and found that for large Schwarzschild black holes that mainly emit photons and gravitons, $|\gamma_{\rm BH} (\tau_{\text{Page}})|/|\gamma_0| \approx 0.60$. As black hole evaporation is irreversible, this ratio should be greater than $0.5$ at the Page time. Since we can approximate the evaporating black hole in the BL model as being Schwarzschild, for which $m^2_{\rm BH} \propto |\gamma_{\rm BH}|$, we can compute the area ratio by using the mass ratio $|\gamma_{\rm BH} (\tau_{\text{Page}})|/|\gamma_0| = (m_\text{BH}(\tau_{\text{Page}})/m_0)^2$, even though we don't know $|\gamma_0|$. For the $n=3$ model, the ratio $m_{\rm BH} (\tau_{\text{Page}})/m_0 = 4.15/5 = 0.83$, and thus $|\gamma_{\rm BH} (\tau_{\text{Page}})|/|\gamma_0 \approx 0.69$, which is greater than $0.5$. This provides a useful check on the validity of the model.

\subsection[Brill-Lindquist evaporation model: n = 4]{Brill-Lindquist evaporation model: $n=4$}
\label{sec:n4}
\begin{figure}[ht]
    \centering
    \begin{subfigure}[b]{0.45\textwidth}
        \centering
        \includegraphics[width = \textwidth]{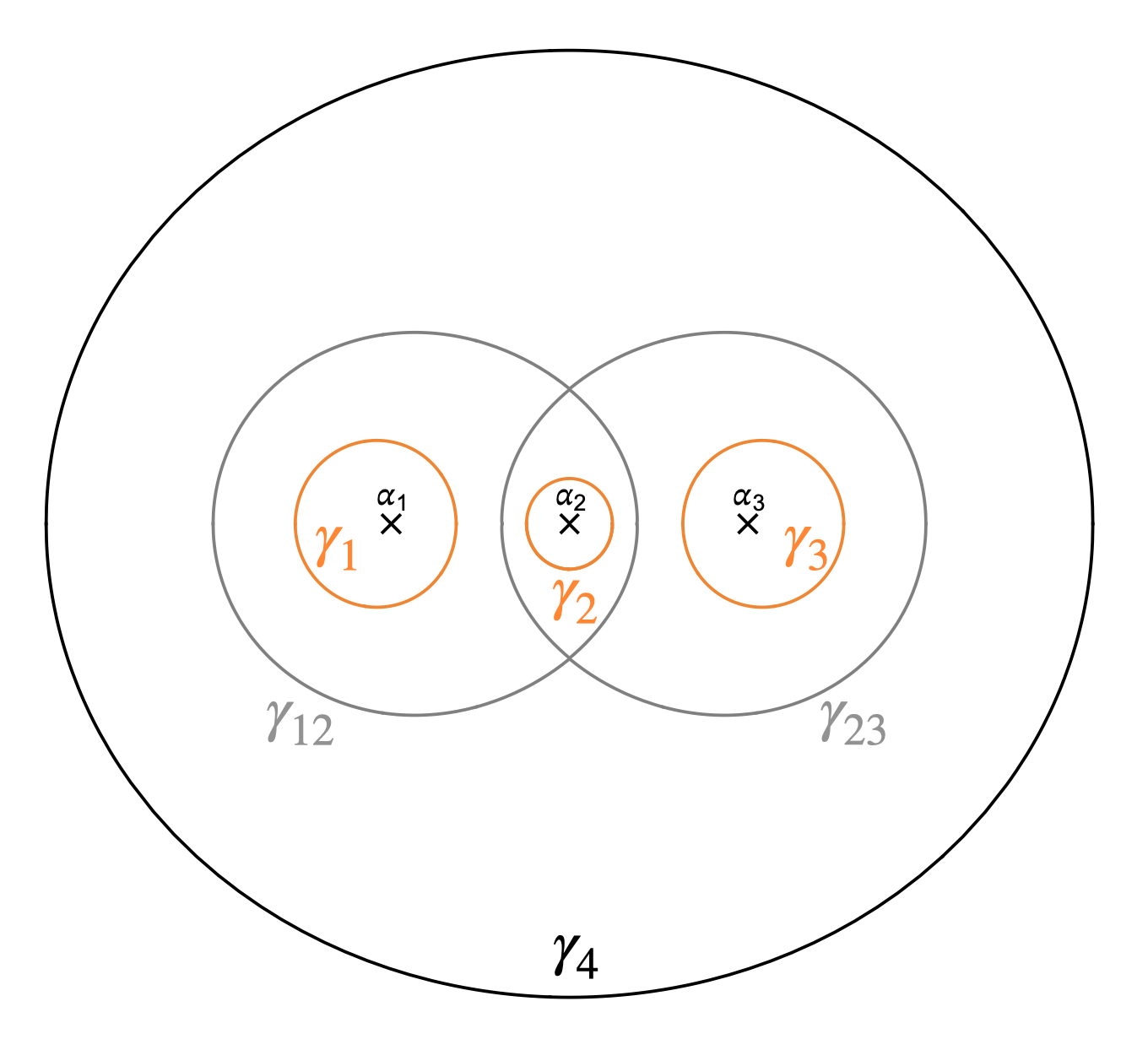}
    \end{subfigure}\hfill
    \begin{subfigure}[b]{0.45\textwidth}
        \centering
        \includegraphics[width=\textwidth]{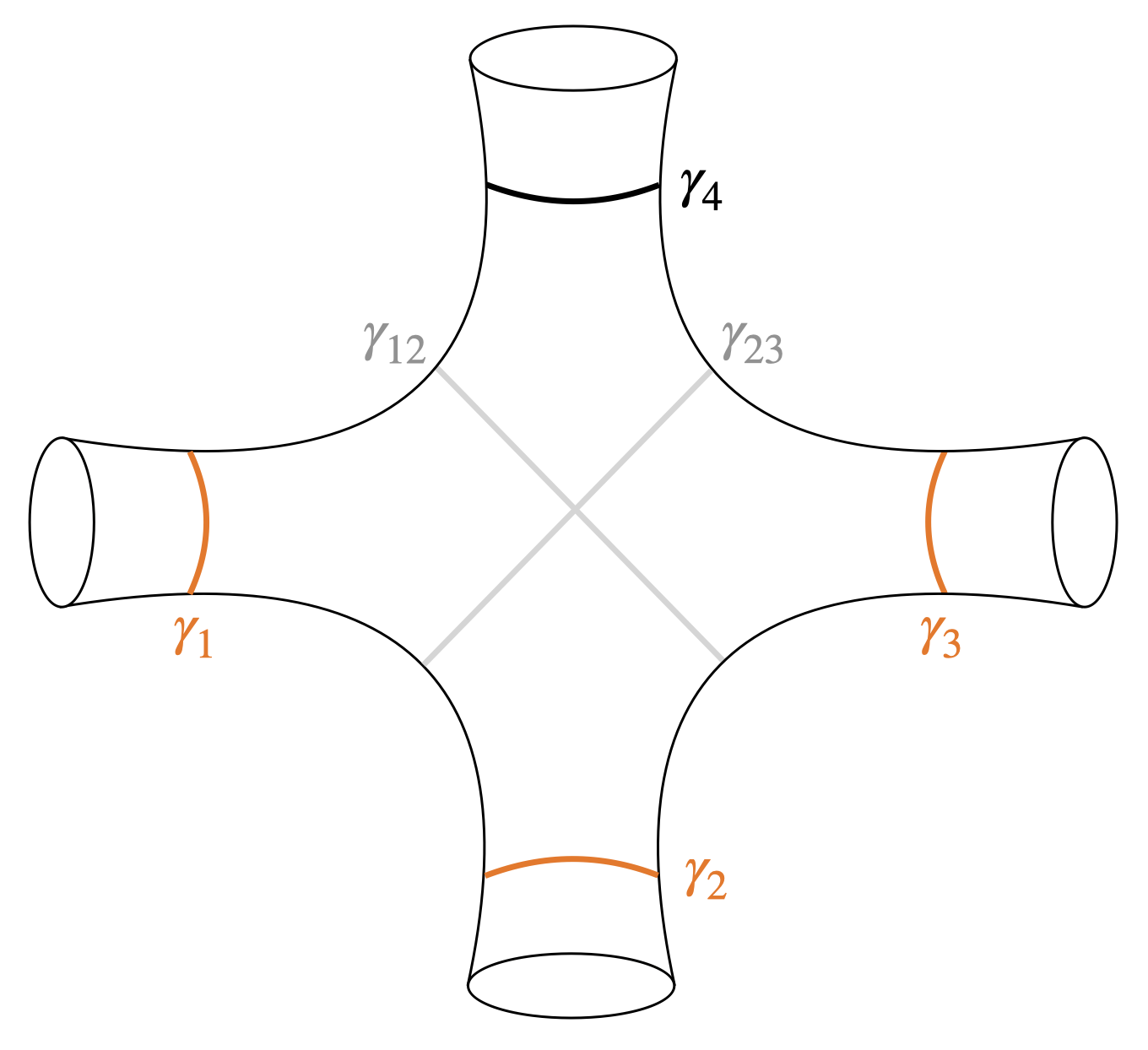}
    \end{subfigure}
    \caption{Minimal surfaces for $n=4$ BL geometry in (left) $\vec{r}$ coordinates and (right) the wormhole geometry.}
    \label{fig:n4 minimal}
\end{figure}

Using an analogous procedure to the one described above for $n=3$ BL wormholes, a similar model can be constructed using $n=4$ BL wormholes. \cite{gupta2025entangleduniverses} provides a more complete discussion of the geometry; here I will discuss the basic features necessary for the model. The geometry is similar to a four-sided AdS wormhole, with four asymptotic regions $a_i$ and a minimal surface $\gamma_i$ associated to each region. The region $a_4 \rightarrow A$ is taken to be the ``head'' of the octopus with the identifications $\gamma_4 \rightarrow \gamma_{\rm BH}, \gamma_1 \cup \gamma_2 \cup \gamma_3 \rightarrow \gamma_{\rm R}$. There are also additional `crossing minimal' surfaces $\gamma_{12},\gamma_{23}$ as shown in Fig. \ref{fig:n4 minimal}.

\begin{figure}[ht]
    \centering
    \begin{subfigure}[b]{0.45\textwidth}
        \centering
        \includegraphics[width = \textwidth]{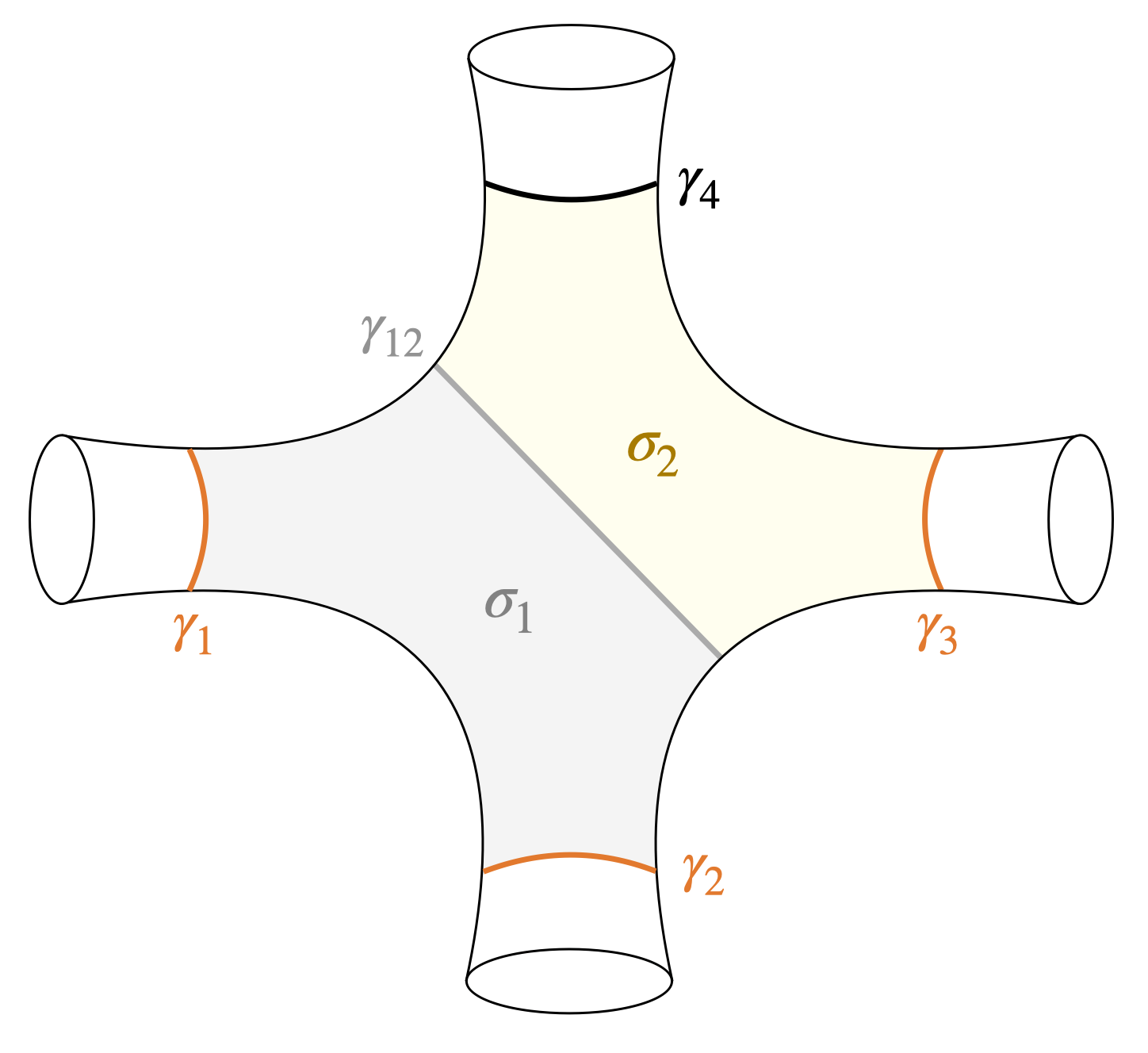}
    \end{subfigure}\hfill
    \begin{subfigure}[b]{0.45\textwidth}
        \centering
        \includegraphics[width=\textwidth]{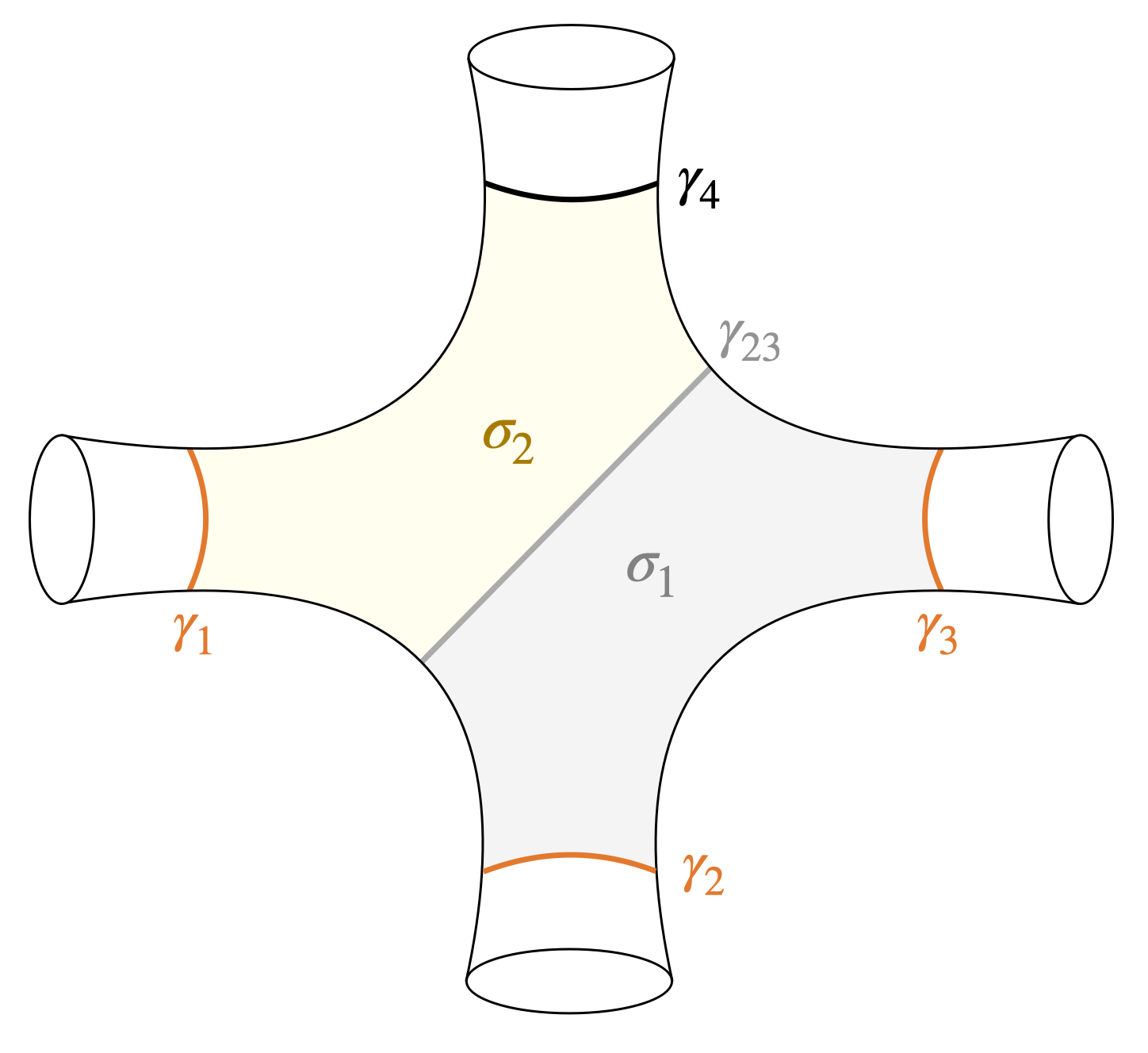}
    \end{subfigure}
    \caption{Choice of non-intersecting set of minimal surfaces including (left) $\gamma_{12}$ or (right) $\gamma_{23}$. I will choose $\gamma_{12}$, yielding the setup on the left.}
    \label{fig:n4 choose gamma12}
\end{figure}

Since the surfaces are intersecting, we must choose one of them so that we have a set of non-intersecting, adjacent minimal surfaces. Setting $m_1 = m_3$ fixes $\alpha_1 = \alpha_3 =: \alpha$, which yields a configuration symmetric under the exchange of $\gamma_1$ and $\gamma_3$. Thus the surfaces are identical and either choice yields the same results. As shown in Fig. \ref{fig:n4 choose gamma12}, I will choose $\gamma_{12}$ which separates the geometry into the Cauchy subregions $\sigma_1,\sigma_2$.

I will once again start by computing the evolution of the areas of the minimal surfaces before moving on to the PLC. Setting $m_1 = m_2 = m_3$ we also have the simplification $|\gamma_1|=|\gamma_2|=|\gamma_3|$. Aside from $\gamma_{\rm BH}, \gamma_{\rm R}$ we now have the additional minimal surface $\gamma_{\rm M} := \gamma_{12} \cup \gamma_{3}$ that is also a candidate RT surface. I set $m_{\rm total} = 10$, $r_{12} = r_{23}$ and the time co-ordinate is chosen to be $t = \alpha_2/r_{12}$. A connected geometry is found for $r_{12} \approx 3.2 \alpha_2$\footnote{A complete analysis of the critical separation(s) for $n=4$ BL geometries remains incomplete; this is only an approximate value found for this model.}, setting the start time $t_0 \approx 0.31$. For the evaporation to end in finite time, I once again define $\tau = \arctan{t}$ to track the evolution.

\begin{figure}[ht]
    \centering
    \includegraphics[width=0.8\textwidth]{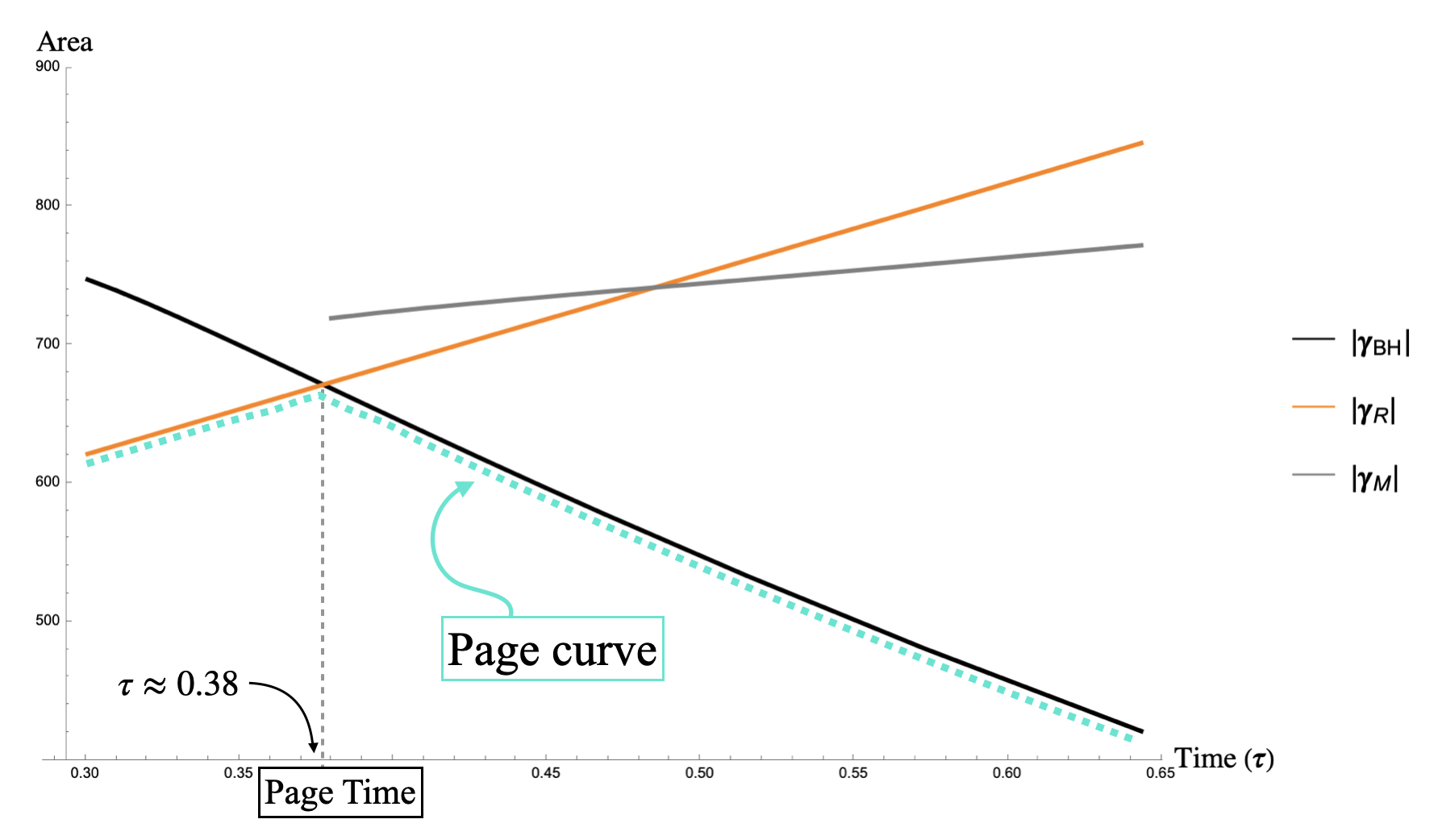}
    \caption{Minimal area (cyan) following the Page curve for $n=4$ evaporation model.}
    \label{fig:BL evaporation N=4}
\end{figure}

Areas of the surfaces as shown in Fig. \ref{fig:BL evaporation N=4} once again recover the Page curve for the model. Interestingly, the new minimal surface $\gamma_{\rm M} = \gamma_{12} \cup \gamma_3$ is never the RT surface, and in fact seems to form only at the Page time (this can be verified with greater accuracy by employing numerical methods to find the critical separation required for $\gamma_{12}$ to form).

The mutual information(s) are more complicated than for $n=3$. Prior to the Page time, both $I(1:2)$ and $I(1:3)$ are 0. After the Page time we expect to have,
\begin{align}
    4I(1:2) &= 2|\gamma_1| - |\gamma_{12}| \\
    4I(1:3) &= 2|\gamma_1| - |\gamma_2|-|\gamma_4| = |\gamma_1| - |\gamma_4|
\end{align}
However, the surfaces $\gamma_{12}$ and $\gamma_2 \cup \gamma_4$ are not minimal immediately at the Page time. They are only minimal some time after, until which we still have $I(1:2) = I(1:3) = 0$. When they do become positive, both these quantities seemingly increase with time.

The mutual information that actually transitions to be non-zero at the Page time is $I(12:3)$. Prior to the Page time it is 0, and after it is $4 I(12:3) = 3|\gamma_1| - |\gamma_4|$, which is again an increasing quantity as expected.\footnote{It should be noted that the expressions for the mutual information may undergo another transition as each asymptotic region has 3 candidate minimal surfaces. Regardless of the transitions, I expect them to be increasing once they are non-zero.}

\begin{figure}[ht]
    \centering
    \begin{subfigure}[b]{0.5\textwidth}
        \centering
        \includegraphics[width = \textwidth]{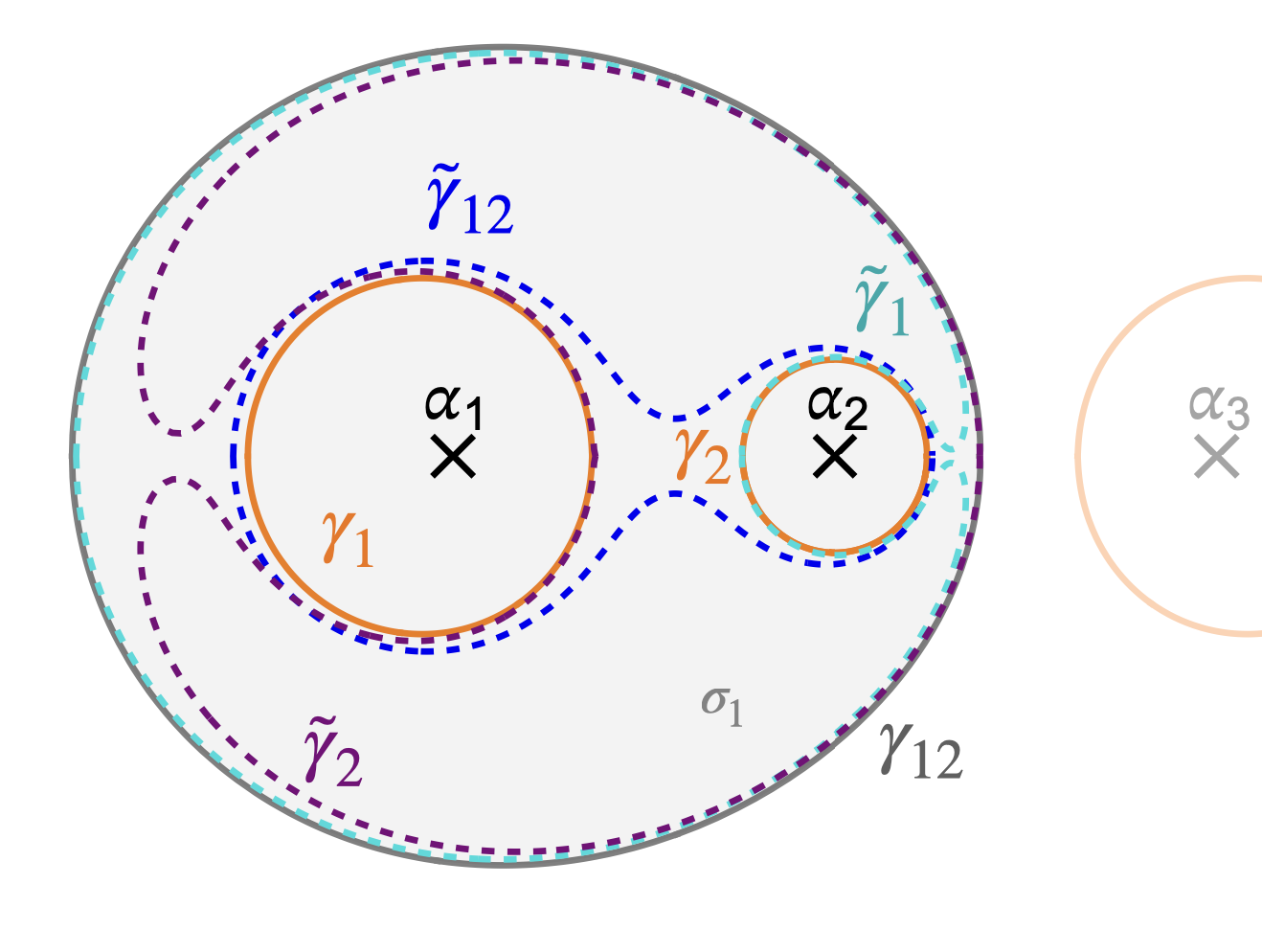}
    \end{subfigure}\hfill
    \begin{subfigure}[b]{0.4\textwidth}
        \centering
        \includegraphics[width=\textwidth]{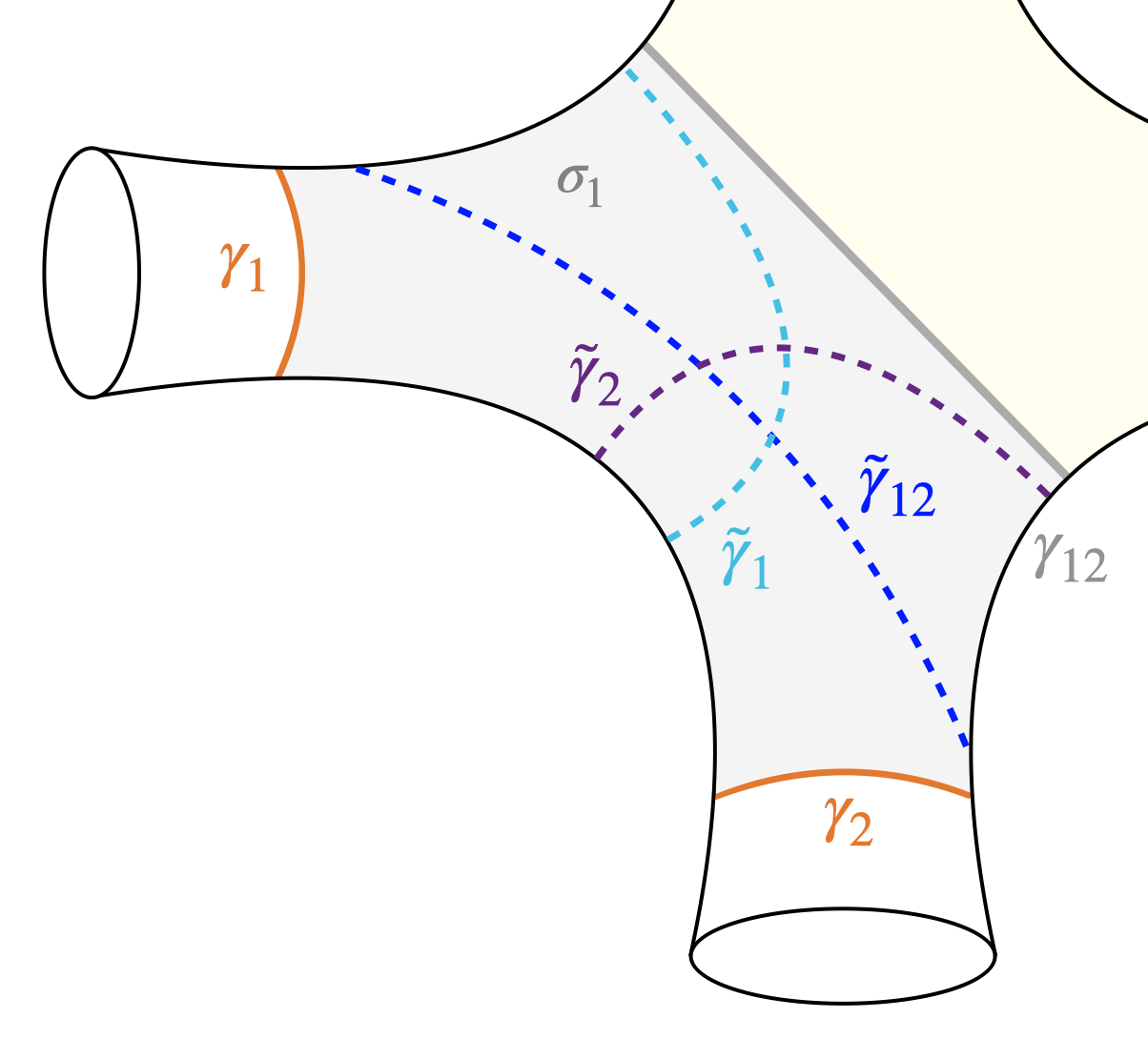}
    \end{subfigure}
    \caption{Index-1 surfaces in subregion $\sigma_1$ in (left) $\vec{r}$ coordinates and (right) the wormhole geometry.}
    \label{fig:sigma1 index1}
\end{figure}

The computation of the python's lunch is also far more complex than for $n=3$ as we have multiple adjacent minimal surfaces and multiple index-1 surfaces in each subregion. From \eqref{eq:complexity multiple lunches multiple index} we see that we must first find the minimal index-1 surface within each subregion $\sigma_1,\sigma_2$. The candidate bulges in $\sigma_1$ are the surfaces $\{\tilde\gamma_{1} \cup \gamma_2 \cup \gamma_3, \gamma_1 \cup \tilde\gamma_2 \cup \gamma_3, \tilde\gamma_{12} \cup \gamma_3\}$ as shown in Fig. \ref{fig:sigma1 index1}. For late times $\tilde\gamma_{1} \rightarrow \gamma_2 \cup \gamma_{12}, \tilde\gamma_2 \rightarrow \gamma_1 \cup \gamma_{12}$ and $\tilde\gamma_{12} \rightarrow \gamma_1 \cup \gamma_2$ (the surfaces approach each other). Thus the areas go as
\begin{align}
    |\tilde\gamma_{1} \cup \gamma_2 \cup \gamma_3| &\sim 3|\gamma_1| + |\gamma_{12}| \\
    |\gamma_{1} \cup \tilde\gamma_2 \cup \gamma_3| &\sim 3|\gamma_1| + |\gamma_{12}| \\
    |\tilde\gamma_{12} \cup \gamma_3| &\sim 3|\gamma_1|,
\end{align}
where we can see that $\tilde\gamma_{12} \cup \gamma_3$ is minimal and thus should be the candidate bulge for $\sigma_1$.

\begin{figure}[ht]
    \centering
    \begin{subfigure}[b]{0.45\textwidth}
        \centering
        \includegraphics[width = \textwidth]{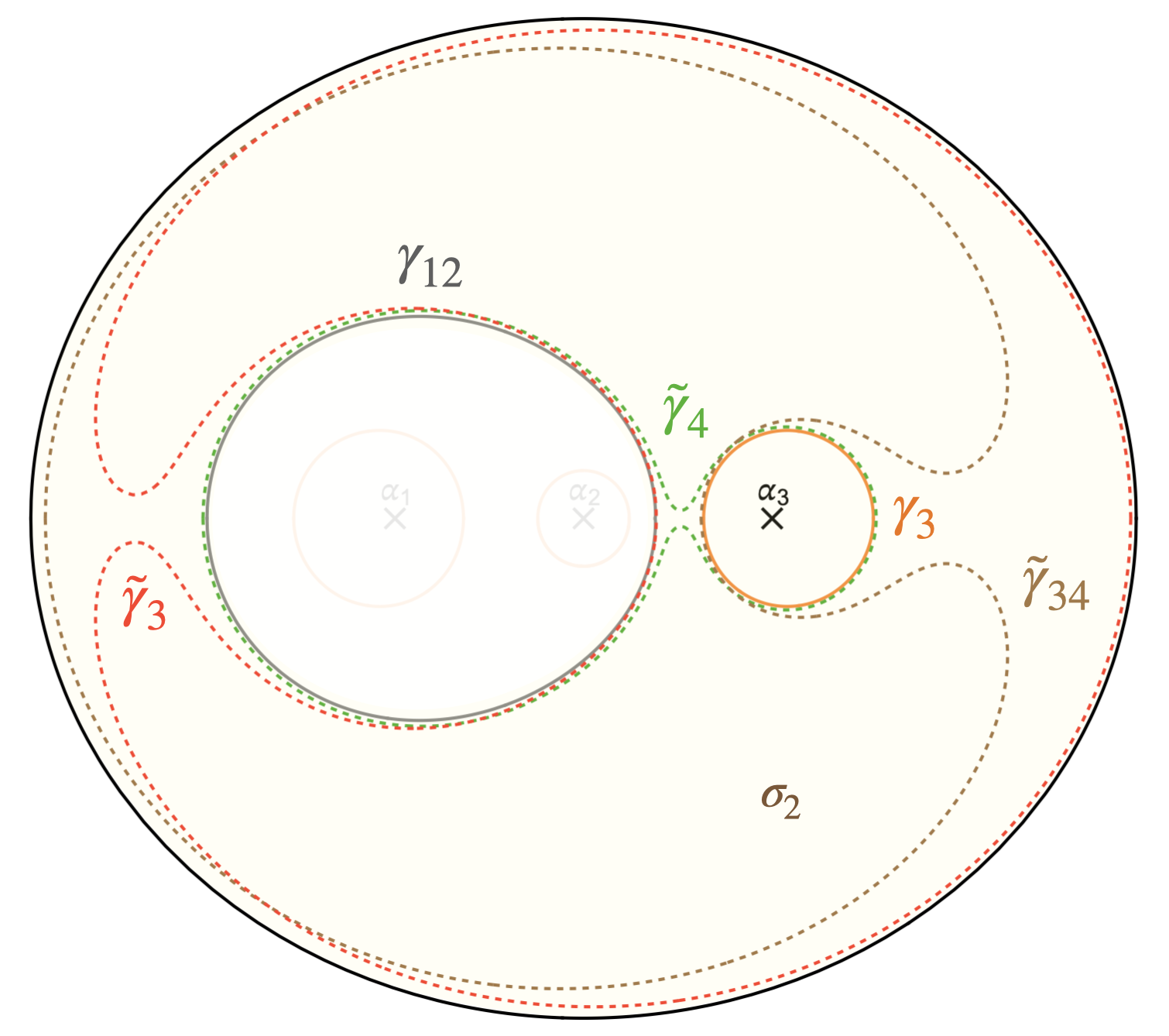}
    \end{subfigure}\hfill
    \begin{subfigure}[b]{0.45\textwidth}
        \centering
        \includegraphics[width=\textwidth]{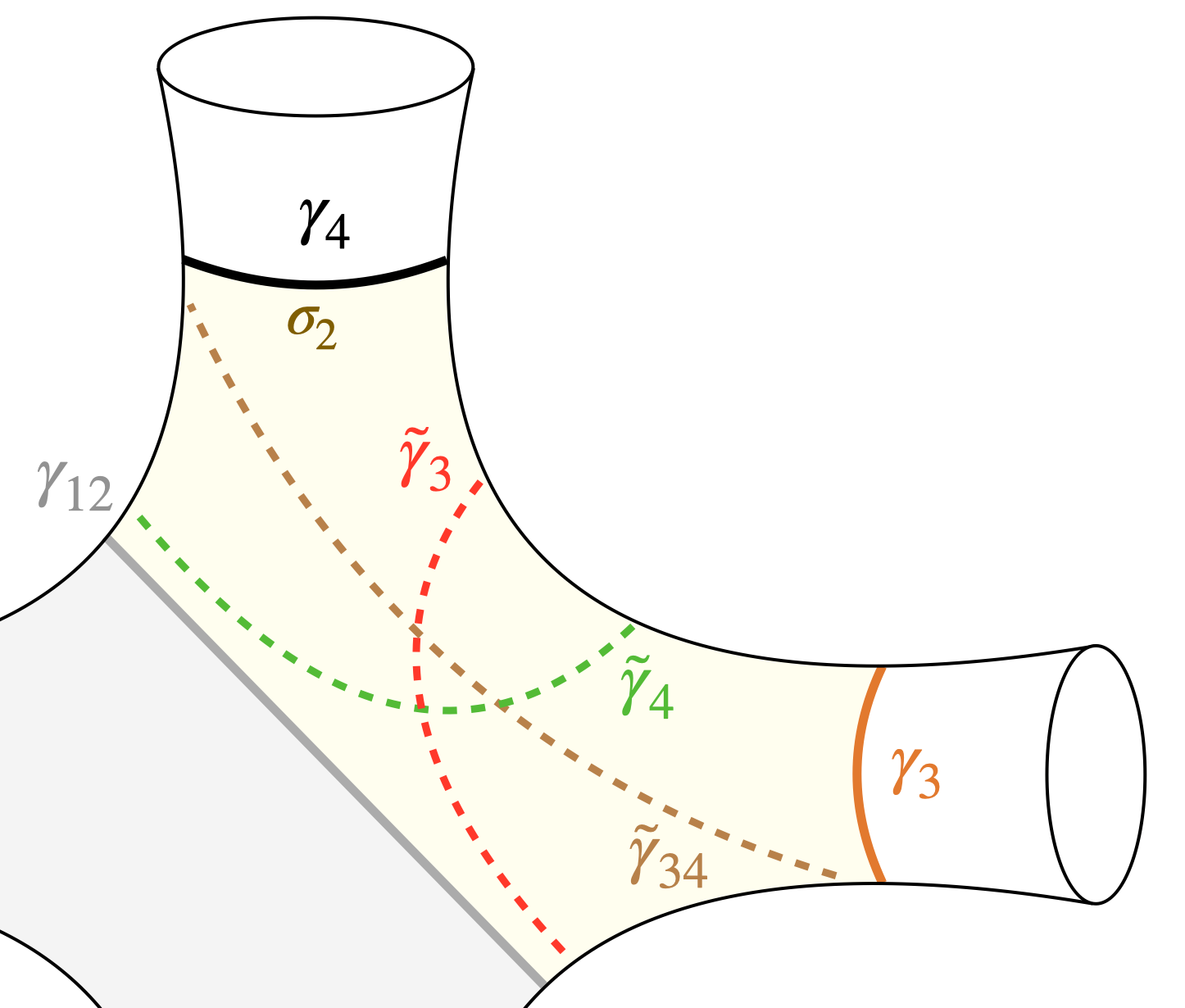}
    \end{subfigure}
    \caption{Index-1 surfaces in subregion $\sigma_2$ in (left) $\vec{r}$ coordinates and (right) the wormhole geometry.}
    \label{fig:sigma2 index1}
\end{figure}

The candidate bulges in $\sigma_2$ are the surfaces $\{\tilde\gamma_{4}, \gamma_{12} \cup \tilde\gamma_3, \tilde\gamma_{34} \cup \gamma_3\}$ as shown in Fig. \ref{fig:sigma2 index1}. To find the minimal area index-1 surface we note that for late times $\tilde\gamma_{4} \rightarrow \gamma_{12} \cup \gamma_{3}, \tilde\gamma_3 \rightarrow \gamma_{12} \cup \gamma_{4}$ and $\tilde\gamma_{34} \rightarrow \gamma_3 \cup \gamma_4$. Thus the areas go as
\begin{align}
    |\tilde\gamma_{4}| &\sim |\gamma_1| + |\gamma_{12}| \\
    |\gamma_{12} \cup \tilde\gamma_3| &\sim |\gamma_4| + 2|\gamma_{12}| \\
    |\tilde\gamma_{34} \cup \gamma_3| &\sim 2|\gamma_1| + |\gamma_4|,
\end{align}
where the asymptotic behavior is not entirely clear.

\begin{figure}[ht]
    \centering
    \includegraphics[width=0.8\textwidth]{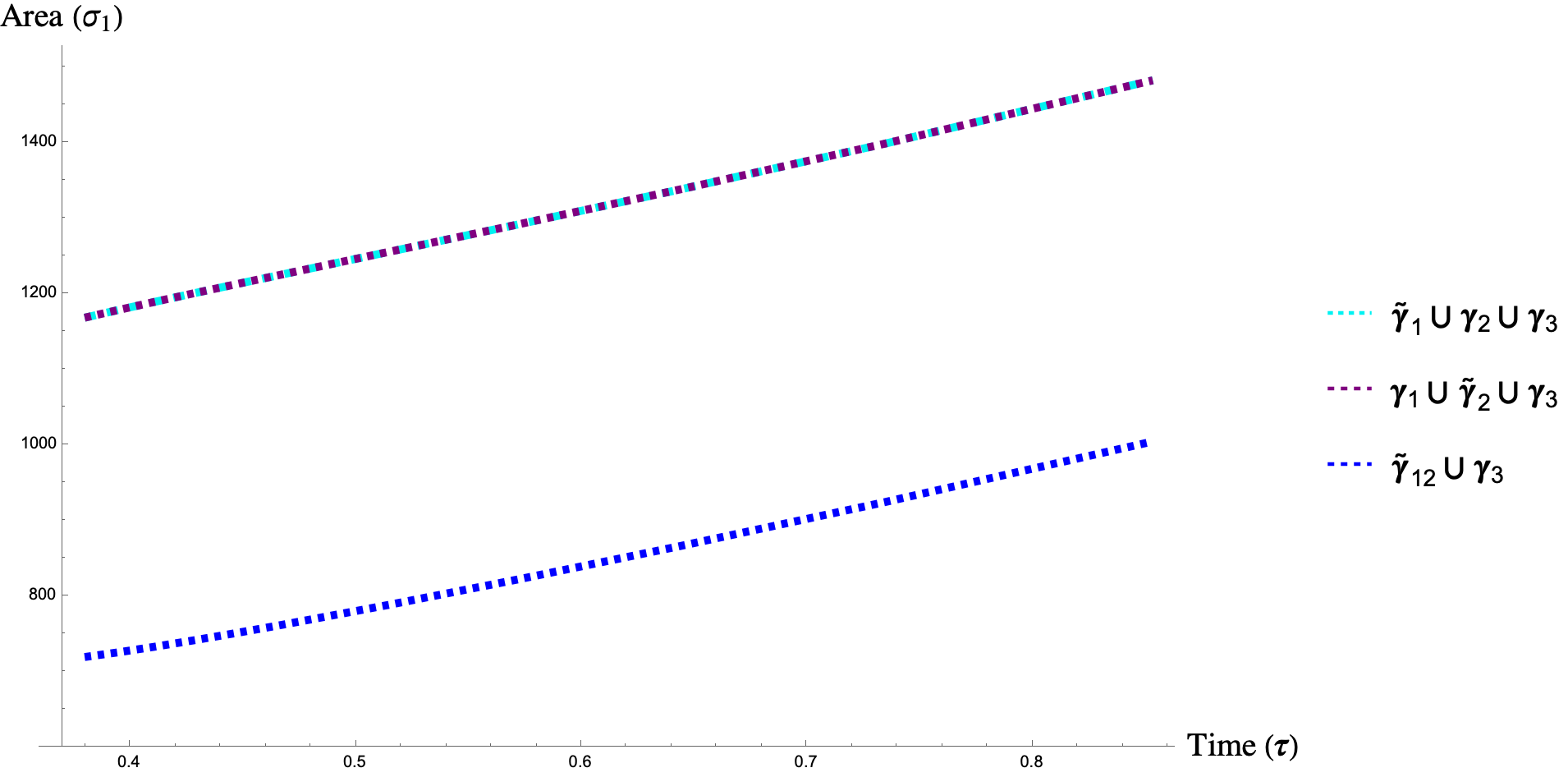}
    \caption{Areas of index-1 surfaces in $\sigma_1$, with the minimal area surface being $\tilde\gamma_{12} \cup \gamma_3$. The cyan and purple curves are overlapping.}
    \label{fig:sigma1 areas}
\end{figure}

\begin{figure}[ht]
    \centering
    \includegraphics[width=0.8\textwidth]{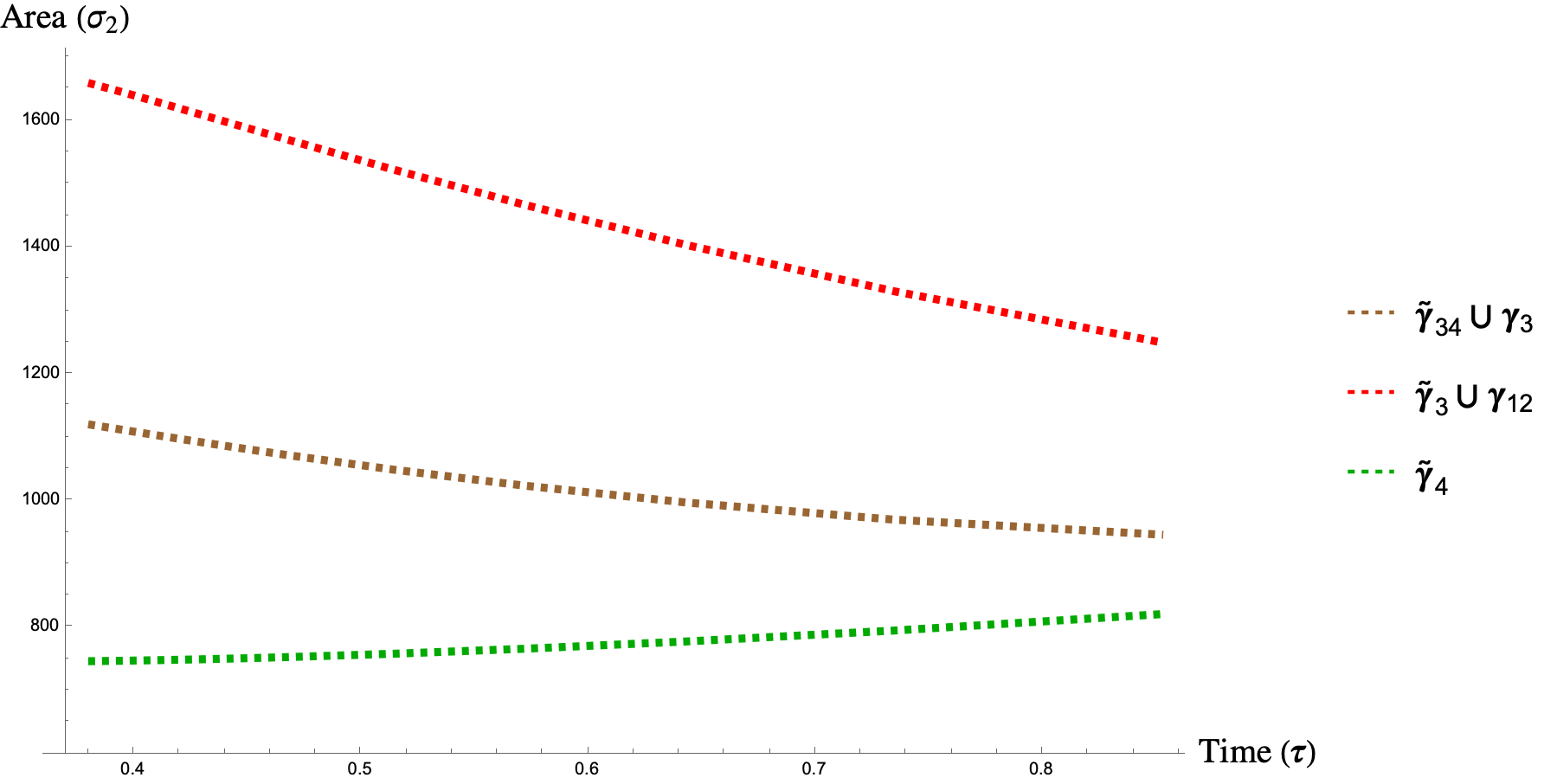}
    \caption{Areas of index-1 surfaces in $\sigma_2$, with the minimal area surface being $\tilde\gamma_4$.}
    \label{fig:sigma2 areas}
\end{figure}

We can instead find the minimal surface in each subregion numerically. Fig. \ref{fig:sigma1 areas} verifies our analysis for $\sigma_1$, where we find $\gamma^b_{1} = \tilde\gamma_{12} \cup \gamma_3$. From Fig. \ref{fig:sigma2 areas} we see that for $\sigma_2$, $\tilde\gamma_4$ has the smallest area; we thus have $\gamma^b_2 = \tilde\gamma_4$. Note that the graphs are for times after the Page time, which is when all the surfaces in $\sigma_1$ form; however the surfaces in $\sigma_2$ form at earlier times. Regardless, the analysis in this section will be regarding their behavior after the Page time as this is when the predictions of the (generalized) PLC apply.


\begin{figure}[ht]
    \centering
    \includegraphics[width=0.85\linewidth]{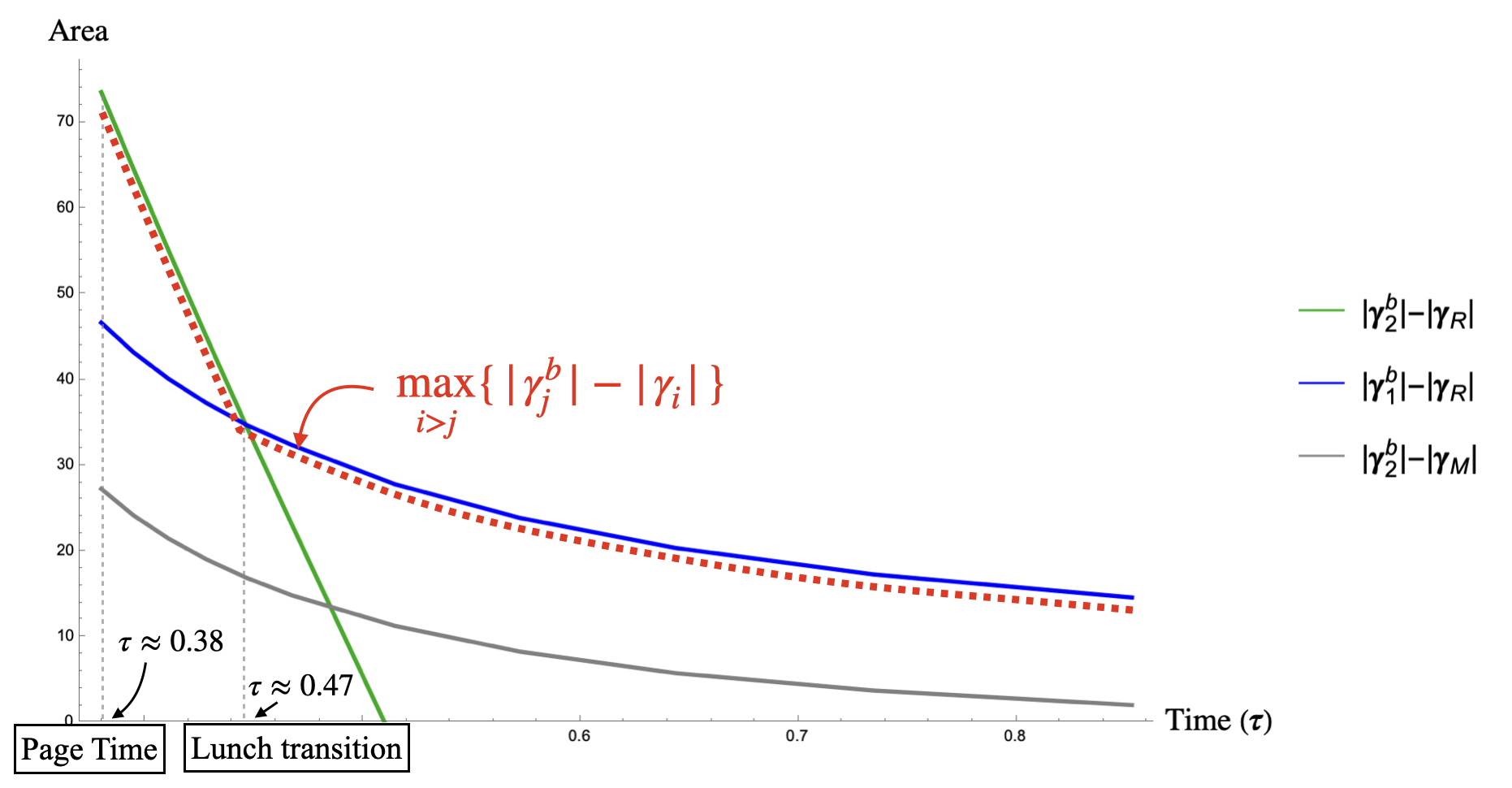}
    \caption{Time evolution of area differences $\mathcal{A}$. $|\gamma^b_2 |-|\gamma_{\rm R}|$ becomes negative, and the plot has been truncated to positive area differences. Red curve is the maximal area difference, which appears as the argument of the exponent in \eqref{eq:complexity multiple lunches}.}
    \label{fig:N=4 python's lunch}
\end{figure}

With a unique bulge $\gamma^b_1, \gamma^b_2$ in each subregion, we can now compute the complexity by applying the maximization procedure in \ref{eq:complexity multiple lunches multiple index}. The maximization is done over the set $\mathcal{A} = \{|\gamma^b_2 |-|\gamma_{\rm M}|, |\gamma^b_2 |-|\gamma_{\rm R}|,|\gamma^b_1 |-|\gamma_{\rm R}|\}$. The evolution of the areas with time is shown in Fig. \ref{fig:N=4 python's lunch}.

At the Page time (when all the candidate bulges have formed), the maximal difference is $|\gamma^b_2 |-|\gamma_{\rm R}|$; thus we have $\log{\mathcal{C}(t)} \sim |\gamma^b_2 |-|\gamma_{\rm R}|$. At $\tau \approx 0.47$, the bulge surface switches to $\gamma^b_1$ and we have $\log{\mathcal{C}(t)} \sim |\gamma^b_1|-|\gamma_{\rm R}|$. It is tempting to conclude that the lunch transition here must correspond to the transition outlined in \cite{Brown:2019rox} between the forward and reverse sweep surface since the transition involves the lunch surface changing from $\tilde\gamma_{4}$ to $\tilde\gamma_{12}\cup \gamma_3$, with the constriction remaining the radiation horizon $\gamma_1 \cup \gamma_2 \cup \gamma_3$. However this is not immediately clear, as the wormhole geometries are different: in the $n=4$ BL evaporation model there is an additional minimal surface $\gamma_{\rm M}$ and the bulge surfaces lie in different Cauchy subregions, while in the python's lunch setup only $\gamma_{\rm BH}, \gamma_{\rm R}$ are present and the forward and reverse sweep surfaces are in the same subregion.
\\\\
Fig. \ref{fig:N=4 python's lunch} suggests the late time complexity is controlled by $|\gamma^b_1|-|\gamma_{\rm R}|$, which goes to 0. This is once again evident from looking at snapshots of the evolution.
\begin{figure}[ht]
    \centering
    \begin{subfigure}[b]{0.49\textwidth}
        \centering
        \includegraphics[width = \textwidth]{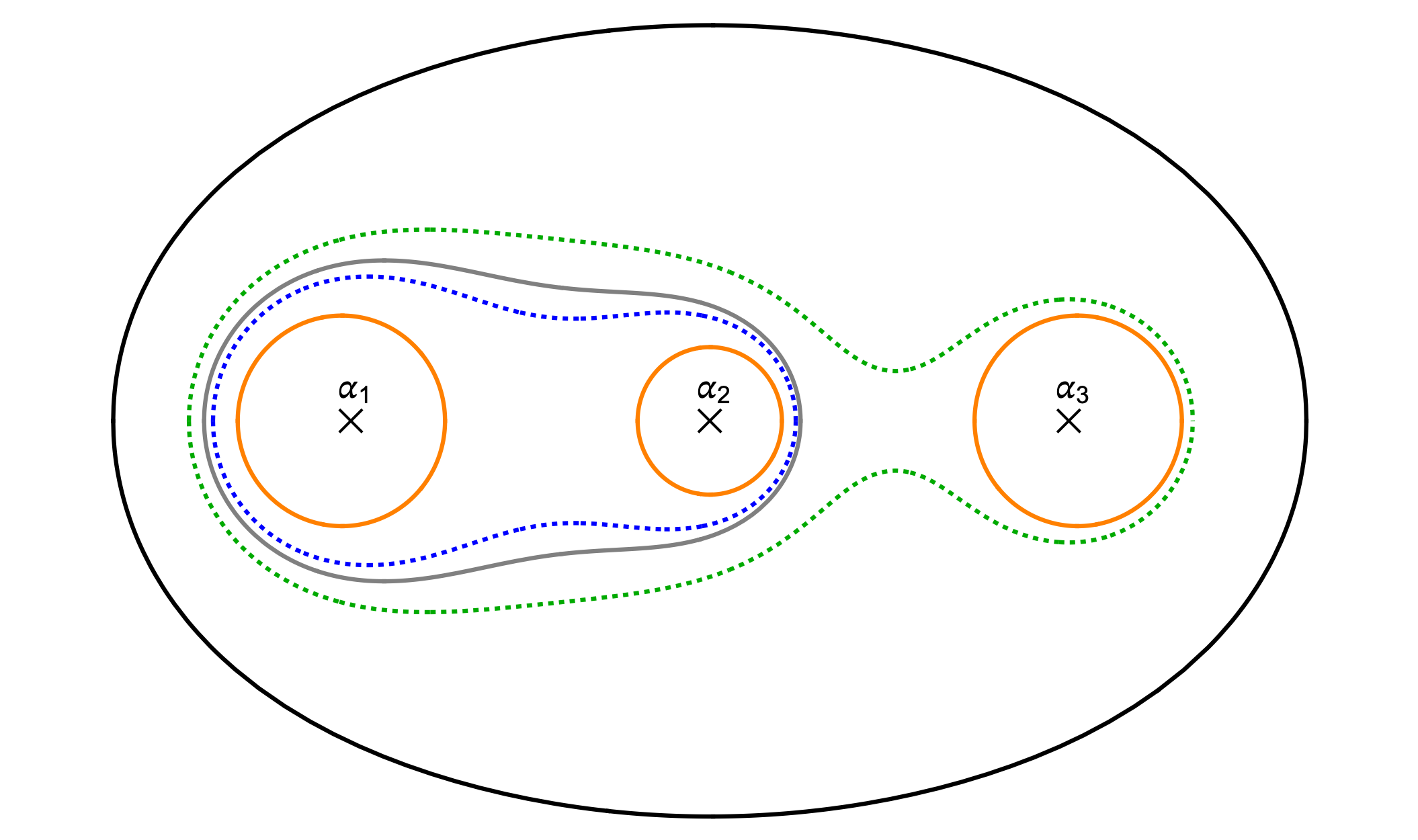}
        \caption*{$\tau = \tau_{\text{Page}} = 0.38$}
        \label{fig:a2/m = 1.26}
    \end{subfigure}
    \vspace{0.3 cm}
    \begin{subfigure}[b]{0.49\textwidth}
        \centering
        \includegraphics[width =\textwidth]{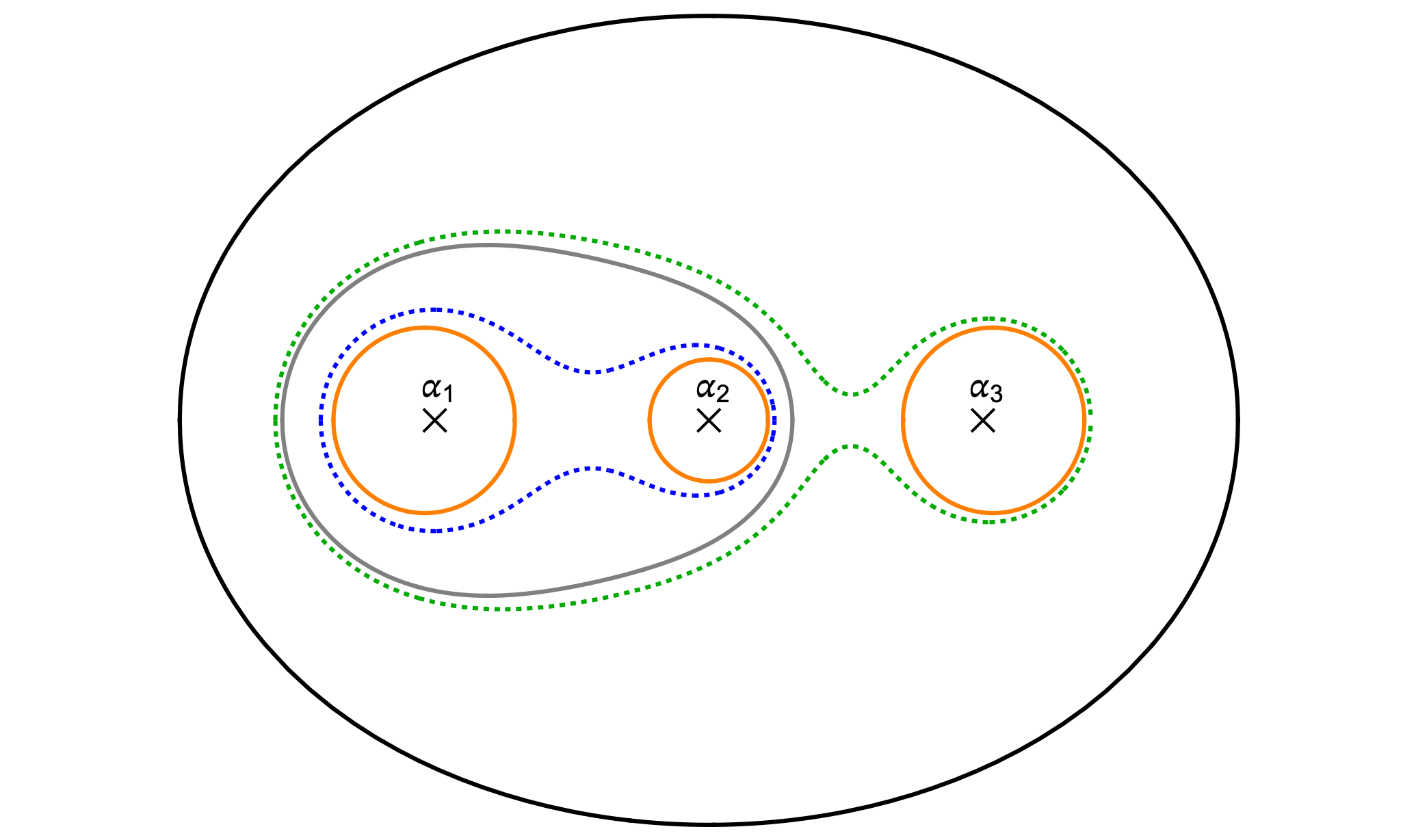}
        \caption*{$\tau = 0.45$}
        \label{fig:a2/m2 = 1.04}
    \end{subfigure}\\[\smallskipamount]
    \begin{subfigure}[b]{0.49\textwidth}
        \centering
        \includegraphics[width = \textwidth]{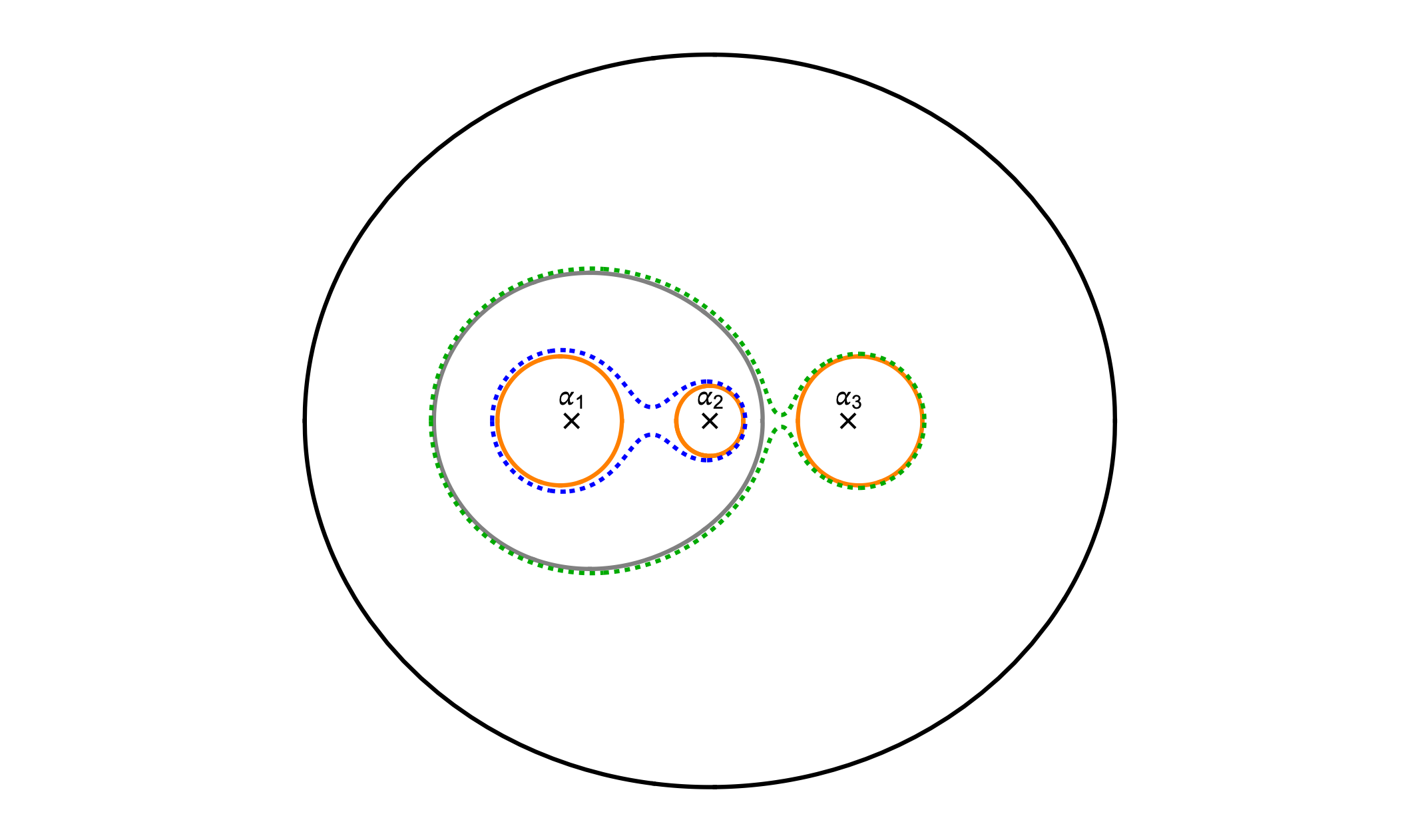}
        \caption*{$\tau = 0.64$}
        \label{fig:a2/m2 = 0.67}
    \end{subfigure}
    \begin{subfigure}[b]{0.49\textwidth}
        \centering
        \includegraphics[width = \textwidth]{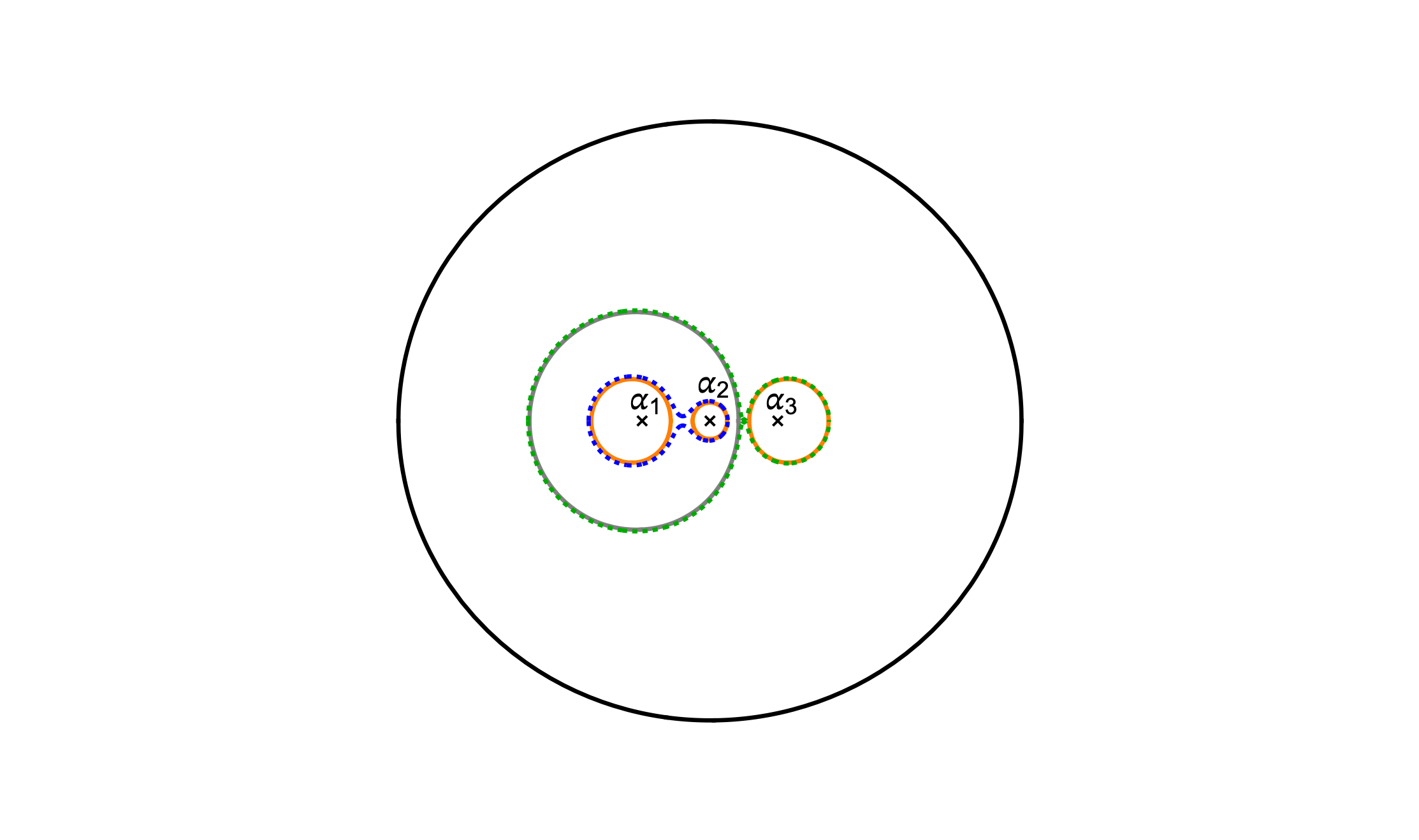}
        \caption*{$\tau =  0.85$}
        \label{fig:a2/m2 = 0.44}
    \end{subfigure}
    \vspace{0.2 cm}
    \caption{Snapshots of the $n=4$ wormhole geometry after the Page time. The black surface is $\gamma_{\rm BH}$, the gray surface is $\gamma_{12}$, dotted blue is $\tilde\gamma_{12}$, dotted green is $\tilde\gamma_4$, and the orange surfaces are $\gamma_1,\gamma_2,\gamma_3$ ($\gamma_{\rm R} = \gamma_1 \cup \gamma_2 \cup \gamma_3$). Observe how $\tilde\gamma_4$ `hugs' $\gamma_{12} \cup \gamma_3$ and $\tilde\gamma_{12}$ `hugs' $\gamma_1 \cup \gamma_2$ closer as the evaporation progresses.}
    \label{fig:N=4 snapshots}
\end{figure}
As can be seen in Fig. \ref{fig:N=4 snapshots}, as we approach $\tau \rightarrow \pi/2$ the surfaces $\tilde\gamma_{12} \rightarrow \gamma_1 \cup \gamma_2, \tilde\gamma_4 \rightarrow \gamma_{12} \cup \gamma_3$ approach each other. Thus regardless of whether their relative sizes change, both $|\gamma^b_2|-|\gamma_{\rm M}|, |\gamma^b_1| - |\gamma_{\rm R}| \rightarrow 0$ and the complexity becomes polynomial as expected.

\begin{figure}[ht]
    \centering
    \includegraphics[width=0.8\textwidth]{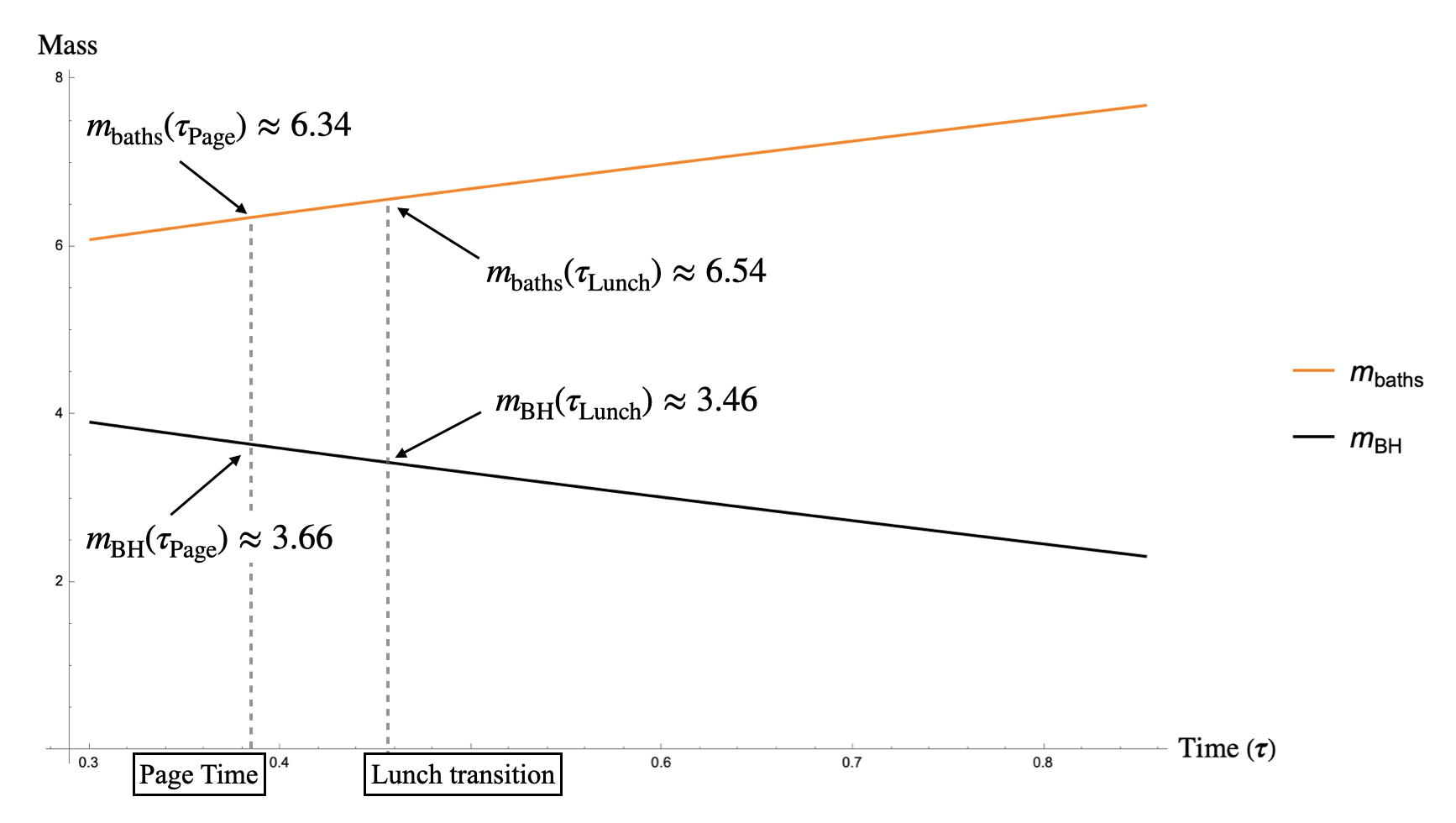}
    \caption{Evolution of black hole mass $m_{\rm BH} = m_4$ and radiation bath masses $m_{\rm baths} = m_1 + m_2 + m_3$ during evaporation.}
    \label{fig:Evaporation mass N=4}
\end{figure}

We also observe how the ADM masses evolve in Fig. \ref{fig:Evaporation mass N=4}. There are 2 important transition points to check. The first is the Page time, where similar to the $n=3$ model we compute $|\gamma_{\rm BH}(\tau_{\text{Page}})|/|\gamma_0| = (m_\text{BH}(\tau_{\text{Page}})/m_0)^2 = (3.66/5)^2 \approx 0.54$. This is lower than the value for $n=3$ (0.69) and Page's computed value (0.60); however it is still greater than 0.5 as we expect for black hole evaporation. Furthermore, \cite{Brown:2019rox} conjectures that the lunch transition from the forward to reverse sweep surface occurs at $|\gamma_{\rm BH}(\tau_{\text{Lunch}})|/|\gamma_0| = 0.5$, where $\tau_{\text{Lunch}}$ is the time when the lunch surface switches from the forward to reverse sweep surface. For the $n=4$ BL model, the lunch transition is at $|\gamma_{\rm BH}(\tau_{\text{Lunch}})|/|\gamma_0| = (m_\text{BH}(\tau_{\text{Lunch}})/m_0)^2 = (3.46/5)^2 \approx 0.48$. This value is close to the predicted forward-reverse sweep transition value and thus seemingly provides evidence that the lunch transition in the $n=4$ BL model is indeed between a forward and reverse sweep bulge surface. It is also interesting to note that while the BL geometry involves the additional minimal surface $\gamma_{\rm M}$, the area of this surface is not involved in the expression for the complexity at any point in time. However, higher $n$ BL evaporation models will have more minimal and index-1 surfaces and thus may have more transitions, which should be checked to determine whether lunch transitions in BL evaporation models correspond to the forward-reverse sweep transition.

\begin{figure}[ht]
    \centering
    \begin{subfigure}[b]{0.49\textwidth}
        \centering
        \includegraphics[width=\textwidth]{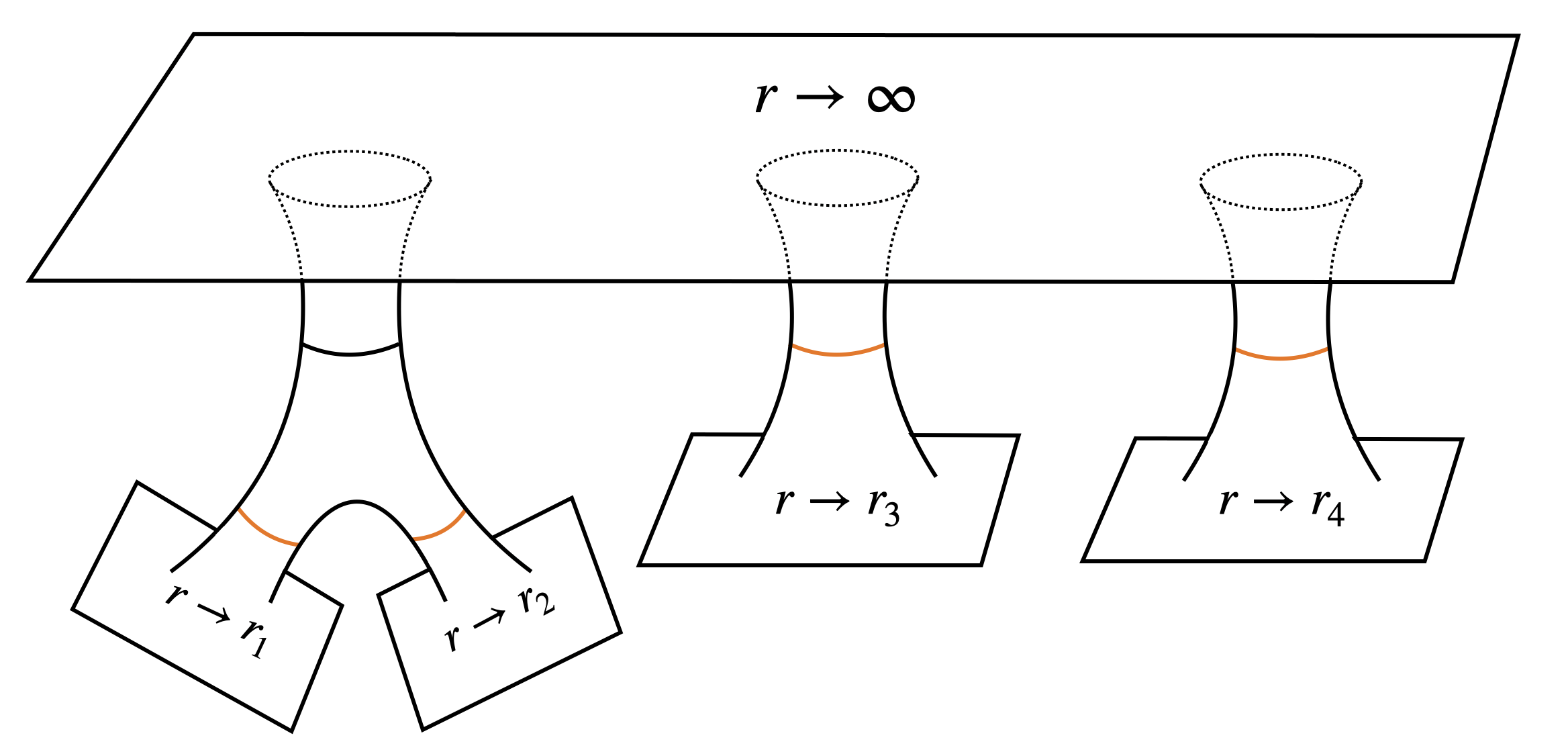}
        \label{fig:Evaporation pre gluing}
    \end{subfigure}\hfill
    \begin{subfigure}[b]{0.49\textwidth}
        \centering
        \includegraphics[width=\textwidth]{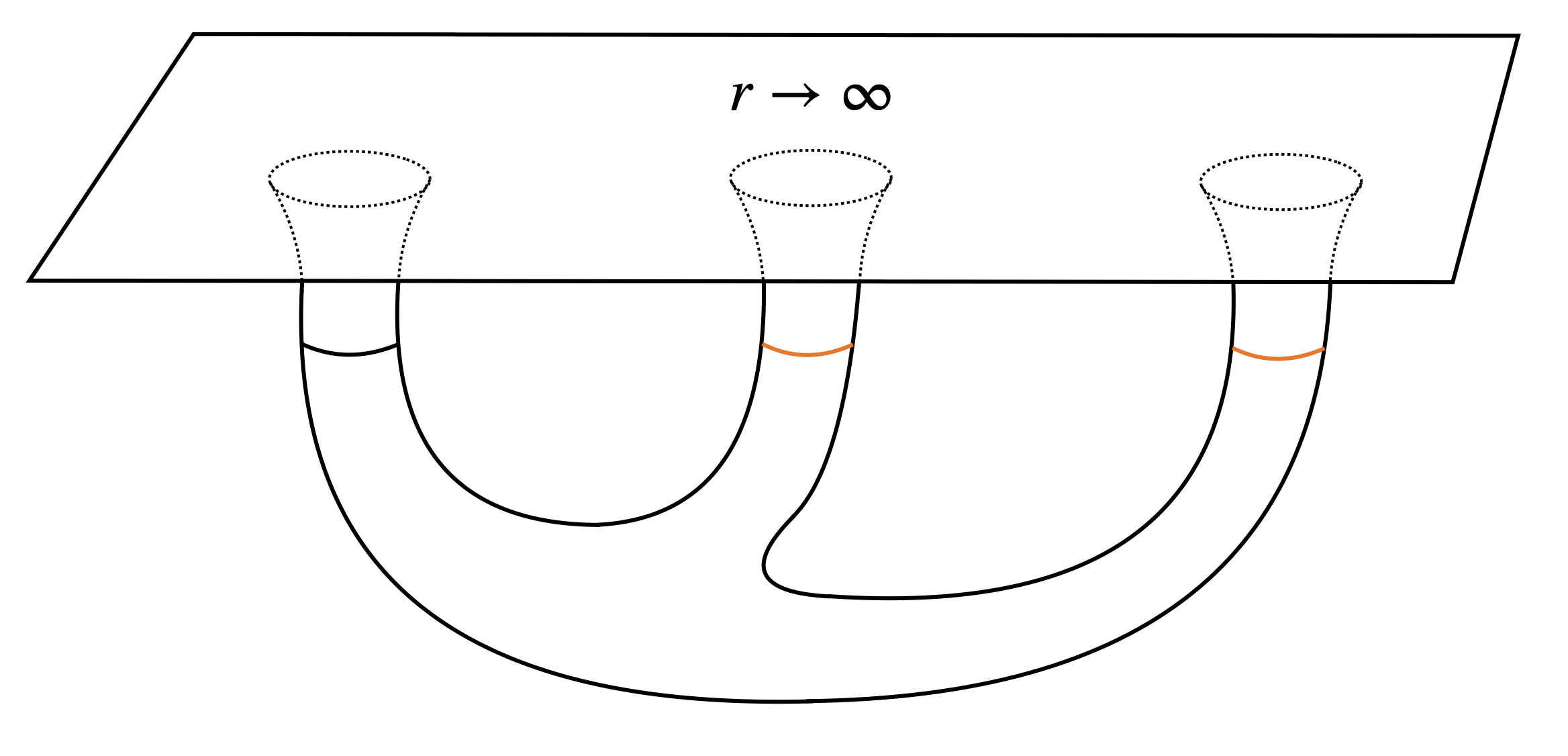}
        \label{fig:Evaporation post gluing}
    \end{subfigure}
    \caption{We start with an $n=5$ BL geometry, with two punctures $\alpha_1, \alpha_2$ close enough to form a connected geometry and the rest well-separated. Choosing parameters such that the areas of the orange minimal surfaces are equal and the extrinsic curvatures match, we can ``cut off" the asymptotic regions and glue along minimal surfaces. The black minimal surface is then the black hole horizon, which emits radiation into the other 2 legs (attached to the same $\|r\| \rightarrow \infty$ asymptotic region), and the orange minimal surfaces are the radiation horizons.}
    \label{fig:Evaporation gluing}
\end{figure}

A few general points should be noted about these BL evaporation models: 
\begin{itemize}
    \item For late times, the black hole horizon $\gamma_n \rightarrow \gamma_{\rm BH}$ is expected to be minimal. In the limit $\alpha_i \gg r_{ij}$ ($t \rightarrow \infty$) \cite{brill-lindquist} showed that $\gamma_n$ can be approximated as a sphere with radius $r = m_n/2 = m_{\rm BH}/2$. Since $m_{\rm BH} \rightarrow 0$ as $t\rightarrow\infty$, $|\gamma_{\rm BH}| \rightarrow 0$, and the black hole entirely evaporates. Furthermore, provided $\gamma_n$ is the minimal surface, the entanglement entropy $S(A) \rightarrow 0$ in accordance with the Page curve.
    \item The BL metric at each snapshot is time-reflection symmetric, while the true evaporation geometry is clearly not. This is because the BL metric merely represents a snapshot of the evolution of the evaporation process. A time-reflection symmetric metric can be used due to the same reasoning as in \cite{Akers:2019nfi}, as we have simply chosen here to cancel the natural Lorentzian evolution, which would only make the wormhole interior grow. Since the evolution can also be described by the product of unitary operators living on distinct AdS boundaries glued to the BL asymptotic regions \cite{gupta2025entangleduniverses}, the Lorentzian evolution does not affect the Page curve and can be canceled. This may affect the areas of the surfaces which are in the wormhole interior; however I expect the qualitative features of the relevant quantities to remain the same i.e. which bulge surface is the python's lunch, which minimal surface is the RT surface etc.
    \item In these models, the radiation is collected in baths that live in a distinct universe from the black hole, whereas physical Hawking radiation should be in the same universe as the evaporating black hole. Here we are making the same assumption as in \cite{erEpr}, where even though the wormhole connecting the black hole and radiation (baths) must connect regions of the same universe, if these are taken to be far apart we may approximate them as being in disjoint universes. One could directly try to construct such a `single universe' model of black hole evaporation using the BL metric, where the wormhole legs are cut along minimal surfaces and connected to form wormholes ``feeding into" the same universe (see Fig. \ref{fig:Evaporation gluing}). The metric will not exactly match on either side of the gluings, but for sufficiently separated punctures the error in the gluing will be small. However, in this large separation limit the interactions between the regions being glued together are negligible, which is why we can simply take them to be in separate asymptotic regions.
    \item Another implicit assumption in these models is that all relevant extremal surfaces lie on the time symmetric slice. This may not be the case, as in \cite{Engelhardt:2023bpv} explicit examples are constructed where the maximinimax procedure picks out a bulge surface off the symmetric slice. This would force us to use the full covariant evolution of the evaporation geometry to extract predictions regarding the reconstruction complexity. It would also suggest that the methods used in this paper, rather than providing a sharp computation of the complexity, instead provide a bound. However, it remains unclear whether such constructions exist more generally or what their proper interpretation is. I will make the simplifying assumption that all relevant extremal surfaces lie on the symmetric slice.
\end{itemize}

\section{Discussion}
\label{sec:discussion}
In this paper, I have proposed a model of black hole evaporation using Brill-Lindquist wormholes to model the entanglement geometry. The entanglement entropy can be computed using an extension of the RT formula proposed in \cite{gupta2025entangleduniverses}; from this perspective this work can be considered a continuation of \cite{gupta2025entangleduniverses}, where I take the RT formula in asymptotically $\text{Mink}_D$ spacetimes \eqref{eq:flat RT} to be the necessary ingredient to construct this model of black hole evaporation. The entanglement entropy is numerically computed for the $n=3,4$ BL wormhole geometries where I show that the models reproduce the Page curve for the entanglement entropy, thus providing a unitary model of black hole evaporation. I also identify index-1 surfaces in the geometries as candidate bulges which can be studied using the (generalized) python's lunch conjecture. In this section I will discuss possible extensions and future work involving the model.

A core idea of this paper is that the RT formula (and at a deeper level, ER=EPR) provides sufficient justification to construct this model. One can justify this by noting that, as discussed in \cite{gupta2025entangleduniverses}, the RT formula is sufficient to form a tensor network (TN) description of the wormhole geometry \cite{Swingle:2009bg,Bao:2018pvs}. A TN description of the BL geometry with asymptotic regions $A_1,\cdots, A_n$ should exist, representing a state $\psi^{A_1 \cdots A_n} \in \mathcal{H}$ in the tensor product Hilbert space $\mathcal{H} = \mathcal{H}_{A_1} \otimes \cdots \otimes \mathcal{H}_{A_n}$. The description provides a discretization of the geometry along (non-intersecting) RT surfaces and many of the ideas involved in this model, including the relationship between the exponential decoding complexity and bulge surfaces in the python's lunch conjecture, can then be reinterpreted via TN toy models using the same arguments as their AdS counterparts. 

\begin{figure}[ht]
    \centering
    \begin{subfigure}{0.45\textwidth}
        \centering
        \includegraphics[width=\textwidth]{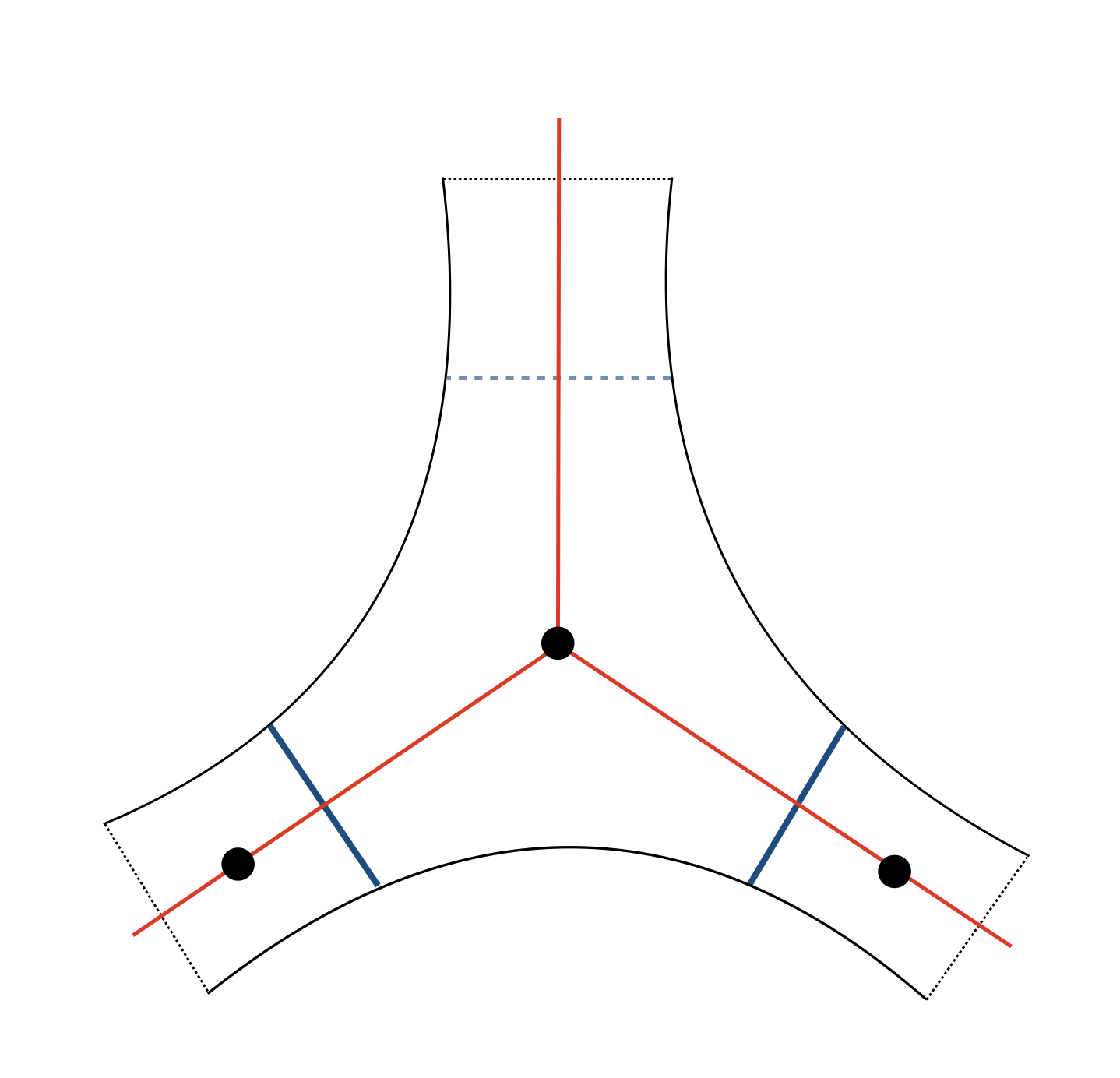}
    \end{subfigure}
    \begin{subfigure}{0.45\textwidth}
        \centering
        \includegraphics[width=\textwidth]{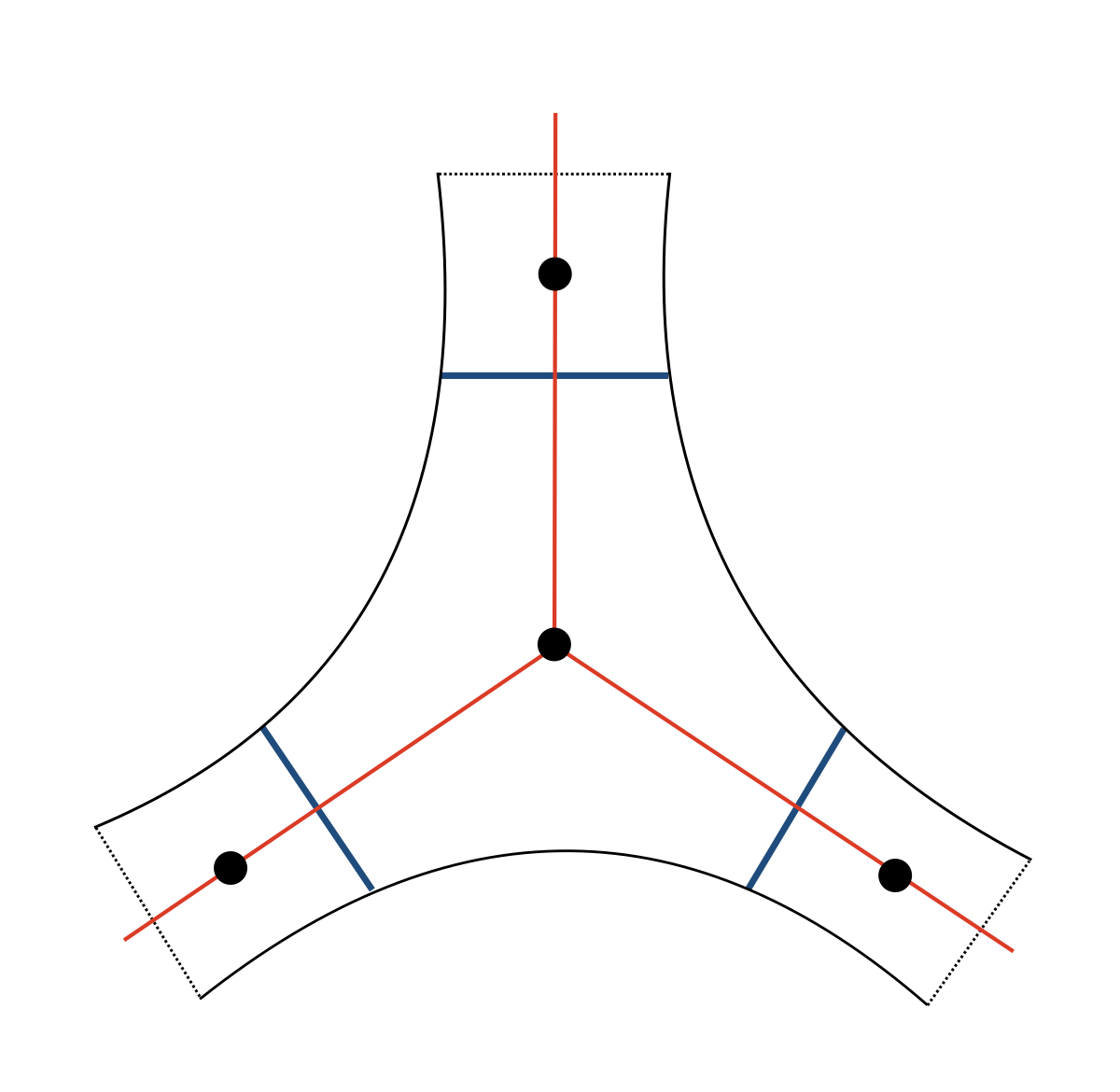}
    \end{subfigure}
    \caption{Tensor network for $n=3$ Brill-Lindquist geometry. The geometry is partitioned by RT surfaces (left) before the Page time, where $\gamma_3$ does not contribute to the TN and (right) after the Page time, where $\gamma_3$ appears.}
    \label{fig:BL_TN}
\end{figure}

It is interesting to note that for the BL evaporation geometry prior to the Page time, $\gamma_n$ is not the RT surface for the boundary region associated to the head of the octopus, $A_n$. By the assumed symmetry wherein $m_i = m_j$, we also have $|\gamma_i| = |\gamma_j|$ for all $i,j \leq n-1$. Thus $\gamma_n$ is not the RT surface (nor a component of the RT surface) for any combination of $A_1,\cdots,A_n$. This is explicitly evident from the TN description, where since it is not an RT surface it does not partition the geometry prior to the Page time. We can consider the $n=3$ geometry as an example in Fig. \ref{fig:BL_TN}.

Here we see that prior to the Page time, the discretization does not involve $\gamma_3$. The network thus involves one fewer node than the TN after the Page time, when $\gamma_3$ becomes an RT surface and partitions the geometry. 

The tensor network for the $n=4$ geometry is more complicated, with the same behavior before the Page time. After the Page time, there is a range of parameters where only $\gamma_4$ is an RT surface, not $\gamma_{12}$ or $\gamma_{23}$. However at a later time the RT surface for $A_1 \cup A_2$ is $\gamma_{12}$, and similarly for $A_2 \cup A_3$ it is $\gamma_{23}$ \cite{gupta2025entangleduniverses}. Since $\gamma_{12}, \gamma_{23}$ are intersecting RT surfaces, it is unclear if the TN can be constructed.

Directions for future work include various generalizations of the model. One could verify the results for higher $n$ BL geometries, where the structure of the minimal surfaces becomes more complex, which would provide a useful check on the validity of the model. Particularly it would be important to check that a Page curve is reproduced even for the higher $n$ geometries. Furthermore, the index-1 surfaces present in those geometries should also reproduce key features of the PLC, such as the complexity becoming polynomial as the black hole evaporates. It would also be interesting to see if higher $n$ geometries have more lunch transitions, and if any of these bear characteristics of the forward-reverse sweep transition in \cite{Brown:2019rox}.

Another generalization includes relaxing the constraint that the radiation baths have equal masses. This would break the exchange symmetries in the $n=3,4$ models, which would require the PLC expression \eqref{eq:complexity multiple lunches multiple index} to be slightly generalized. For $n=4$, the exchange symmetry between $\gamma_1,\gamma_3$ meant that the intersecting minimal surfaces $\gamma_{12},\gamma_{23}$ were identical and we could simply pick one of them to partition the geometry into Cauchy subregions. However, in the case where the symmetry is broken and the surfaces are not identical, the 2 possible choices of our set of non-intersecting minimal surfaces $\mathcal{S}$ as shown in Fig. \ref{fig:n4 choose gamma12} are no longer identical. The complete PLC \cite{Arora:2024edk} then generalizes to
\begin{equation}
    \label{eq:complexity general}
    \mathcal{C}(\mathcal{R}) \sim \left(\max_{\sigma}\right) \min_{\mathcal{S}}\max_{i>j} \min_{k}\left\{\exp\left(\frac{|\gamma_{j,k}^b| - |\gamma_i|}{8G_{\rm N}}\right)\right\},
\end{equation}
where we must now minimize over the possible choices of the set $\mathcal{S}$. For complete generality, I have also included the maximization over the choice of Cauchy slice $\sigma$ that must be done \cite{Brown:2019rox}; for the BL model this was ignored as the time reflection symmetry allowed us to work with a fixed slice.

It would also be interesting to compare the model dynamics to other (unitary) models of black hole evaporation. The octopus model of \cite{Akers:2019nfi} is not directly comparable due to the system dynamics differing; however the model could be modified to have dynamics similar to this model i.e. a fixed number of asymptotic AdS regions that exchange mass. It would be interesting to see if such dynamics still yield a Page curve for the \cite{Akers:2019nfi} model. \cite{Abdolrahimi_2019} provides an entirely different perspective, computing the back-reaction to the Schwarzschild metric from a stress tensor approximating the Hawking radiation, which forms a time-dependent metric for an evaporating black hole in an asymptotically flat spacetime. Applying the generalized entropy formula \eqref{eq:quantum rt} to the geometry can provide an alternate ansatz for computing the entanglement entropy, forming a useful comparison to the proposed BL model.

Furthermore, quantum corrections to the BL model could also be considered. \cite{gupta2025entangleduniverses} proposed a generalization of the RT formula to asymptotically flat spacetimes, but the question of how the complete quantum corrected formula generalizes remains open. Using such a formula for the proposed BL model may yield further insight into the evaporation dynamics, especially on the lack of a connected geometry for early times.

\acknowledgments

This work is a continuation of previous work in \cite{gupta2025entangleduniverses}. I'd like to thank my co-authors Matthew Headrick and Martin Sasieta for their useful feedback and collaboration. I'd also like to thank Ning Bao, Roberto Emparan, Netta Engelhardt, Guglielmo Grimaldi, Robie Hennigar, Gary Horowitz, Veronika Hubeny, Albion Lawrence, Pratik Rath, Brian Swingle, Zhencheng Wang, Chris Akers and Tom Faulkner for useful discussions.

\appendix

\bibliographystyle{jhep}
\bibliography{bibliography}

\end{document}